\documentclass[a4paper,11pt]{article}
\pdfoutput=1 

\usepackage{jheppub} 

\usepackage[T1]{fontenc} 

 \usepackage[disable]{todonotes}
\newcommand{\ckd}{\todo{checked :-)}}

\newcommand{\mA}{\mathcal{A}}

\newcommand{\mC}{\mathcal{C}}

\newcommand{\mS}{\mathcal{S}}

\newcommand{\mJ}{\mathcal{J}}

\newcommand{\mR}{\mathcal{R}}
\newcommand{\mT}{\mathcal{T}}
\newcommand{\mK}{\mathcal{K}}
\newcommand{\mN}{\mathcal{N}}
\newcommand{\mO}{\mathcal{O}}
\newcommand{\mW}{\mathcal{W}}

\newcommand{\mV}{\mathcal{V}}

\newcommand{\scz}{\text{\texttt{z}}}
\newcommand{\sct}{\text{\texttt{t}}}

\newcommand{\sctf}{\sct^+(z,x)}
\newcommand{\sctfm}{\sct^-(z,x)}
\newcommand{\tsctf}{\tilde{\sct}^+(\tz,\tx)}
\newcommand{\tsctfm}{\tilde{\sct}^-(\tz,\tx)}
\newcommand{\sctbdy}{\sct^{bdy}(x)}
\newcommand{\tz}{\tilde{z}}
\newcommand{\ttt}{\tilde{t}}
\newcommand{\tx}{\tilde{x}}
\newcommand{\txp}{\tilde{x}^+}
\newcommand{\txm}{\tilde{x}^-}

\newcommand{\Poincare}{Poincar\'e}

\usepackage{bbm}
\usepackage{physics}

\newcommand{\ep}{\sigma}

\newcommand{\qed}{\nobreak \ifvmode \relax \else\ifdim\lastskip<1.5em
\hskip-\lastskip\hskip1.5em plus0em minus0.5em \fi \nobreak\vrule height0.75em
width0.5em depth0.25em\fi}

\newcommand{\be}{\begin{eqnarray}}
\newcommand{\ee}{\end{eqnarray}}

\def\>{\rangle}
\def\<{\langle}

\newcommand{\executeiffilenewer}[3]{%
\ifnum\pdfstrcmp{\pdffilemoddate{#1}}%
{\pdffilemoddate{#2}}>0%
{\immediate\write18{#3}}\fi%
}

\usepackage{subcaption}
\graphicspath{{plots/}}

\usepackage{hyperref}
\hypersetup{
	pdftitle    = {WdW-patches in AdS3 and complexity change under conformal transformations II},
	pdfauthor   = {Mario Flory}
}

\title{\boldmath WdW-patches in AdS$_{3}$ and complexity change under conformal transformations II}

\author{Mario Flory}

\affiliation{Institute of Physics, Jagiellonian University, \\
\L{}ojasiewicza 11, 30-348 Krak\'ow, Poland}

\emailAdd{mflory@th.if.uj.edu.pl}

\abstract{We study the null-boundaries of Wheeler-de Witt (WdW) patches in three dimensional \Poincare-AdS, when the selected boundary timeslice is an arbitrary (non-constant) function, presenting some useful analytic statements about them. Special attention will be given to the piecewise smooth nature of the null-boundaries, due to the emergence of caustics and null-null joint curves. 
	This is then applied, in the spirit of one of our previous papers, to the problem of how the complexity of the CFT$_2$ groundstate changes under a small local conformal transformation according to the action (CA) proposal. In stark contrast to the volume (CV) proposal, where this change is only proportional to the second order in the infinitesimal expansion parameter $\sigma$, we show that in the CA case we obtain terms of order $\sigma$ and even $\sigma\log(\sigma)$. This has strong implications for the possible field-theory duals of the CA proposal, ruling out an entire class of them.}

\begin{document} 
\maketitle
\flushbottom

\setcounter{tocdepth}{1}
\thispagestyle{empty}

\section{Introduction}
\label{sec::Intro}

In the past years, a consensus has formed that quantum information theory has an important role to play in the understanding of the AdS/CFT correspondence and quantum gravity. One comparably recent avenue of study is the investigation of conjectured holographic measures of \textit{complexity}. From a quantum information theoretic perspective, the \textit{circuit complexity} of a unitary operator $U$ would be the minimal number of quantum-gates (picked from a given gateset) needed to implement the operation $U$ to within a specified error tolerance $\epsilon$. Similarly, the (relative) complexity of a state $\left|\psi_U\right\rangle$ with respect to the reference state $\left|\mR\right\rangle$, $\mC(\mR,\psi_U)$, would be identified with the minimal complexity of any operator $U$ that satisfies the equation 
\begin{align}
\left|\psi_U\right\rangle=U\left|\mR\right\rangle. 
\label{eqn:DefU}
\end{align}
In \cite{2005quant.ph..2070N,2006Sci...311.1133N}, a proposal was formulated to calculate the complexity of an operator $U$ in geometric terms, choosing a distance measure on the space of unitary operators and equating the complexity of $U$, $\mC(U)$, as the (minimal) distance between $U$ and the identity operator $\mathbbm{1}$ according to this distance function. 

Ideas relating to such notions of complexity entered the holography literature in \cite{Harlow:2013tf,Susskind:2013aaa,Susskind:2014rva}, see \cite{Susskind:2018pmk} for a recent overview. Curiously, there are more than one proposal for what bulk quantity might be a measure of complexity in AdS/CFT.  The first is the \textit{volume proposal} \cite{Susskind:2013aaa,Susskind:2014rva,Susskind:2014moa,Stanford:2014jda,Susskind:2014jwa},
according to which the complexity $\mC$ of a field theory state with a smooth holographic dual geometry should be measured by the volumes $\mV(\Sigma)$ of certain spacelike extremal codimension-one bulk hypersurfaces $\Sigma$, i.e.   
\begin{align}
\mC\propto \frac{\mV(\Sigma)}{LG_N},
\label{complexity}
\end{align} 
wherein a length scale $L$ has to be introduced into equation \eqref{complexity} for dimensional reasons which is usually picked to be the AdS scale \cite{Stanford:2014jda,Susskind:2014jwa,Brown:2015bva,Brown:2015lvg}. The second proposal is the \textit{action proposal} \cite{Brown:2015bva,Brown:2015lvg} 
\begin{align}
\mC=\frac{\mA(\mW)}{\pi\hbar}
\label{action}
\end{align}
wherein $\mA(\mW)$ is the bulk action over a certain (codimension zero) bulk region, the Wheeler-de Witt patch $\mW$. A third, less utilised proposal, is the \textit{volume 2.0} proposal of of \cite{Couch:2016exn}. It suggests that holographic complexity may be given by the volume of the WdW-patch,
\begin{align}
\mC\propto\mV(\mW).
\label{Vol2.0}
\end{align}

Despite sparking a flurry of activity from the AdS/CFT community, these  proposals are on much less firm ground as for example the famous RT and HRT proposals for holographic entanglement entropy, simply because in the case of complexity even the precise definition of the quantity of interest on the field-theory side is somewhat uncertain. 
However, some progress has been made to ease this predicament. Field theory techniques for defining and calculating complexity where investigated in \cite{Jefferson:2017sdb,Yang:2017nfn,Hackl:2018ptj,Ali:2018fcz,Chapman:2018hou,Magan:2018nmu,Khan:2018rzm,Caputa:2018kdj,Bhattacharyya:2018bbv,Guo:2018kzl} following the geometric ideas of 
\cite{2005quant.ph..2070N,2006Sci...311.1133N}, in \cite{Caputa:2017urj,Caputa:2017yrh,Czech:2017ryf,Bhattacharyya:2018wym} following path integral methods and in \cite{Yang:2018nda,Yang:2018tpo} following an axiomatic approach. 
A fascinating connection with group theory was investigated in \cite{Lin:2018cbk}. 
See also \cite{Chapman:2017rqy,Hashimoto:2017fga,Hashimoto:2018bmb,Abt:2017pmf,Abt:2018ywl,Yang:2018cgx} for other relevant works. 
Comparisons between field theory calculations and holographic calculations of complexity where attempted in \cite{Reynolds:2017jfs,Kim:2017qrq,Fu:2018kcp,Agon:2018zso,Flory:2018akz,Numasawa:2018grg}, however, in the holographic proposals \eqref{complexity}, \eqref{action}, and \eqref{Vol2.0} it is not clear what choices of reference state, gate set and error tolerance might be needed to fix ambiguities of the dual field theory definition of complexity. 
If a field theory definition of complexity corresponding to \eqref{complexity}, \eqref{action}, or \eqref{Vol2.0} is to depend on such choices, they appear to be \textit{implicit} in the holographic dictionary. This, and the fact that with currently developed techniques the calculation of complexity in field-theories sometimes requires the assumption of weak coupling or even free theories, the comparisons attempted in \cite{Reynolds:2017jfs,Kim:2017qrq,Fu:2018kcp,Agon:2018zso,Flory:2018akz,Numasawa:2018grg} are somewhat limited to a rather qualitative level. 

For example, in \cite{Flory:2018akz} we studied how in AdS$_3$/CFT$_2$ complexity, according to the volume proposal \eqref{complexity}, changes under infinitesimal local conformal transformations from the groundstate. The rationale behind this is that on the CFT side the conformal transformations can be written to be generated by unitary operators with a very simple form in terms of the Virasoro generators or the energy-momentum tensor, irrespectively of whether the central charge is large or not. 
Our hope was hence that for such a transformation, the holographic results on the change of complexity might be somewhat universal among $1+1$ dimensional CFTs, allowing for a potentially easier and more meaningful comparison to field theory models in which computations of complexity are possible. 
In fact, in \cite{Belin:2018bpg} a certain proposal was made for what the field theory definition of $\mC$ in \eqref{complexity}  should be, finding precise agreement with our results of \cite{Flory:2018akz}.\footnote{Another conjectured holographic dual of bulk volumes is the so called \textit{fidelity susceptibility} \cite{MIyaji:2015mia}, see however \cite{Moosa:2018mik} for a recent critique of this proposal.} 

The main goal of this paper is hence to extend the results of \cite{Flory:2018akz} from the volume proposal \eqref{complexity} to the action proposal \eqref{action}. Hence, we will calculate how the complexity of the state of a holographic two dimensional conformal field theory (CFT$_2$)  dual to \Poincare-AdS$_3$ changes under an infinitesimal local conformal transformation. 
The structure of our paper is as follows: 
In section \ref{sec::GS} we present in detail the calculation of complexity, according to the proposal \eqref{action}, for the case of \Poincare-AdS in $2+1$-dimensions. This serves as an introduction of some relevant concepts and notation, and will be used as a reference for our later more non-trivial results. Section \ref{sec::SGD} is devoted to an explanation of how we will study conformal transformations in AdS$_3$/CFT$_2$, following the lines of our previous paper \cite{Flory:2018akz}.
Our novel results then start in section \ref{sec::WdWs}, where we discuss the features of generic WdW-patches in  \Poincare-AdS$_3$. Based on this, we will then calculate contributions to the action on the WdW-patch term by term, starting with the bulk term in section \ref{sec::bulk}, and then moving on to the surface terms (section \ref{sec::surfaceresults}), the parametrization of the null-rays constituting the null-boundaries in section 
\ref{sec::affine}, joint-terms in sections \ref{sec::TNjoints} and \ref{sec::NNjoints}, and finally the so called counter terms in section \ref{sec::countersigma}. We close with a summary and conclusion in section \ref{sec::conc}.
Further technical details will be relegated to the appendices \ref{sec::equations} and \ref{sec::details}. 

\section{Complexity of the groundstate}
\label{sec::GS}

\subsection{WdW-patch}
\label{sec::WdWpatch}

We start by considering the vacuum state of a large-$c$ CFT$_2$ living in $1+1$ dimensional Minkowski space, with coordinates $t,x$ ($-\infty<t,x<+\infty$), respectively lightcone-coordinates  $x^\pm = t \pm x$. 
If this CFT has a holographic dual, the bulk geometry dual to the vacuum state will most easily be given by the \Poincare-patch of AdS$_3$:\footnote{In contrast to the notations and convention of \cite{Mandal:2014wfa,Flory:2018akz}, we are using a coordinate $z$ instead of $\lambda$, with $\lambda=1/z^2$.  }
\begin{align}
ds^2 =\frac{L^2}{z^2}\left(-dt^2+dx^2+dz^2\right)  =\frac{L^2}{z^2}\left(dz^2-dx^+dx^-\right).  
\label{Poincare} 
\end{align} 
Clearly, this metric is conformally flat, however, it has an asymptotic boundary at $z=0$ where one would commonly define a cutoff at $z=\epsilon$ with $\epsilon=const.\ll1$. 
In this section, we will revise the calculation of the complexity \eqref{action} for the state described by the metric \eqref{Poincare}, following the outline and conventions of \cite{Chapman:2016hwi}.  Readers who are already well familiar with this material may safely skip this section, however it will serve to setup our conventions and notation, and it will give us the opportunity to remark on a few details that will become important later on again. 

\begin{figure}[htb]
	\centering
	\def\svgwidth{0.4\columnwidth}
\executeiffilenewer{CD2.svg}{CD2.pdf}%
{inkscape -z -D --file=CD2.svg %
--export-pdf=CD2.pdf --export-latex}%
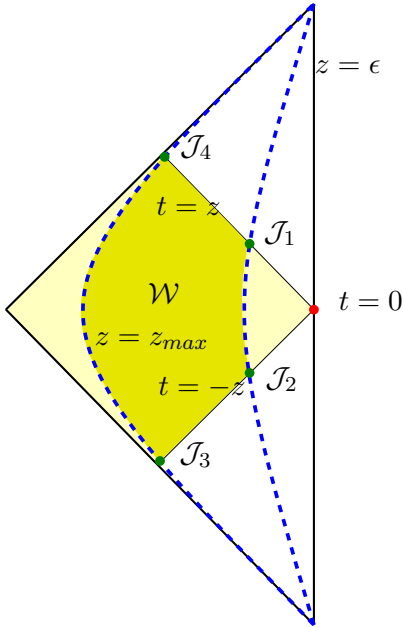%

	\caption{WdW-patch for the $t=0$ boundary slice in the \Poincare-patch. Technically, the WdW-patch would be the lightly shaded square region between the lightfronts $t=\pm z$ and the \Poincare-horizon. However, we introduce the field-theory UV cutoff $z=\epsilon$ and the IR cutoff $z=z_{max}$ near the \Poincare-horizon, shown as dashed (blue) lines. Hence, the integration-domain $\mW$ for the action proposal, which we will still refer to as WdW-patch, is the darkly shaded region. We also mark the locations of the four spacelike joints $\mJ_1$-$\mJ_4$.}
	\label{fig::WdW}
\end{figure}

First of all, what do we mean by ``$\mA$" in \eqref{action}? $\mA$ is meant to be \cite{Brown:2015bva,Brown:2015lvg} the integral of the bulk action over the \textit{Wheeler-de Witt (WdW) patch} $\mW$. This codimension-0 region of the bulk is defined as the region enclosed by future and past \textit{lightfronts}\footnote{We use this term instead of \textit{lightsheet}, as a priori the lightfronts bounding a WdW-patch do not have to satisfy the necessary requirements to be lightsheets according the the definition of \cite{Bousso:1999xy}.} emanating from a chosen equal-time slice on the asymptotic boundary. Consequently, the spacetime-points inside of $\mW$ are not in causal contact with the chosen boundary timeslice, while the points outside of $\mW$ can be reached by at least one causal curve from at least one point on the boundary slice. 
For the boundary time-slice $t\equiv0$, the WdW-patch in the \Poincare-patch is bounded by the \Poincare-horizon at $z\rightarrow\infty$ and by the two lightfronts
\begin{align}
\sct^+(z,x)=+z,\ \sct^-(z,x)=-z,
\end{align} 
to the future and the past, respectively. In order to avoid divergences, a cutoff surface has to be imposed near the boundary at $z=\epsilon$. Similarly, a cutoff can be imposed at $z=z_{max}$, with $z_{max}\rightarrow\infty$. As pointed out in \cite{Lehner:2016vdi}, it is generically \textit{not} possible to calculate the contributions from a null boundary to the action via a limiting procedure from family of timelike or spacelike boundaries, with the exception being the case where the null-boundary in question is a Killing horizon. This is the case for the \Poincare-horizon. 
Another intricacy arises in defining the WdW-patch in the presence of a small cutoff $\epsilon$, see appendix D.4 of \cite{Chapman:2016hwi} and also \cite{Carmi:2016wjl}. 
Roughly speaking, it might make a difference whether the null-boundaries of the WdW-patch are defined to emanate from the cutoff-surface at $z=\epsilon$, or whether they are defined to emanate from a time-slice of the exact asymptotic boundary at $z=0$, and are hence intersected by the cutoff surface. We will pick the latter convention, which appears to be the overall more common one in the literature. It was also shown in \cite{Chapman:2016hwi,Carmi:2016wjl} that, for many interesting questions, these two possible choices lead to the same results in the limit $\epsilon\rightarrow 0$. 
See figure \ref{fig::WdW} for an illustration of the WdW-patch.

As ultimately worked out in detail in \cite{Lehner:2016vdi}, the action is (see also \cite{Hilbert:1915tx,1917SPAW.......142E,PhysRevLett.28.1082,PhysRevD.15.2752,PhysRevD.47.3275,Brill:1994mb,Parattu:2015gga,Reynolds:2016rvl}, we mostly follow \cite{Chapman:2016hwi,Carmi:2016wjl,Reynolds:2016rvl}\footnote{See also footnote 7 in \cite{Chapman:2018dem} for a remark on the sign of the term $\propto\kappa$.})
\begin{align}
\mA=&\frac{1}{16\pi G_N}\int_{\mW} \left(R-2\Lambda\right)\sqrt{-g}d^3x 
\label{CAbulk}
\\
&+\frac{1}{8\pi G_N}\sum_{\mT_i}\int_{\mT_i} K \sqrt{-\gamma}d^2x+\frac{1}{8\pi G_N}\sum_{\mS_i}\int_{\mS_i} K \sqrt{\gamma}d^2x+\frac{1}{8\pi G_N}\sum_{\mN_i}\int_{\mN_i} \kappa d\lambda \sqrt{\rho}dx
\label{CAbdy}
\\
&+\frac{1}{8\pi G_N}\sum_{\mJ_i}\int_{\mJ_i} \eta_{\mJ_i} \sqrt{\rho}dx
\label{CAjoint}
\\
&+\frac{1}{8\pi G_N}\sum_{\mN_i}\int_{\mN_i} \theta \log(|\theta \ell_c|) d\lambda \sqrt{\rho}dx
\label{counter}
\end{align} 
where we have included the appropriate surface, boundary, joint and counter terms. Of course, $G_N$ stands for Newton's constant. This form of the action was derived by demanding not only a well-defined variational principle under Dirichlet boundary conditions, but also additivity of the action under joining of bulk regions and independence of the value of the action under reparametrisation of the generators of the null-boundaries.

We will now go through these terms one by one.

\subsection{Bulk term}
\label{subsec::bulk}

The bulk term is the integral of the bulk Einstein-Hilbert action \cite{Hilbert:1915tx,1917SPAW.......142E} over the codimension $0$ region $\mW$,
\begin{align}
\mA_{bulk}=\frac{1}{16\pi G_N}\int_{\mW} \left(R-2\Lambda\right)\sqrt{-g}d^3x, 
\end{align} 
with $\Lambda=-\frac{1}{L^2}$ in $2+1$ dimensions. The volume-element is as usual $\sqrt{-g}d^3x$ with $g$ being the determinant of the bulk metric. Due to \eqref{Poincare} being a vacuum solution of three dimensional gravity, the integrand reads $R-2\Lambda=\frac{-4}{L^2}$, with $L$ being the AdS radius which we will generally set equal to one in the later sections. We hence find \ckd
\begin{align}
\mA_{bulk}=\frac{-L}{4\pi G_N}\int_{\epsilon}^{z_{max}}dz \int_{-z}^{z}dt \int_{-\infty}^\infty dx \frac{1}{z^3}=\frac{-LV}{2\pi G_N}\left(\frac{1}{\epsilon}-\frac{1}{z_{max}}\right),
\label{actionresult}
\end{align} 
where we have set $\int_{-\infty}^\infty dx \equiv V$.

\subsection{Surface terms}
\label{sec::surface}

There are potentially three types of codimension-one surfaces, namely timelike ones $\mT_i$, spacelike ones $\mS_i$, and null ones $\mN_i$. From figure \ref{fig::WdW}, we see that we will have to deal with two null boundaries, two timelike boundaries, and no spacelike boundaries. Let us begin with the null ones, discussed only recently in \cite{Parattu:2015gga,Lehner:2016vdi,Chapman:2016hwi}. 
The null boundaries are generated by null rays with (possibly affine) parameter $\lambda$, and the measure $\sqrt{\rho}dx$ comes from integrating over all the different null rays constituting the lightfront. 
The integrand $\kappa$ is fixed by the equation \cite{Lehner:2016vdi,Chapman:2016hwi}
\begin{align}
k^\mu \nabla_\mu k_\nu \equiv \kappa k_\nu,
\label{kappa}
\end{align}
and measures the failure of $\lambda$ to be an affine parameter. Hereby, $k_\mu$ is the null normal to the lightfront, directed out of $\mW$. 
It is common to choose $k^\mu$ such that $\kappa=0$ and that $k\cdot \hat{t}\big|_{z=0}=\pm1$ (the sign depending on the orientation of $k$) where $\hat{t}=\partial_t=\delta^\mu_t\partial_\mu$ is a future pointing  vector at the boundary \cite{Lehner:2016vdi,Chapman:2016hwi,Carmi:2016wjl}.\footnote{In fact,  $\hat{t}=\delta^\mu_t\partial_\mu$ is a timelike Killing vector, which defines the units in which we measure boundary time. This kind of consideration also played a role in \cite{Yang:2016awy}.}
For the upper lightfront $\sct^+=z$, we find that $k_\mu dx^\mu\equiv d(t-\sct^+)=dt-dz$ has just the desired properties $\kappa=0$ and $k\cdot \hat{t}=1$. 
Similar statements hold for the past lightfront $\sct^-$.

The choice $\kappa=0$ clearly makes the contribution from the null-boundary vanish in \eqref{CAbdy}. 
We are hence left with the terms for the timelike boundaries, which are just the well known Gibbons-Hawking type boundary terms \cite{PhysRevLett.28.1082,PhysRevD.15.2752}. Then $\gamma$ is the determinant of the induced metric on the surface $z=const.$, and the extrinsic curvature can be easily calculated (see appendix \ref{sec::equations}) to be $K=\frac{2}{L}$\ckd at $z=\epsilon$ and $K=-\frac{2}{L}$ at $z=z_{max}$. Hence\ckd
\begin{align}
\mA_{surface}
&=\frac{V}{8\pi G_N}\left(\int_{-\epsilon}^{\epsilon} \frac{2}{L} \frac{L^2}{\epsilon^2} dt+\int_{-z_{max}}^{z_{max}} \frac{2}{L} \frac{L^2}{z_{max}^2} dt\right)=\frac{LV}{2\pi G_N}\left(\frac{1}{\epsilon}-\frac{1}{z_{max}} \right).
\end{align}

\subsection{Joint terms}
\label{sec::joints}

We are left with the four codimension-two joint terms $\mJ_1$-$\mJ_4$ that arise where two of the codimension-one boundaries come together \cite{Lehner:2016vdi,Carmi:2016wjl}:
\begin{align}
\mA_{joint}=\frac{1}{8\pi G_N}\sum_{i=1}^{4}\int_{\mJ_i} \eta_{\mJ_i} \sqrt{\rho}dx.
\end{align}
Herein $\sqrt{\rho}dx$ is the induced line-element on the joints $\mJ$, which are one-dimensional spacelike submanifolds. In principle, there might be timelike-timelike, spacelike-spacelike, timelike-spacelike, timelike-null, spacelike-null or even null-null joints. From figure \ref{fig::WdW}, it is apparent that so far we will only have to deal with timelike-null type joints. 
The integrand for this case is then defined as \cite{Lehner:2016vdi,Carmi:2016wjl}
\begin{align}
\eta_{\mJ_i} =-sign(k\cdot s)sign\left(k\cdot \check{t}\right)\log\left(|k\cdot s|\right),
\label{etaJ}
\end{align}
with the null normal $k_\mu$ defined in section \ref{sec::surface}, the unit normal form $s$ of the timelike boundary surface $\mT_i$ (defined to point out of $\mW$) and $\check{t}$, a normalized timelike vector living in the tangent space of the boundary $\mT_i$ and normal to the joint surface, pointing away from $\mW$. See e.g.~\cite{Carmi:2016wjl} for details, and note that $\check{t}\neq\hat{t}$.
Let us focus on the joint $\mJ_1$ first. We find $sign(k\cdot s)=+1,\ sign\left(k\cdot \check{t}\right)=+1$, $k\cdot s=\frac{\epsilon}{L}$ and $\sqrt{\rho}=L/\epsilon$, thus
\begin{align}
\mA_{\mJ_1}=\frac{V}{8\pi G_N}\frac{L}{\epsilon}\log\left(\frac{L}{\epsilon}\right).
\label{J1}
\end{align}
Similarly, at $\mJ_2$, $sign(k\cdot s)=+1,\ sign\left(k\cdot \check{t}\right)=+1$, $k\cdot s=\frac{\epsilon}{L}$ and 
\begin{align}
\mA_{\mJ_2}=\frac{V}{8\pi G_N}\frac{L}{\epsilon}\log\left(\frac{L}{\epsilon}\right).
\label{J2}
\end{align}
On the other hand, at both $\mJ_3$ and $\mJ_4$, we find $sign(k\cdot s)=-1,\ sign\left(k\cdot \check{t}\right)=+1$, $k\cdot s=-\frac{z_{max}}{L}$ and
\begin{align}
\mA_{\mJ_{3,4}}=\frac{-V}{8\pi G_N}\frac{L}{z_{max}}\log\left(\frac{L}{z_{max}}\right).
\label{J34}
\end{align}
Consequently\ckd
\begin{align}
\mA_{joint}=\frac{VL}{4\pi G_N}\left[\frac{1}{\epsilon}\log\left(\frac{L}{\epsilon}\right)-\frac{1}{z_{max}}\log\left(\frac{L}{z_{max}}\right)\right].
\label{vacJoints}
\end{align}

\subsection{Counter terms}
\label{sec::counter}

Lastly, we are dealing with the term
\begin{align}
\mA_{counter}=\frac{1}{8\pi G_N}\sum_{\mN_i}\int_{\mN_i} \theta \log(|\theta \ell_c|) d\lambda \sqrt{\rho}dx,
\label{counter2}
\end{align}
which has been introduced already in \cite{Lehner:2016vdi}, but the importance of which was pointed out in \cite{Reynolds:2016rvl} (see also \cite{Chapman:2018dem,Chapman:2018lsv} for the importance of these terms, but also \cite{Jubb:2016qzt}).
Again, the null boundaries are generated by null rays with affine (due to $\kappa=0$, see section \ref{sec::surface}) parameter $\lambda$, and the measure $\sqrt{\rho}dx$ comes from integrating over all the different null rays constituting the lightfront.
The reason why these terms are called \textit{counter-terms} is that they make sure that the value of the action remains the same under reparametrisations of the affine parameter $\lambda$ parametrising the lightrays that make up the null boundaries \cite{Lehner:2016vdi,Reynolds:2016rvl}. As we can see, however, this comes at the price of introducing an arbitrary lengthscale $\ell_c$.\footnote{One might be tempted to set this lengthscale equal to the AdS-scale $L$ as e.g.~\cite{Reynolds:2016rvl}, however in general this tends to simplify the results for complexity almost \textit{too} much. So we leave $\ell_c$ to be arbitrary in this paper.}

With the equations of appendix \ref{sec::equations} in mind, we could now proceed to directly evaluate \eqref{counter2}, however, we will first simplify the expression following \cite{Reynolds:2016rvl}. To do so, we remind ourselves firstly that the \textit{expansion} $\theta$ is given by (see appendix \ref{sec::equations})
\begin{align}
\theta=\frac{1}{\sqrt{\rho}}\partial_\lambda \sqrt{\rho}.
\label{theta1}
\end{align}
Hence \eqref{counter2} can be rewritten as
\begin{align}
\mA_{counter}&=\frac{1}{8\pi G_N}\int_{\lambda_{min}}^{\lambda_{max}} \int_{-\infty}^{+\infty} \left(\partial_\lambda \sqrt{\rho}\right) \log(|\theta \ell_c|) d\lambda dx
\\
&=\frac{1}{8\pi G_N}\int_{-\infty}^{+\infty}\left[\sqrt{\rho}\log(|\theta \ell_c|) \right]\Big|_{\lambda_{min}}^{\lambda_{max}}dx
-\frac{1}{8\pi G_N}\int_{\lambda_{min}}^{\lambda_{max}} \int_{-\infty}^{+\infty} \frac{\partial_\lambda\theta}{\theta}d\lambda \sqrt{\rho}dx.
\end{align}
This is as far as \cite{Reynolds:2016rvl} went, but we can make an additional step by using Raychaudhuri's equation \eqref{R}, which in a $2+1$-dimensional vacuum bulk-spacetime boils down to $\frac{\partial_\lambda\theta}{\theta}=-\theta$, and hence, using \eqref{theta1} again, 
\begin{align}
\mA_{counter}&=\frac{1}{8\pi G_N}\int_{-\infty}^{+\infty}\left[\sqrt{\rho}\log(|\theta \ell_c|) \right]\Big|_{\lambda_{min}}^{\lambda_{max}}dx
+\frac{1}{8\pi G_N}\int_{\lambda_{min}}^{\lambda_{max}} \int_{-\infty}^{+\infty} \partial_\lambda \sqrt{\rho}d\lambda dx
\\
&=\frac{1}{8\pi G_N}\int_{-\infty}^{+\infty}\left[\sqrt{\rho}\log(|\theta \ell'_c|) \right]\Big|_{\lambda_{min}}^{\lambda_{max}}dx,
\label{newcounter}
\end{align}
where we have redefined the arbitrary lengthscale $\ell_c$ such that $\log(\ell_c)+1=\log(\ell_c')$. We have now achieved to rewrite the term \eqref{counter2} as a term to be evaluated solely on the joints $\mJ_i$. 

For the upper lightfront, $\sct^+=z$ and $k_\mu dx^\mu\equiv d(t-\sct^+)=dt-dz$. The surfaces of constant $\lambda$ along this lightfront are codimension 2 spacelike slices defined by $t\equiv z\equiv const.$ with induced line-element $\sqrt{\rho}=L/z$. Hence (see \cite{Reynolds:2016rvl} and appendix \ref{sec::equations}) $\lambda=L^2/z$ and $\theta=z/L^2$. The upper lightfront gives the term
\begin{align}
\mA_{counter,+}=\frac{VL}{8\pi G_N}\left[\frac{1}{\epsilon}\log\left(\frac{\ell_c'\epsilon}{L^2}\right)-\frac{1}{z_{max}}\log\left(\frac{ \ell_c'z_{max}}{L^2}\right)\right].
\label{counter3}
\end{align}
A similar contribution comes from the lower lightfront $\sct^-=-z$: 
Here $k_\mu dx^\mu\equiv d(t-\sct^-)=-dt-dz$, $\sqrt{\rho}=L/z$, $\lambda=L^2/z$ and $\theta=z/L^2$. Hence
\begin{align}
\mA_{counter,-}=\frac{VL}{8\pi G_N}\left[\frac{1}{\epsilon}\log\left(\frac{\ell_c'\epsilon}{L^2}\right)-\frac{1}{z_{max}}\log\left(\frac{ \ell_c'z_{max}}{L^2}\right)\right].
\label{counter4}
\end{align}

\subsection{End result}

Taking the results from the previous subsections together, we find\footnote{See also \cite{Alishahiha:2018lfv} for related, but more general results.}
\begin{align}
\mA&=\frac{VL}{4\pi G_N}\left[\frac{1}{\epsilon}\log\left(\frac{\ell_c'}{L}\right)-\frac{1}{z_{max}}\log\left(\frac{\ell_c'}{L}\right)\right],
\label{endresult}
\end{align} 
which is exclusively given by $\mA_{joint}$ and $\mA_{counter,\pm}$, as the bulk and surface terms cancel precisely. Furthermore, it is noteworthy that even without this cancellation, all the terms involving $z_{max}$ vanish independently in the limit $z_{max}\rightarrow\infty$. This is a consequence of the special properties of the \Poincare-horizon, which when mapping the WdW-patch to global AdS would collapse to a pair of lightrays, emanating from what would be the ``point at infinity" for the \Poincare-patch. Similarly, the joints $\mJ_3$ and $\mJ_4$, after taking the limit $z_{max}\rightarrow\infty$, would be mapped to two caustic \textit{points} which do not contribute to the action, see appendix B of \cite{Chapman:2016hwi}.
Usually, it would not be consistent to calculate the contributions from a null surface by taking a limit of spacelike surfaces, however in the case of the \Poincare-horizon, this is possible \cite{Lehner:2016vdi}. 
Following \cite{Reynolds:2016rvl}, it is also interesting to point out that due to the inclusion of the counter terms \eqref{counter}, the overall result \eqref{endresult} diverges only as $\frac{V}{\epsilon}$ with the $x$-Volume $V$ and the UV-regulator $\epsilon$, as opposed to a divergence $\frac{V}{\epsilon}\log(\epsilon)$ indicated by \eqref{vacJoints}. This however comes at the price of introducing the ambiguous lengthscale $\ell_c'$.

\section{Conformal transformations in AdS$_3$/CFT$_2$}
\label{sec::SGD}

\subsection{Solution generating diffeomorphisms}
\label{sec::SGDsub}

Let us revise some of the details about how to implement local conformal transformations in AdS$_3$/CFT$_2$, discovered in \cite{Banados:1998gg}, but following the outline and notation of  \cite{Mandal:2014wfa,Flory:2018akz}. 
We start with equation \eqref{Poincare}. Local conformal transformations can now be implemented by applying global bulk diffeomorphisms which act nontrivially near the boundary \cite{Banados:1998gg}, see also \cite{Mandal:2014wfa}. These diffeomorphisms map solutions of the equations of $2+1$ dimensional AdS gravity to new solutions which will be physically inequivalent, hence describing distinct CFT-states. 
They can thus be called \textit{solution generating diffeomorphisms (SGDs)} \cite{Mandal:2014wfa}. 
For example, holographically calculating the expectation value of the energy-momentum tensor of the boundary theory by the method of \cite{Balasubramanian:1999re} after applying an SGD will give a result different from zero (which we would get from the metric \eqref{Poincare}), which however agrees with the formula for the energy-momentum tensor of a CFT after a conformal transformation due to the Schwarzian derivative \cite{Mandal:2014wfa}. 
The resulting metrics, due to their discovery in \cite{Banados:1998gg}, are called \textit{Ba\~{n}ados geometries} and have been studied in more detail for example in  \cite{Mandal:2014wfa,Sheikh-Jabbari:2014nya,Sheikh-Jabbari:2016unm,Sheikh-Jabbari:2016znt,Bhaseen:2013ypa,Erdmenger:2017gdk}. 

The SGDs are of course only defined up to a residual diffeomorphism which is trivial at the boundary. Following \cite{Mandal:2014wfa}, we will write them as\footnote{This is different from the convention used in \cite{Banados:1998gg}. The convention used here and in  \cite{Mandal:2014wfa,Flory:2018akz}, while leading to a somewhat more involved expression for the line element, has the advantage of presenting the SGDs in a simpler form. This will not affect our physical endresults. } 
\begin{align}
z &=\tilde{z}\sqrt{G'_+(\tilde x^+) G'_-(\tilde x^-)},
\label{SGD1}\\
x^+ &= G_+(\tilde x^+), \\
x^- &= G_-(\tilde x^-),
\label{SGD3}
\end{align}
where $G_\pm$ are some functions with $G'_\pm>0$. The line element in the new coordinates $\tilde{z},\tilde{x}^\pm$ is \cite{Mandal:2014wfa}\ckd
\begin{align}
ds^2 &= \frac{1}{\tz^2} d\tz^2 -\frac{1}{\tz^2} d\tilde x^+\cdot d\tilde x^-+\left( A_+ d\tilde x^+ + A_- d\tilde x^-\right)^2 - \frac{2}{\tz}   d\tz \cdot \left(A_+ d\tilde x^+ +A_- d\tilde x^-  \right),
\label{afterSGD}
\\
A_\pm &= -\frac{1}{2}\frac{G_\pm{}''(\tilde x^\pm)}{G_\pm{}'(\tilde x^\pm)}.
\end{align}

\begin{figure}[htb]
	\centering
	\def\svgwidth{0.4\columnwidth}
\executeiffilenewer{ConformalDiagram.svg}{ConformalDiagram.pdf}%
{inkscape -z -D --file=ConformalDiagram.svg %
--export-pdf=ConformalDiagram.pdf --export-latex}%
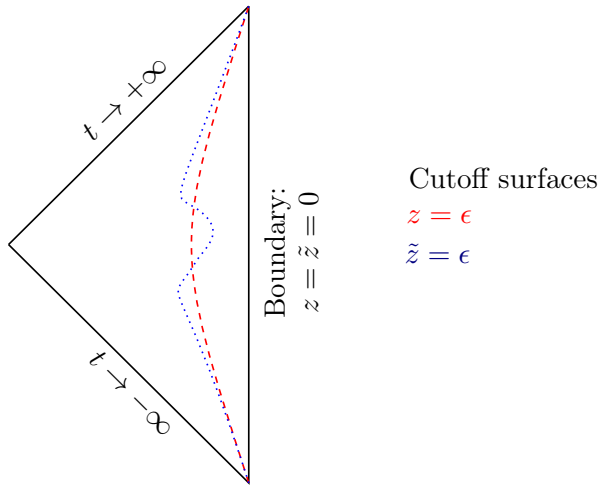%

	\caption{A conformal diagram of the \Poincare-patch of AdS$_3$. The vertical line is the asymptotic boundary while the two diagonal lines are the two \Poincare-horizons where $t\rightarrow\pm\infty$. The two cutoff surfaces $z=\epsilon$ and $\tilde{z}=\epsilon$ are shown as dashed (red) and dotted (blue) lines, respectively. The figure is taken from \cite{Flory:2018akz}.}
	\label{fig:CutOffSurfaces}
\end{figure}

There are two possible equivalent viewpoints from which we can approach these geometries. The first would be to just take \eqref{afterSGD} and treat it like any other solution in holography. In order to calculate quantities like the expectation value of the energy-momentum tensor or entanglement entropies, we would introduce the natural
\todo{Remember \cite{Brown:1986nw,vanAlbada:2016nss}} 
cutoff $\tz=\epsilon$ with $\epsilon=const.\ll1$. We call this choice of cutoff \textit{natural} because $\tz$ takes the role of the (inverse) radial coordinate in \eqref{afterSGD}, and the induced line element on the cutoff surface reads
\begin{align}
ds^2_{ind}=-\frac{1}{\epsilon^2} d\tilde x^+\cdot d\tilde x^-+\mO(\epsilon^0),
\label{firstinduced}
\end{align}
i.e.~the dual CFT lives on flat space. By \eqref{SGD1}, this coice of cutoff would correspond to deforming the cutoff in the old coordinates:
\begin{align}
\tilde z=\epsilon \Leftrightarrow
z = \epsilon\sqrt{G_+{}'(\tilde{x}^+) G_-{}'(\tilde{x}^-)}.
\label{cutoff}
\end{align}
This motivates the second (equivalent) perspective that we can take, namely that in the old coordinates of the \Poincare-patch, the  SGDs actively shift the position of the cutoff surface according to \eqref{cutoff}, which in the holographic calculation of CFT quantities then leads to the changes expected for a conformal transformation \cite{Mandal:2014wfa}.\footnote{Something similar happens in AdS$_2$-holography: There, the family of physically inequivalent solutions to the bulk equations is given by the set of curves defining different cutoff-surfaces near the boundary of AdS$_2$ \cite{Maldacena:2016upp}.
} 
This is shown in Figure \ref{fig:CutOffSurfaces}. In the coordinates of \eqref{Poincare}, the induced line element on this cutoff surface \eqref{firstinduced} then reads
\begin{align}
ds^2_{ind}=-\frac{G^{(-1)}_+{}'(x^+) G^{(-1)}_-{}'(x^-)}{\epsilon^2} d x^+ d x^-+\mO(\epsilon^0)=
-\frac{1}{\epsilon^2}\frac{d\tx^+}{dx^+}\frac{d\tx^-}{dx^-} d x^+ d x^-+\mO(\epsilon^0),
\end{align}
which is of course consistent with the way the metric transforms under conformal transformations, acquiring an overall prefactor.
Throughout this paper, we will switch between these two perspectives, depending on what is easier for the given task at the time.

Following \cite{Flory:2018akz}, we will again consider a small SGD
\begin{align}
x^+=G_+(\tilde x^+) &= \tilde x^+ +\ep ~ g_{+}(\tilde x^+) ,  
\label{smallSGD1}
\\
x^-=G_-(\tilde x-) &= \tilde x^- +\ep ~  g_{-}(\tilde x^-),
\label{smallSGD2}
\end{align}
with the expansion parameter $\ep \ll 1$. Just as in \cite{Flory:2018akz}, we will throughout the paper assume that the functions $g_\pm$ as well as their derivatives are smooth, bounded, and fall off to zero at infinity. 
The line-element \eqref{afterSGD} can similarly be expanded, yielding\ckd
\begin{align}
ds^2 &= \frac{1}{\tz^2}\left(-d\ttt^2+d\tx^2+d\tz^2\right)\label{metric} \\
&~~~+\frac{\ep}{\tilde z} \Big[ \Big(g_{+}''\left(\tilde{t}+\tilde{x}\right)+g_{-}''\left(\tilde{t}-\tilde{x}\right)\Big)d\tilde t
+\Big(g_{+}''\left(\tilde{t}+\tilde{x}\right)-g_{-}''\left(\tilde{t}-\tilde{x}\right)\Big)d\tilde x\Big]d\tilde z +\mO(\ep^2),
\nonumber
\end{align}
where we have switched from lightcone coordinates $\tilde{x}_\pm$ to standard coordinates $\tilde{t},\tilde{x}$ on the boundary. In this paper, as in \cite{Flory:2018akz}, we will be interested in terms up to and including order $\mO(\ep^2)$, however we have not written out the terms of this order in the line-element above as they are rather cumbersome. It is a trivial exercise to derive them from \eqref{afterSGD}.

\subsection{Towards complexity change under conformal transformations}
\label{sec::towards}

The SGDs \eqref{SGD1}-\eqref{SGD3} not only wrap the cutoff-surface as explained in section \ref{sec::SGD} and sketched in figure \ref{fig:CutOffSurfaces}, they also lead to a change of the definition of equal-time slice, as clearly $t\equiv const.$ and $\ttt\equiv const.$ are two inequivalent conditions. Our goal is to holographically calculate the complexity of the state after applying an SGD, which is naturally understood to live on an equal time-slice of the new coordinates, $\ttt\equiv \ttt_0=const.$ How will this time-slice look like in the old, untilded coordinates? 

In general, it will not be possible to exactly invert the transformations in \eqref{smallSGD1}, \eqref{smallSGD2}. However, when working perturbatively in $\ep$, we can make use of the inverse transformations\ckd
\begin{align}
\tx^+&=G^{(-1)}_+(x^+) \approx  x^+ -\ep ~ g_{+}(x^+)+\ep^2g_{+}(x^+)g'_{+}(x^+)+\mO(\ep^3) ,  
\label{smallSGD1inv}
\\
\tx^-&=G^{(-1)}_-(x^-) \approx x^- -\ep ~  g_{-}(x^-)+\ep^2g_{-}(x^-)g'_{-}(x^-) +\mO(\ep^3).
\label{smallSGD2inv}
\\
\tz &= z \sqrt{G^{(-1)}_+{}'(x^+) G^{(-1)}_-{}'(x^-)}.
\label{smallSGD3inv}
\end{align}

Consequently, the equal-time boundary-slice in the new coordinates, $\ttt=\frac{1}{2}(\txp+\txm)\equiv t_0, \tz=0$, when mapped back to the old \Poincare-patch coordinates takes the (approximate) form\ckd 
\begin{align}
\sct^{bdy}(x)&=t_0+\frac{\ep}{2}\left[g_+(t_0+x)+g_-(t_0-x)\right]
\label{tbdygen}
\\
&-\frac{\ep^2}{4}\left[g_-(t_0-x)-g_+(t_0+x)\right]\left[g'_-(t_0-x)-g'_+(t_0+x)\right]+\mO(\ep^3), 
\nonumber
\\
z&=0.
\end{align}
From now on, unless explicitly specified otherwise, we will generally assume  
\begin{align}
t_0\equiv0.
\label{t0}
\end{align}
Given the time-translation invariance of the background \eqref{Poincare} from which we start, this is possible without loss of generality. However, in order to simplify our calculations, we will also generally assume 
\begin{align}
g_-(t_0-x)=g_+(t_0+x),
\label{gpm}
\end{align}
which yields
\begin{align}
\sct^{bdy}(x)&=t_0+\ep g_+(t_0+x)+\mO(\ep^3).
\label{tbdy}
\end{align}
This now sets the stage for what we have to do in the rest of the paper. In order to compute the change of the complexity \eqref{action} due to an infinitesimal conformal transformation, we have to calculate the WdW-patch for the state after the transformation. This could be tried in the tilded coordinates, where the line-element is given by \eqref{metric}. 
We would then be faced with the task of solving for generic lightcones or null geodesics in such a metric with $\ttt$ and $\tx$-dependent components. An alternative approach would be to work in the old coordinates, where the background spacetime \eqref{Poincare} is manifestly conformally flat. In this setup, we hence know all lightcones and null-geodesics trivially, however we will need to find the WdW-patch for a boundary-slice of the form \eqref{tbdy}. This is indeed what we will do in the following sections.

\section{General WdW-patches in AdS$_3$}
\label{sec::WdWs}

Due to its definition, which inherently relates the shape of the WdW-patch $\mW$ to the causal connectivity of the spacetime in question, the boundary of $\mW$ will, apart from cutoff surfaces which we have artificially introduced or bulk-horizons, consist of null surfaces generated by lightrays emanating from the boundary slice, see figure \ref{fig::WdW}. 
How can we calculate these null-surfaces for a general boundary slice like \eqref{tbdy}? Assuming that in the coordinates of \eqref{Poincare} (with $L=1$ from now on), the future\footnote{Of course the treatment of the past boundary will be almost identical, so we will not spell it out in every step in the following.}  null-boundary can be expressed as a function $t=\sct^+(z,x)$, we can easily calculate the induced metric on such a general surface. As a \textit{null} surface, the determinant of this metric should then vanish, 
and demanding this leads to the PDE
\begin{align}
(\partial_z\sct^+(z,x))^2+(\partial_x\sct^+(z,x))^2=1.
\label{pde}
\end{align}
This will be the central equation defining the null-boundaries of $\mW$ in the \Poincare-patch, subject to the boundary condition
\begin{align}
\sct^+(0,x)=\sct^{bdy}(x).
\label{pdebdy}
\end{align}
A similar but more cumbersome equation can be derived for the embedding $\ttt=\tilde{\sct}^+(\tz,\tx)$ of the lightfront in the tilded coordinates. In appendix \ref{sec::details}, we will give a numerical scheme for obtaining solutions and a discussion of some generic properties of such solutions. Here, we will just state some of the most important observations for later.

First of all, in the case $\sct^{bdy}(x)=0$, equation \eqref{pde} is trivially solved by the lightfronts $\sct^+(z,x)$ found in section \ref{sec::GS}. These lightfronts are well behaved all the way from the boundary to the \Poincare-horizon. However, for general boundary conditions $\sct^{bdy}(x)$, equation \eqref{pde} does only allow for \textit{piecewise} smooth solutions. 
A physicist's proof for this can be given by the use of the focusing theorem, which generically implies \textit{caustics} to emerge at finite $z\sim\mO(1/\ep)$, see the discussion in appendix \ref{sec::caustics}. These caustics will be the starting point of null-null joints, where two piecewise smooth parts of the function $\sct^+(z,x)$ will meet in a non-smooth manner. These joints will then give rise to extra contributions to the action, which we will discuss in sections \ref{sec::NNjoints} and \ref{sec::counterjoint}. 

\begin{figure}[htbp]
	\centering
	\includegraphics[width=0.46\textwidth]{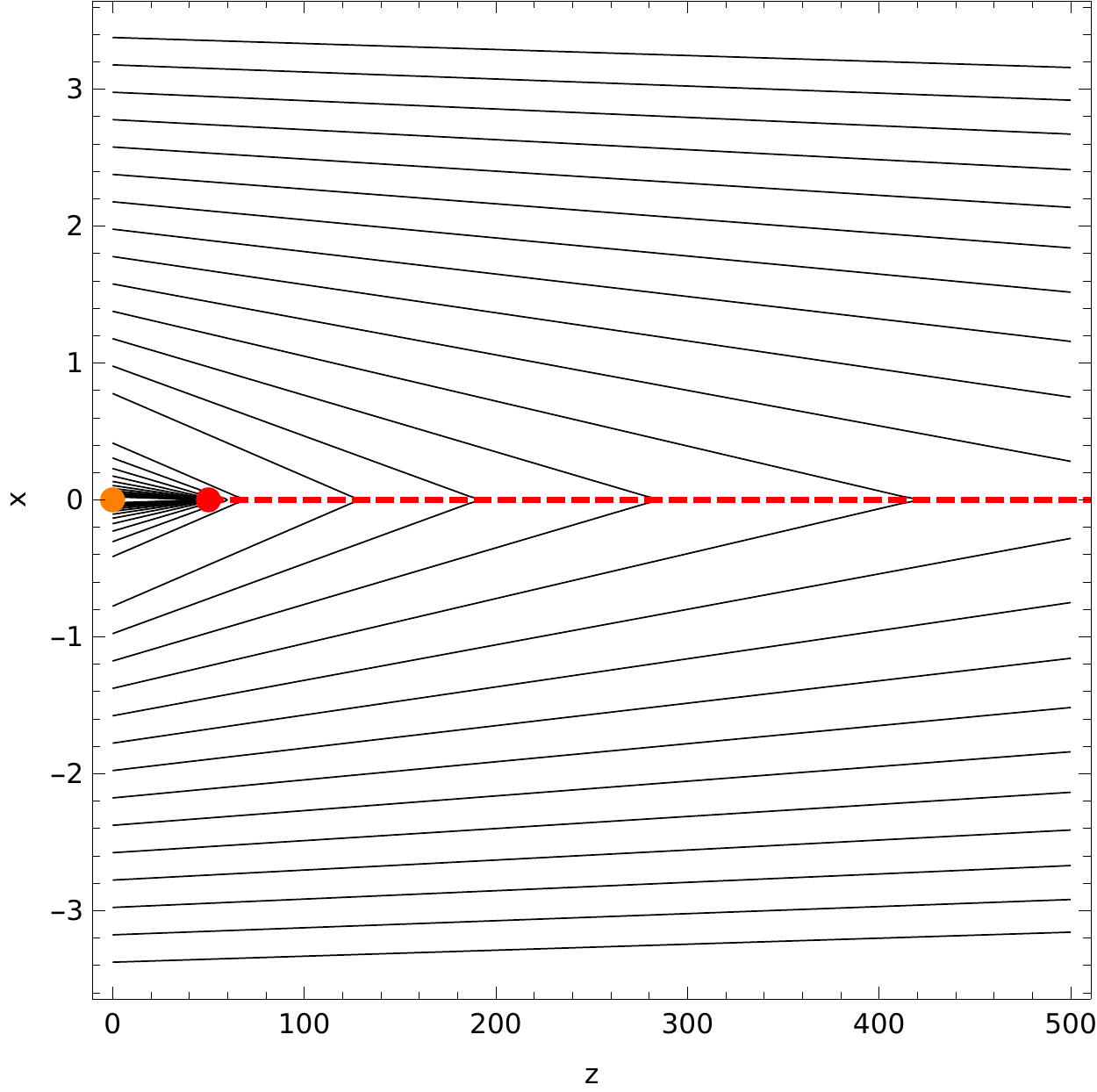}
	\includegraphics[width=0.46\textwidth]{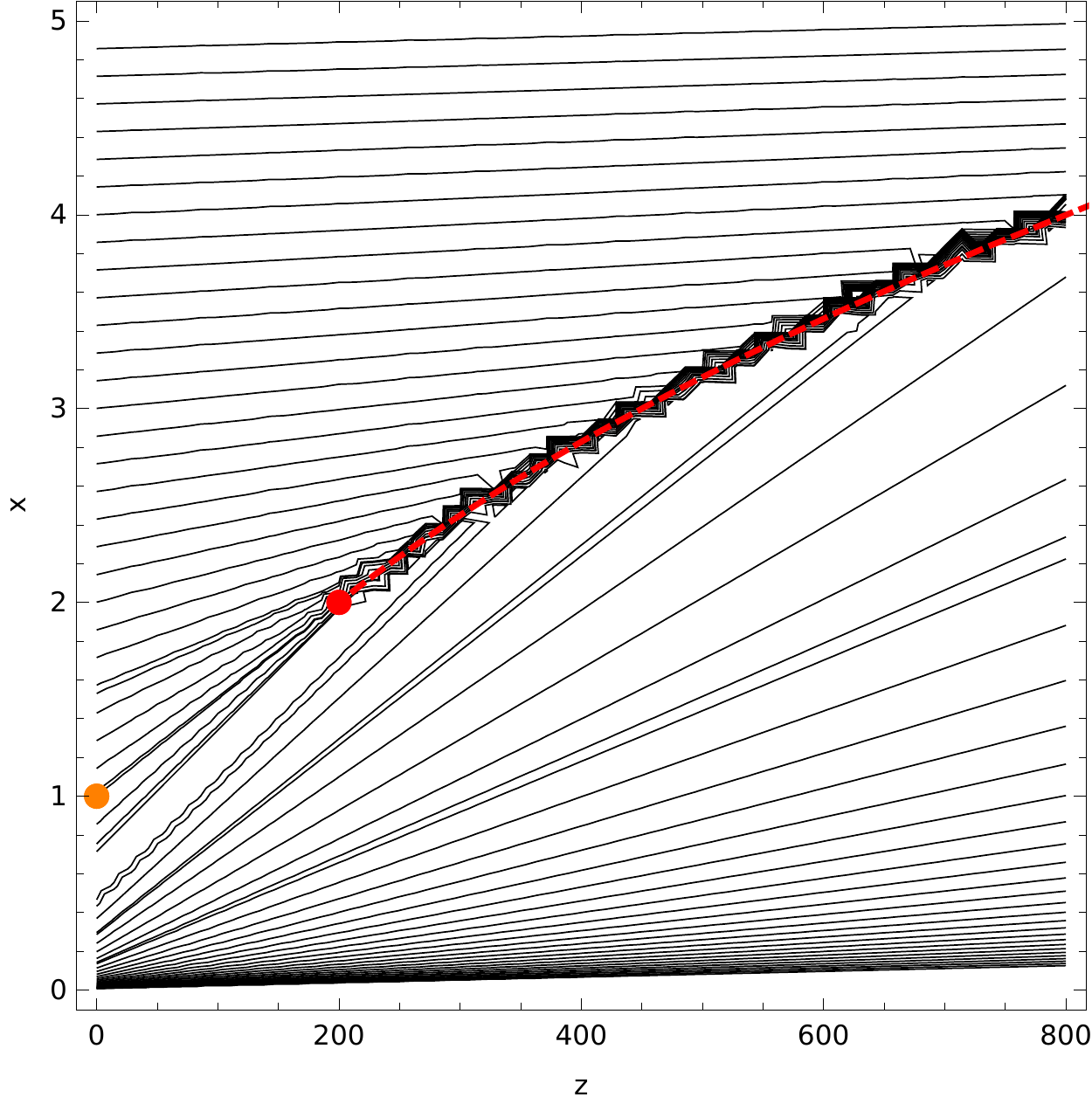}
	\\
	\includegraphics[width=0.46\textwidth]{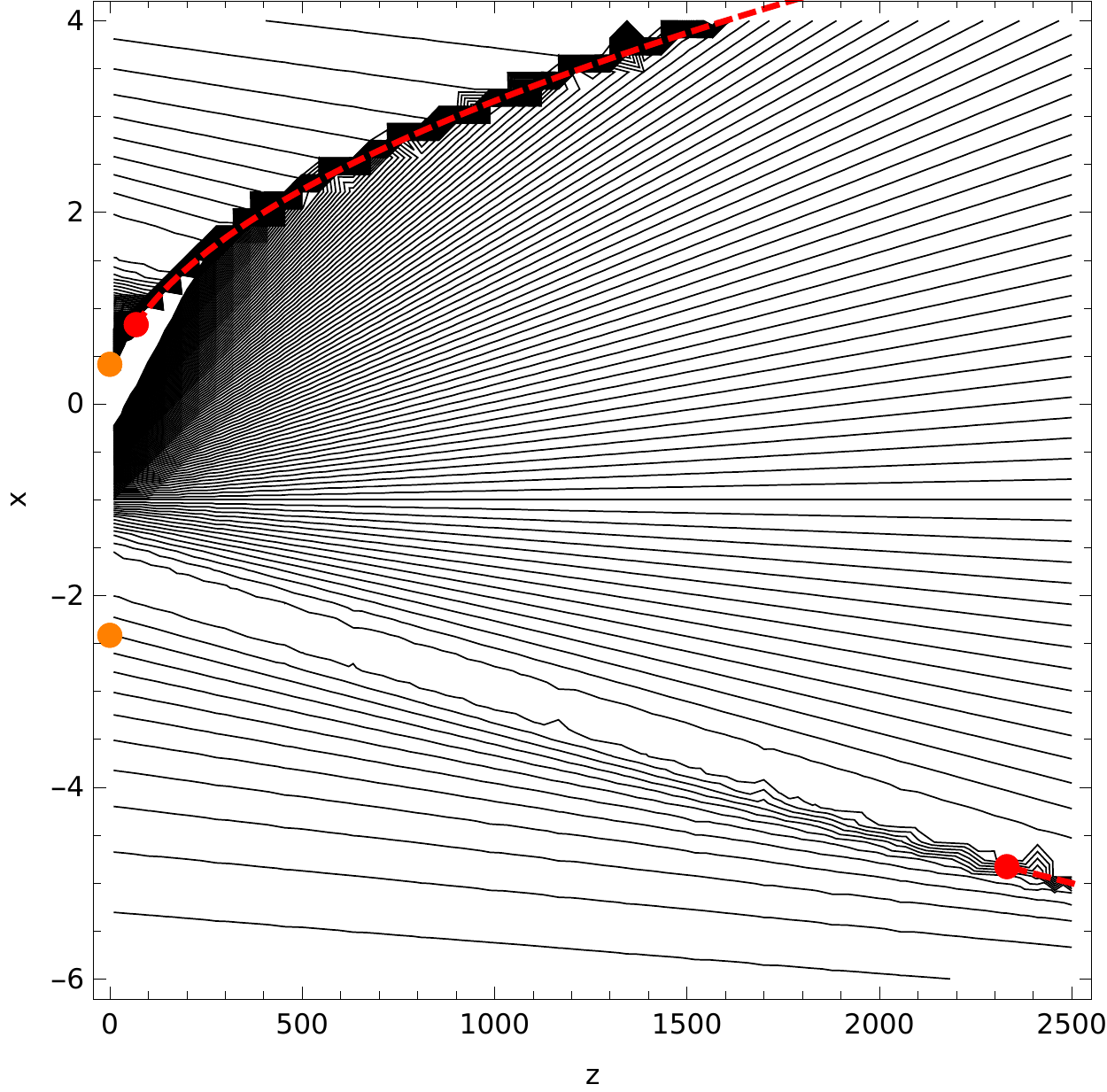}
	\includegraphics[width=0.46\textwidth]{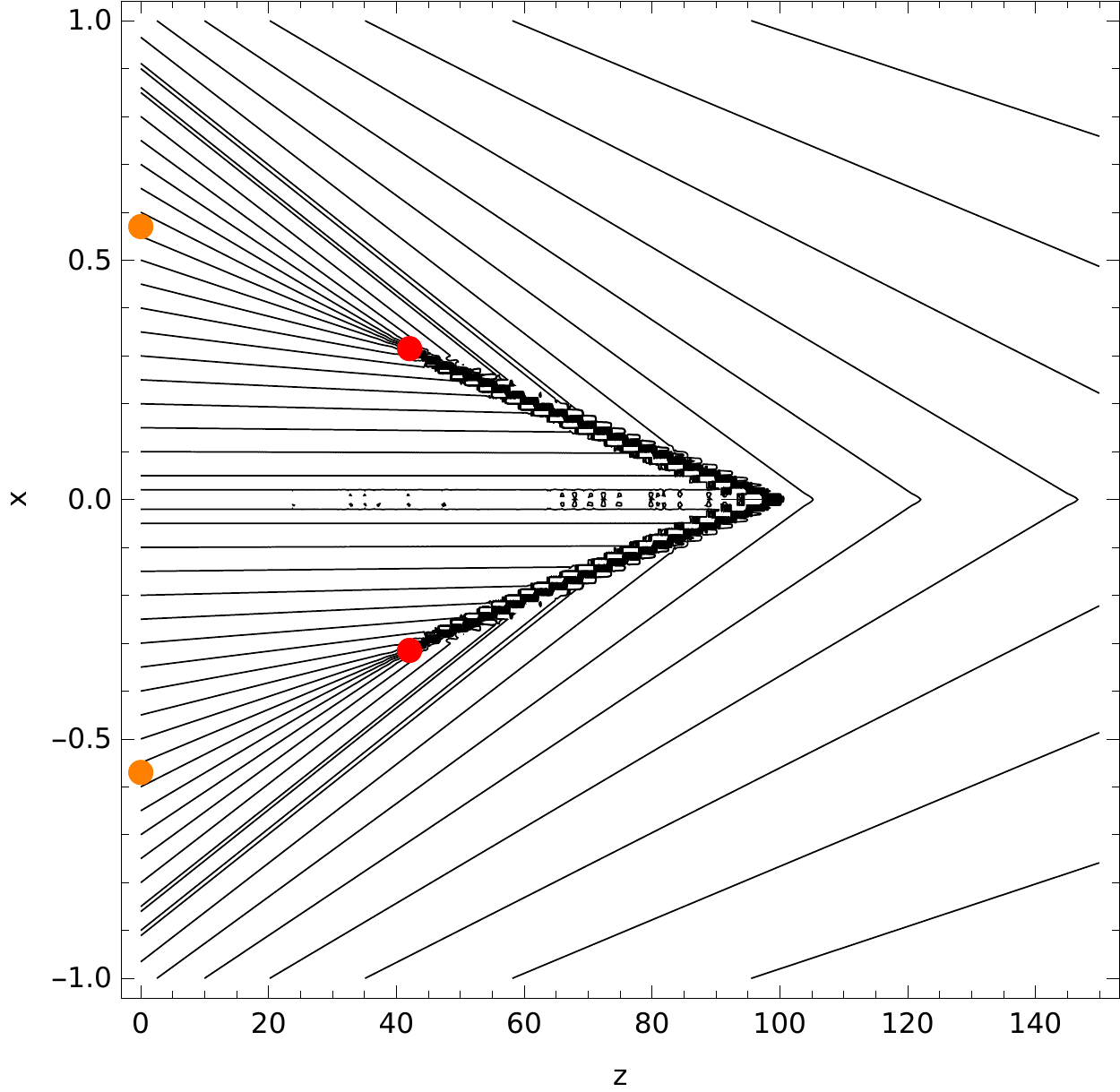}
	\caption{Contour plots for the functions $\partial_z\sct^+(z,x)$ respectively $\partial_x\sct^+(z,x)$ (up to numerical errors, the contours for both expressions are identical) for various boundary conditions $\sctbdy$.	 \textit{Top left}: $\sctbdy=\frac{0.01}{1+x^2}$. \textit{Top right}: $\sctbdy=\frac{-0.01}{1+x^2}$. \textit{Bottom left}: $\sctbdy=\frac{0.01\cdot x}{1+x^2}$. \textit{Bottom right}: $\sctbdy=\frac{0.01}{1+x^4}$. 
		The black lines are projections of the null rays forming the lightfront down to the $x,z$-plane, and should hence be perfectly straight. Any deviation from straight line behaviour is due to numerical inaccuracies. 
		The orange points at the boundary ($z=0$) are what we called \textit{hyperbolic points} in section \ref{sec::caustics}, while the red points in the bulk are \textit{caustics}, which are generated by the hyperbolic points. These caustics are generally the starting point of \textit{creases} or \textit{null-null joints} on which the function $\sctf$ is not smooth (leading to increased numerical problems). Starting from a caustic, these creases will extend from there towards the \Poincare -horizon. Those creases that we could determine analytically are marked by a dashed red line, see the discussion later in section \ref{sec::creases}. In the case shown on the bottom right, we see that generically, creases may collide and merge into one. }
	\label{fig::rays}
\end{figure}

Secondly, due to the conformal flatness of \eqref{Poincare}, the lightrays that foliate the surface $\sctf$ are straight lines of unit slope in the coordinate system spanned by $t,z,x$. Hence, \ckd along each of these lightrays, the expressions $\partial_z\sct^+(z,x)$ and $\partial_x\sct^+(z,x)$ will be constant. 
Drawing the lines in the $z,x$-plane along which these quantities are constant\footnote{For example using the \texttt{ContourPlot[...]} command of \textit{Wolfram Mathematica}.} will hence be an easy method to draw the projections to the $z,x$-plane of the lightrays which foliate the null front, given a numerical solution of $\sctf$. In figure \ref{fig::rays}, we show the corresponding figures for some simple choices of $\sct^{bdy}(x)$. 

Thirdly, apart from numerical approaches, we can also try to solve \eqref{pde} iteratively in $\ep$, starting with the $\ep=0$ result $\sctf=+z$. To second order, this yields\footnote{Similar expansions of general lightfronts in the $z$ coordinate were done for example in \cite{Carmi:2016wjl,Reynolds:2016rvl,Moosa:2017yiz}.}
\begin{align}
\sct^+(z,x)&\approx +z+\sct^{bdy}(x) -\frac{1}{2} z \sct^{bdy}{}'(x)^2+\mO(\ep^3),
\label{tseries}
\\
\sct^-(z,x)&\approx -z+\sct^{bdy}(x) +\frac{1}{2} z \sct^{bdy}{}'(x)^2+\mO(\ep^3).
\label{tseriesm}
\end{align}
Hereby, we have assumed $t_0=0$ and hence $\sct^{bdy}(x)\sim\mO(\ep)$, $\sct^{bdy}{}'(x)^2\sim\mO(\ep^2)$, see \eqref{tbdy}.
As we already pointed out in section \ref{sec::GS}, it is generally not correct to evaluate the terms of the action coming from null boundaries by a limiting procedure of boundary-terms on space- or timelike surfaces. 
Similarly, we cannot evaluate such null-boundary terms directly from the approximate solutions \eqref{tseries}, \eqref{tseriesm}, however, in the calculation of the bulk term and timelike boundary terms near the asymptotic boundary this approximation will be useful later on. 
It should also be pointed out that \eqref{tseries} takes on a series-expansion form not only in $\ep$, but also in $z$. This stays true even in higher orders. In fact, it is clear that even with arbitrarily higher order terms in $\ep$, the expression \eqref{tseries} will have a finite convergence radius in $z$ for fixed $\sct^{bdy}(x)$. The reason for this is that in the iterative procedure for deriving the terms of \eqref{tseries} for any additional order of $\ep$, the resulting term will always be smooth by construction as long as 
$\sct^{bdy}(x)$ is smooth. However, as discussed above and in appendix \ref{sec::caustics}, the focusing theorem implies that even for smooth but otherwise generic $\sct^{bdy}(x)$, the function $\sctf$ cannot be smooth for large enough $z$. This is also clearly visible in figure \ref{fig::rays}. Hence expressions of the form \eqref{tseries} can only be a good approximation close to the boundary. 
As the caustics will only appear at coordinates of order $z\sim1/\ep$, we will from now on assume the solutions \eqref{tseries}, \eqref{tseriesm} to be valid up to $z\lesssim\mO\left(1/\ep\right)$.

In the following sections, we will now evaluate the action on the WdW-patch after a conformal transformation perturbatively in $\ep$ up to second order, subject to the simplifying assumptions \eqref{gpm} and $t_0=0$. Whenever possible, we will try to work with analytical expressions as much as possible, only using numerical solutions of $\sctf$ for specific examples of $\sct^{bdy}(x)$ when necessary. As mentioned already earlier, we will switch between the coordinate systems of \eqref{Poincare} and \eqref{metric} depending on what is more convenient in the given situation.

\section{Bulk term}
\label{sec::bulk}

To calculate the bulk term of \eqref{CAbulk}, we need to know the surfaces by which the WdW-patch $\mW$ is bounded. To the future and the past, this will be the lightfronts $\sct^\pm(z,x)$, which we can calculate numerically as explained in section \ref{sec::WdWs} and appendix \ref{sec::details}, and for which we also possess the approximate solutions \eqref{tseries}, \eqref{tseriesm} valid close to the boundary, for coordinates $z\lesssim\mO\left(1/\ep\right)$. By our assumptions, the function $\sctbdy$ is bounded and fluctuates around $t_0=0$ with an amplitude of order $\ep$, so $|\sctbdy|\leq A\ep$ with some $\mO(1)$ constant $A$. Consequently, due to causality, we know 
\begin{align}
z+A\ep&\geq\sct^+(z,x)\geq z-A\ep,
\label{b1}
\\
-z+A\ep&\geq\sct^-(z,x)\geq -z-A\ep,
\label{b2}
\end{align}
for any $z$. This will be of use shortly.

\begin{figure}[htb]
	\centering
	\def\svgwidth{0.9\columnwidth}
\executeiffilenewer{bounds2.svg}{bounds2.pdf}%
{inkscape -z -D --file=bounds2.svg %
--export-pdf=bounds2.pdf --export-latex}%
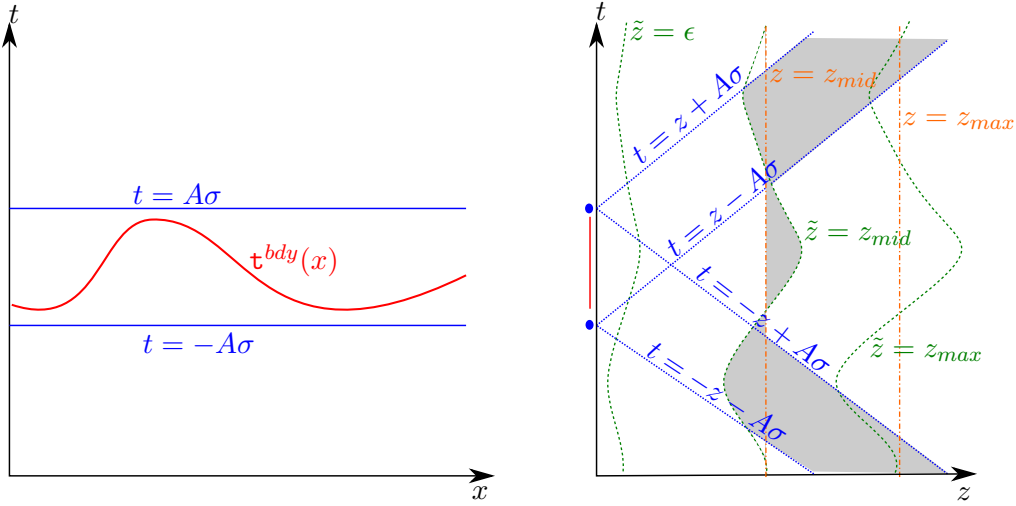%

	\caption{Bounds relevant for the calculation of the bulk integral, not to scale. Left: Asymptotic boundary. Right: A $x=const.$ slice of the bulk, in \Poincare-patch coordinates of \eqref{Poincare}. }
	\label{fig::bounds}
\end{figure}

Towards the asymptotic boundary, $\mW$ will be bounded by the cutoff surface $\tz=\epsilon$ ($\epsilon\ll1$), as explained in section \ref{sec::SGDsub}, see also figure \ref{fig:CutOffSurfaces}. This surface will be timelike (i.e.~$1+1$-dimensional), and is most conveniently described in the new, tilded, coordinates. In section \ref{sec::WdWpatch}, see also figure \ref{fig::WdW}, we introduced a timelike IR-cutoff surface $z=z_{max}$ near the \Poincare-horizon. 
In the tilded coordinates, it might now seem most natural to employ a cutoff-surface $\tz=z_{max}$, however a problem arises here: Because of the relation \eqref{SGD1}, we know that in the original \Poincare-coordinates, a surface $\tz=const.$ will fluctuate, and the magnitude of these fluctuations will be $\propto const.$, see figure \ref{fig::bounds} for an illustration. For some $const.\sim\mO(1/\ep)$, these fluctuations will become so strong that the surface defined by $\tz=const.$ is not everywhere timelike anymore. So instead of $\tz=z_{max}$, we will introduce an IR-cutoff surface at $z=z_{max}$, even for the cases after a conformal transformation. This is no problem, as we are only interested in taking the limit $z_{max}\rightarrow\infty$, and as in this limit the IR cutoff-surfaces approach the \Poincare-horizon, we expect that the end result will be independent of the specific family of cutoff surfaces with which this limit was taken \cite{Lehner:2016vdi}. 
We will also introduce a $z_{mid}\sim\mO(1/\ep)$, 
which we assume to be small enough such that the series expansions of \eqref{tseries}, \eqref{tseriesm} still is a good approximation up to this point.     

To summarise, for the calculation of the bulk term, we take the WdW-patch $\mW$ to be bounded by the surfaces $\tz=\epsilon,t=\sct^+,t=\sct^-,z=z_{max}.$ Furthermore, we split the integration domain into two parts, $\mW=\mW_1+\mW_2$, where $\mW_1$ is bounded by the surfaces $\tz=\epsilon,t=\sct^+,t=\sct^-,\tz=z_{mid}$ and $\mW_2$ is bounded by the surfaces $\tz=z_{mid},t=\sct^+,t=\sct^-,z=z_{max}$, with $\epsilon\ll1$, $z_{mid}\sim\mO(1/\ep)$, and $z_{max\rightarrow\infty}$. Clearly then
\begin{align}
\mA_{bulk}(\mW)=\mA_{bulk}(\mW_1)+\mA_{bulk}(\mW_2).
\end{align}

We will first look at the term $\mA_{bulk}(\mW_1)$. This will be easiest to do in the tilded coordinates, as then the integration bounds $\tz=\epsilon$ and $\tz=z_{mid}$ will not depend on the other coordinates, see figure \ref{fig:CutOffSurfaces} and \eqref{cutoff}. The approximate expressions for the lightfronts are given in \Poincare-coordinates in \eqref{tseries}, \eqref{tseriesm}, but they can just as well be calculated in tilded coordinates. The result is a little bit more cumbersome, and given in equation \eqref{tildetseries} of appendix \ref{sec::aux}. We are dealing with vacuum solutions of Einstein's equations, hence $R-2\Lambda=-4$ (setting $L=1$) exactly, and from \eqref{metric} one can show $\sqrt{-g}\approx \frac{1}{\tz^3}+\mO(\ep^3)$. Consequently
\begin{align}
\mA_{bulk}(\mW_1)=\frac{-1}{4\pi G_N}\int_{\epsilon}^{\tz_{max}}d\tz \int_{-\infty}^\infty d\tx\int_{\tsctfm}^{\tsctf}d\ttt \frac{1}{\tz^3}
=\frac{-1}{4\pi G_N}\int_{\epsilon}^{z_{mid}}d\tz \int_{-\infty}^\infty d\tx \frac{\tsctf-\tsctfm}{\tz^3}.
\end{align} 
Expanding $\tsctf-\tsctfm$ in $\ep$, we find that the $\mO(\ep^0)$-term is identical to \eqref{actionresult} under the replacement $z_{max}\rightarrow z_{mid}$. As can be seen from \eqref{tildetseries}, the $\mO(\ep^1)$-term of $\tsctf-\tsctfm$ vanishes identically. The $\mO(\ep^2)$-term of $\tsctf-\tsctfm$ is more complicated, and so for the moment we obtain
\begin{align}
\mA_{bulk}(\mW_1,\ep)&=\mA_{bulk}(\mW_1, 0)+\frac{-1}{4\pi G_N}\int_{\epsilon}^{z_{mid}}d\tz \int_{-\infty}^\infty d\tx \frac{\tsctf-\tsctfm-2\tz}{\tz^3}
\label{bulkInt}
\\
&\equiv \mA_{bulk}(\mW_1,0)+\ep^2\mA_{bulk, 1}^{(2)}+\mO(\ep^3).
\end{align} 
A series expansion of $\tsctf-\tsctfm-2\tz$ in $\tz$ shows that the term $\mA_{bulk, 1}^{(2)}$ will not contribute any divergences in the limit $\epsilon\rightarrow\infty$. This is as good as our general approach gets. For specific examples similar to the ones evaluated in \cite{Flory:2018akz}, we find (keeping in mind \eqref{gpm} and \eqref{t0} and taking $\epsilon\rightarrow0$)\ckd
\begin{align}
g_{+}(\tilde x^+)=\frac{a\cdot c}{a^2+\tx^2}\ \ \Rightarrow\ \ \mA_{bulk, 1}^{(2)}=\frac{-1}{4\pi G_N}\frac{3c^2\pi}{8|a|^3z_{mid}}+\mO(z_{mid}^{-5}),
\label{bulkex1}
\\
g_{+}(\tilde x^+)=\frac{ c\cdot\tx}{a^2+\tx^2}\ \ \Rightarrow\ \ \mA_{bulk, 1}^{(2)}=\frac{-1}{4\pi G_N}\frac{3c^2\pi}{8|a|^3z_{mid}}+\mO(z_{mid}^{-5}).
\label{bulkex2}
\end{align} 

As explained above, we assume $z_{mid}\sim\mO(1/\ep)$, and hence the combination $\ep^2\mA_{bulk, 1}^{(2)}$ does in general not contribute at order $\mO(\ep^2)$. Consequently, up to and including second order in $\ep$,
\begin{align}
\mA_{bulk}(\mW_1,\ep)\approx\mA_{bulk}(\mW_1,0),
\end{align} 
at least for the examples studied above. We still need to calculate the term $\mA_{bulk}(\mW_2)$, or more specifically the difference
\begin{align}
\mA_{bulk}(\mW_2,\ep)-\mA_{bulk}(\mW_2, 0)=+\frac{-1}{4\pi G_N}\int_{\tz=z_{mid}}^{z=z_{max}}dz \int_{-\infty}^\infty dx \frac{\sctf-\sctfm-2z}{z^3},
\end{align}
which we have now spelled out in (untilded) \Poincare-coordinates. Again, we will argue that this does not contribute at order $\mO(\ep^2)$, in the following way: As said above, the region $\mW_2$ is bounded by the surfaces $\tz=z_{mid},t=\sct^+,t=\sct^-,z=z_{max}$. When replacing $\mA_{bulk}(\mW_2,\ep)$ with $\mA_{bulk}(\mW_2,\ep=0)$, we are instead integrating (the same integrand) over the region bounded by the surfaces $z=z_{mid},t=+z,t=-z,z=z_{max}$. How big is the error that we make by changing the integral bounds? This can be estimated by integrating over the gray-shaded areas in figure \ref{fig::bounds}. Due to the bounds \eqref{b1}, \eqref{b2}, the error $E_1$ introduced by replacing $t=\sctf$ with $t=+z$ and $t=\sctfm$ with $t=-z$ is at most of order
\begin{align}
E_1\propto2\int_{z_{mid}}^{z_{max}\rightarrow\infty}\frac{2A\ep}{z^3}\propto \frac{\ep}{z_{mid}^2}\sim\mO(\ep^3).
\end{align}
Similarly, the error $E_2$ due to integrating from $z=z_{mid}$ instead of $\tz=z_{mid}\Leftrightarrow z=z_{mid}/ \sqrt{G_+{}'(x^+) G_-{}'(x^-)}$ (where we have used \eqref{smallSGD3inv}) is estimated by\footnote{Below, we do not specify the integral bounds in the $\int dt$ integral explicitly, but it is enough to know that by \eqref{b1}, \eqref{b2}, $|t|\lesssim\mO(z_{mid})$. The dependence of the exact integration bounds on the other coordinates does not play a role to lowest order in $\ep$, so we can assume that the integration bounds of the $t$-integral are independent of $x$ and $z$ below.}
\begin{align}
E_2&\propto\int dt \int_{-\infty}^{\infty} dx \int_{z=z_{mid}}^{z=z_{mid}/\sqrt{G_+{}'(x^+) G_-{}'(x^-)}} dz \frac{1}{z^3}
\\
&\propto\int dt \int_{-\infty}^{\infty} dx \frac{\ep}{ z_{mid}^2}\left(g_+'(t+x)-g_-'(x-t)\right)
\\
&\propto\int dt  \ep^3\left(g_+(t+x)-g_-(x-t)\right)\Big|_{x=-\infty}^{x=\infty}.
\end{align}
The last expression vanishes identically, due to our assumption that the functions $g_\pm$ fall off to zero at infinity (see section \ref{sec::SGDsub}). To summarise, we find
\begin{align}
\mA_{bulk}(\mW_2,\ep)\approx\mA_{bulk}(\mW_2,0)
\end{align} 
and consequently
\begin{align}
\mA_{bulk}(\mW,\ep)\approx\mA_{bulk}(\mW_1,\ep)\approx\mA_{bulk}(\mW,0)
\label{bulkresult}
\end{align} 
up to and including $\mO(\ep^2)$ for the examples studied in \eqref{bulkex1}, \eqref{bulkex2}. 
This leads us to the first main result of this paper: For the action proposal \eqref{complexity}, we will still have to take into account the remaining terms \eqref{CAbdy}, \eqref{CAjoint}, \eqref{counter}, however for the volume 2.0 proposal of \cite{Couch:2016exn}, \eqref{Vol2.0}, the result \eqref{bulkresult} is all we need. 
As the gravitational Lagrangian of our spacetime was constant, $R-2\Lambda=-4$, we find $\mA_{bulk}(\mW)\propto\mV(\mW)$. Hence, we have shown that the complexity, according to \eqref{Vol2.0}, does not change under infinitesimal conformal transformations up to order $\mO(\ep^2)$ for the examples studied above. For general $g_+$, there may be a change of order $\mO(\ep^2)$, independent of the UV-cutoff $\epsilon$, that can be calculated by the integral in \eqref{bulkInt}, using \eqref{tildetseries}.

\section{Timelike surface terms}
\label{sec::surfaceresults}

Next we turn to the timelike boundary terms which, as explained in the previous section, we evaluate at the UV and IR cutoff surfaces $\tz=\epsilon$ ($\epsilon\ll1$) and $z=z_{max}$ ($z_{max}\rightarrow\infty$). 
The term at $z=z_{max}$ is the easiest to deal with, which we do in \Poincare-coordinates. Then, just as in section \ref{sec::surface}, we find $K=-2$ and $\sqrt{\gamma}=1/z_{max}^2$. So
\begin{align}
\mA_{surface, IR}\propto \int_{-\infty}^{+\infty} dx \int_{\sct^-(z_{max},x)}^{\sct^+(z_{max},x)} dt\frac{1}{z_{max}^2}
=\int_{-\infty}^{+\infty} dx \left(\frac{2}{z_{max}}+\mO\left(\frac{\ep}{z_{max}^2}\right)\right),
\end{align}
where in the last step we have used the bounds \eqref{b1}, \eqref{b2}. So we see that in the limit $z_{max}\rightarrow\infty$, the variation of the term $\mA_{surface, IR}$ vanishes, just as the $\mO(\ep^0)$ result, which we discussed in section \ref{sec::GS}.

Next we turn to the term to be evaluated at the UV cutoff $\tz=\epsilon\ll1$. The trace of the extrinsic curvature at this surface is $K=2$, independently of $\ep$. The reason for this is simple: The holographic energy-momentum tensor of the dual theory is calculated by the famous equation \cite{Balasubramanian:1999re}
\footnote{Compared to \cite{Balasubramanian:1999re}, we changed the sign of the extrinsic curvature, to conform with our conventions of appendix \ref{sec::equations}.}
\begin{align}
8\pi G_N T_{ij}=\lim_{\epsilon\rightarrow0}\left(-K_{ij}+K\gamma_{ij}-\gamma_{ij}\right).
\end{align}
Now, taking the trace and ensuring $T=0$ for the CFT even after a conformal transformation is equivalent to demanding $K=2+\mO(\epsilon)$, independently of $\ep$. The induced metric and volume element on this surface read
\begin{align}
\gamma_{ij}d\tx^i d\tx^j=\frac{1}{\epsilon^2}\left(-d\ttt^2+d\tx^2\right)+\mO(\ep^2)  ,\  \sqrt{\gamma}=\frac{1}{\epsilon^2}-\frac{\ep^2}{2}g_+''(\tx-\ttt)g_+''(\ttt+\tx)+\mO(\ep^3) . 
\end{align}
Consequently
\begin{align}
\mA_{surface, UV}=&+\frac{1}{8\pi G_N}\int_{-\infty}^{+\infty}d\tx\int_{\tilde{\sct}^-(\epsilon,\tx)}^{\tilde{\sct}^+(\epsilon,\tx)} 2\left(\frac{1}{\epsilon^2}-\frac{\ep^2}{2}g_+''(\tx-\ttt)g_+''(\ttt+\tx)+\mO(\ep^3)\right) d\ttt
\\
&=\mA_{surface, UV}(\ep=0)+\frac{1}{8\pi G_N}\int_{-\infty}^{+\infty}d\tx\ \mO\left(\epsilon\ep^2\right),
\end{align}
where in the last step we have made use of $\tilde{\sct}^+(\epsilon,\tx)-\tilde{\sct}^-(\epsilon,\tx)=\mO(\epsilon)$ (see \eqref{tildetseries}) and the mean value theorem for definite integrals. As $\epsilon\ll1$, we drop all terms of order $\epsilon$, and consequently we see that up to and including order $\mO(\ep^2)$ the divergent (and finite) contribution from $\mA_{surface, UV}$ does not change.

\section{Affine parametrisation of lightrays and normalisation}
\label{sec::affine}

In order to compute the remaining terms, namely the null-surface term, the joint terms and the counter terms, we need to discuss the normalisation of the null normals $k_\mu$ for the lightfronts in question. Without loss of generality, we will focus on the future lightfront, described by the function $\sctf$ in \Poincare-coordinates. Generalising section \ref{sec::surface}, the null-normal $k_\mu$ is given by the equation
\begin{align}
\Phi^{-1}(\ep,t,x,z)k_\mu dx^\mu\equiv d(t-\sctf)=dt-\partial_z \sctf dz-\partial_x \sctf dx.
\label{kmu}
\end{align}
Herein, the function $\Phi(\ep,t,x,z)$ is meant to allow for general local rescalings which of course don't affect the orthogonality of $k_\mu$ to the lightfront or the condition $k_\mu k^\mu=0$, which is equivalent to 
\begin{flalign}
&&(\partial_z\sct^+(z,x))^2+(\partial_x\sct^+(z,x))^2=1. \text{\hspace{4cm} \eqref{pde}}
\nonumber
\end{flalign}
We now have to plug \eqref{kmu} into the equation 
\begin{flalign}
&&k^\mu \nabla_\mu k_\nu \equiv \kappa k_\nu, \text{\hspace{5.5cm} \eqref{kappa}}
\nonumber
\end{flalign}
in order to calculate $\kappa$. Ideally, we would like to be able to set $\kappa=0$, as was also the case in section \ref{sec::surface}. Calculating the Christoffel symbols and covariant derivative in \Poincare-coordinates is an easy exercise, and in fact in the special case $\Phi=1$ we find $\kappa=0$ as a consequence of \eqref{pde} and \eqref{kmu}. In the more general case, we obtain (again using \eqref{pde})
\begin{align}
\kappa=0\ \Leftrightarrow\ \partial_t\Phi(\ep,t,x,z)+\partial_z\sctf \partial_z\Phi(\ep,t,x,z) + \partial_x\sctf \partial_x\Phi(\ep,t,x,z)=0
\label{Phi1}
\end{align} 
Interestingly, there is a large class of general solutions to this equation: If the function $\Phi(\ep,t,x,z)$ only depends on the coordinates $x,z$ via the expressions $\partial_z\sctf,\partial_x\sctf$, i.e.~$\Phi(\ep,t,x,z)=\Phi\left(\ep,\partial_z\sctf,\partial_x\sctf\right)$, then \eqref{Phi1} is implied to vanish identically by \eqref{pde}. So, in a vector-like notation with coordinates $t,x,z$ (in that order), we obtain $\kappa=0$ for
\begin{align}
k_\mu=\Phi\left(\ep,\partial_z\sctf,\partial_x\sctf\right)
\left(
\begin{array}{c}
1 \\
-\partial_x \sctf\\ 
-\partial_z \sctf \\
\end{array}
\right),
\label{kmuresult}
\end{align}
where the remaining function $\Phi(\ep,\cdot,\cdot)$ is still up to our choice. Hence, just as in section \ref{sec::surface}, we will have a vanishing null-surface term,
\begin{align}
\mA_{surface,\ \mN_i}&=\frac{1}{8\pi G_N}\sum_{\mN_i}\int_{\mN_i} \kappa d\lambda \sqrt{\rho}dx
=0.
\end{align}
The result \eqref{kmuresult} is also important because it only depends on the coordinates via the expressions $\partial_z\sctf$, $\partial_x\sctf$, and as discussed in section \ref{sec::WdWs}, these expressions will be \textit{constant} along any lightray that foliates the lightfront. Hence, in \Poincare-coordinates, the components $k_\mu$ of the null normal will be constant along each lightray. Remember that it was the projections of these lightrays to the $x,z$-plane which the plots in figure \ref{fig::rays} show. 
Consequently, even though we do not know the function $\sctf$ analytically for too large coordinates of $z$, as long as we know where the lightray in question starts at the boundary, we can use the approximate solution \eqref{tseries} to calculate the components $k_\mu$ within order $\mO(\ep^2)$ in the region near the boundary. This will be of use later in sections \ref{sec::NNjoints} and \ref{sec::countersigma}.

In section \ref{sec::surface}, we had fixed the overall normalisation of $k_\mu$ by demanding $k\cdot \hat{t}\big|_{z=0}=1$ where $\hat{t}$ is a future pointing  vector at the boundary 
\cite{Lehner:2016vdi,Chapman:2016hwi,Carmi:2016wjl}.\footnote{Of course, the presence of the counter terms \eqref{counter} is designed to make the action reparametrisation invariant \cite{Lehner:2016vdi,Reynolds:2016rvl}, but fixing a specific parametrisation is still convenient in practice.}	
In our more general setting, we will take $\hat{t}=\partial_{\ttt}=\delta^\mu_{\ttt}\partial_\mu$ to be the future pointing vector at the boundary $\tz=z=0$. Ensuring $k\cdot \hat{t}\big|_{z=\tz=0}=1$ then fixes our choice of $\Phi$ as a function of $\ep$ and $x$ at the boundary.
As we know that $\Phi$ has to be constant along each of the lightrays due to \eqref{kmuresult}, $\Phi$ can then be extended from the boundary into the bulk. So at $z=0$, we make the ansatz
\begin{align}
k_\mu\big|_{z=0}\approx\hat{\Phi}\left(\ep,x\right)
\left(
\begin{array}{c}
1 \\
-\ep g_+'(x)\\ 
-1 +\frac{1}{2}\ep^2g_+'(x)^2 \\
\end{array}
\right),
\label{kmuatbdy}
\end{align}
where \eqref{tseries} was used, and 
$\hat{\Phi}\left(\ep,x\right)=\lim_{z\rightarrow0}\Phi\left(\ep,\partial_z\sctf,\partial_x\sctf\right)$.\footnote{Strictly speaking, because of this limit $\hat{\Phi}$ cannot have an arbitrary $x$-dependence, but should be only a function of $g'_+(x)$, $\hat{\Phi}\left(\ep,x\right)=\hat{\Phi}\left(\ep,g'_+(x)\right)$, because as visible in \eqref{kmuatbdy} this is how $\partial_x \sctf$ and $\partial_z \sctf$ depend on $x$ in this limit. 
	We will see shortly that this is indeed satisfied, at least to second order in $\ep$. This is not surprising, as $g'_+(x)\sim\sct^{bdy}{}'(x)$, and at the beginning of section \ref{sec::creases} we will see how some properties of $k_\mu$ at the boundary are only functions of $\sct^{bdy}{}'(x)$.} Also, in \Poincare-coordinates \ckd
\begin{align}
\hat{t}^\mu=\delta^\mu_{\ttt}\big|_{z=0}\approx
\left(
\begin{array}{c}
1+\frac{\ep}{2} \left(g_+'(x^+)-g_+'(-x^-)\right)
-\frac{\ep ^2}{2} \left(g_+(-x^-) g_+''(-x^-)+g_+(x^+) g_+''(x^+)\right) \\
\frac{\ep}{2}  \left(g_+'(-x^-)+g_+'(x^+)
+\frac{\ep ^2}{2} \left(g_+(-x^-) g_+''(-x^-)-g_+(x^+) g_+''(x^+)\right)\right)\\ 
0\\
\end{array}
\right).
\label{madhatter}
\end{align}
Then, we find\footnote{Note that in this equation, evaluating the product at the boundary $z=0$ also implies setting the $t$-coordinate in \eqref{madhatter} to be $t=\sctbdy$, as this is the time-coordinate as a function of $x$ for which the lightfront emanates from the boundary.} 
\begin{align}
1\equiv k\cdot \hat{t}\big|_{z=0}\approx \hat{\Phi}(0,x)+\partial_{\ep} \hat{\Phi}(\ep,x)\big|_{\ep=0}+\ep^2\left(\frac{1}{2}\partial_\ep^2\hat{\Phi}(\ep,x)\big|_{\ep=0}- \hat{\Phi}(0,x)g_+'(x)^2\right)+\mO(\ep^3),
\end{align}
which can be solved by
\begin{align}
\hat{\Phi}(\ep,x)\approx 1+\ep^2 g_+'(x)^2 +\mO(\ep^3)\approx 1+(\partial_x \sctf)^2\big|_{z=0}+\mO(\ep^3),
\end{align}
hence up to order $\mO(\ep^2)$ we can assume
\begin{align}
\Phi\left(\ep,\partial_z\sctf,\partial_x\sctf\right)\approx 1+(\partial_x \sctf)^2+\mO(\ep^3).
\label{Phi}
\end{align}

\section{Timelike-Null joints}
\label{sec::TNjoints}

\ckd

The types of timelike-null joints that we might have to deal with for nonzero $\ep$ will be similar to the joint-terms already studied in section \ref{sec::joints} for the $\ep=0$ case. At the IR-cutoff surface $z=z_{max}$, we will again have a volume element $\sqrt{\rho}\sim 1/z_{max}$ and an integrand $\eta\sim\log\left(|k\cdot s|\right)$ with at most a logarithmic divergence, so these terms will again vanish in the limit $z_{max}\rightarrow\infty$.

We are left with the timelike-null joints at the cutoff surface $\tz=\epsilon$. For simplicity, we will focus on the joint between the cutoff surface and the \textit{future} lightfront $\sctf$, the calculation for the joint with the past lightfront would be analogous. As seen in section \ref{sec::joints}, the joint term takes the form \cite{Lehner:2016vdi,Chapman:2016hwi}:
\begin{align}
\mA_{joint,1}=\frac{1}{8\pi G_N}\int_{\mJ_1} \eta_{\mJ_1} \sqrt{\rho}dx,
\end{align}
with integrand
\begin{flalign}
&&\eta_{\mJ} =-sign(k\cdot s)sign\left(k\cdot \check{t}\right)\log\left(|k\cdot s|\right), \text{\hspace{3cm} \eqref{etaJ}}
\nonumber
\end{flalign}
with the null normal $k_\mu$ now generally defined as in \eqref{kmuresult} with $\Phi$ as in \eqref{Phi}, the unit normal vector $s$ of the timelike boundary surface (defined to point out of $\mW$) and $\check{t}$, a normalized timelike vector living in the tangent space of the timelike boundary. The values of $sign(k\cdot s)=+1$ and $sign\left(k\cdot \check{t}\right)=+1$ had already been calculated in section \ref{sec::joints} for the $\ep=0$ case, and we assume that they stay the same perturbatively.
For the generic cutoff surface defined by $\tz=\epsilon$ with \eqref{smallSGD3inv}, we find that its intersection with the lightfront $\sctf$, described accurately by \eqref{tseries} near the boundary, can be parametrised perturbatively in $\ep$ and $\epsilon$ as
\begin{align}
t^I(x)\approx \sigma  g_+(x)+ \epsilon  \left(1-\sigma ^2 g_+'(x)^2\right)+\sigma  \epsilon ^2 g_+''(x),\ 
z^I(x)\approx \sigma \epsilon  \left(1-\frac{1}{2} \sigma ^2 g_+'(x)^2\right)+ \epsilon ^2 g_+''(x).
\label{TNembedding}
\end{align}
Hence we find the induced volume element on the joint curve
\begin{align}
\sqrt{\rho}\approx \frac{1}{\epsilon }-\ep  g_+''(x)
\label{TNvol}
\end{align}
and the product
\begin{align}
k\cdot s\big|_{\mJ_1} = \epsilon+\mO(\ep^2,\epsilon^2).
\end{align}
Consequently
\begin{align}
\mA_{joint,1}=\frac{-1}{8\pi G_N}\int_{-\infty}^{\infty} \left(\frac{\log (\epsilon )}{\epsilon }-\ep  \log (\epsilon ) g_+''(x) \right)dx+\mO(\ep^3).
\end{align}
The term $\sim \frac{\log (\epsilon )}{\epsilon }$ is the order $\mO(\ep^0)$ result and the term $\sim \int_{-\infty}^{\infty}g_+''(x)dx$ vanishes by our assumption that the function $g_+(x)$ (and hence its derivative) vanishes at large $|x|$. 
We are thus left with
\begin{align}
\delta\mA_{joint,1}=\mO(\ep^3).
\label{TNfinres}
\end{align}

\section{Null-Null joints}
\label{sec::NNjoints}

\ckd

Our next step will be to evaluate the joint-terms corresponding to the null-null joints or ``creases". These terms will be interesting, because they have no analogue in the $\ep=0$ case: 
In section \ref{sec::GS}, there simply were no null-null joints in the lightfronts $\sct^\pm(x,z)$. However, as explained in section \ref{sec::caustics}, such creases will exist whenever $\sctbdy\neq const.$. In figure \ref{fig::rays}, we plotted some examples for different physically interesting choices of $\sctbdy$ (the null-null joints where marked in red), and in section \ref{sec::creases} we explained how these creases can be located perturbatively in $\ep$. 
The most important thing here is to remember that in section \ref{sec::creases} we introduced the coordinate $x^B_1$ \textit{on the crease}, such that for a lightray that meets the crease at this coordinate (from one of its two sides), $x^B_1$ is also the value of the $x$-coordinate at which that lighray started at the boundary\footnote{For the lightray coming to the crease from the other side, we had introduced the coordinate $x^B_2$, which has to be a function of $x^B_1$.}. In this sense, $x^B_1$ has a double meaning. The embedding of the crease into the \Poincare\ ambient-space is thus given by a triplet of functions $t^P(x^B_1),x^P(x^B_1),z^P(x^B_1)$, see e.g.~\eqref{caus1}.
Unfortunately, these calculations were only possible on a case by case basis, so in this section we will only present explicit results for the three cases $\sctbdy=\frac{\pm\ep}{1+x^2},\ \frac{\ep x}{1+x^2}$.

\subsubsection*{Case $\sctbdy=\frac{\ep}{1+x^2}$}

See the upper left figure in figure \ref{fig::rays}. In terms of the coordinate $x^B_1\in[0,+\infty[$, the embedding $t^P(x^B_1)$, $x^P(x^B_1)$, $z^P(x^B_1)$ for this crease is given in \eqref{caus1}, and the induced volume-element on this curve can then be calculated to be
\begin{align}
\sqrt{\rho}dx^B_1=\frac{8 \sigma  x^B_1{}^2 \left(\left(x^B_1{}^2+1\right)^3-\sigma ^2\right)}{\left(x^B_1{}^2+1\right)^6-4 \sigma ^2 x^B_1{}^2 \left(x^B_1{}^2+1\right)^2}dx^B_1
=
\frac{8 \sigma  x^B_1{}^2}{\left(x^B_1{}^2+1\right)^3}dx^B_1+\mO(\ep^3).
\label{volcrease1}
\end{align}
An interesting observation that can be made here is that $\lim_{x^B_1\rightarrow0}\sqrt{\rho}=0$. I.e.~while the crease is overall a spacelike curve, as we approach the caustic point at which it starts, it approaches a null-ray such that the induced volume-element at the caustic point vanishes. This fact will be very important shortly. Another interesting fact is that also $\lim_{x^B_1\rightarrow\infty}\sqrt{\rho}=0$, consequently the overall volume (or more accurately length) of the crease is finite:
\begin{align}
\int_{0}^{+\infty}\sqrt{\rho}dx^B_1\approx\frac{\pi\ep}{2}.
\label{vol1}
\end{align}
Again, this leads us to a very important and general observation: The creases are always spacelike curves starting at a caustic, and as explained in section \ref{sec::caustics}, we always expect the caustics to be located at $z$-coordinates of order $\mO(1/\ep)$. Consequently, it is our generic expectation that the volume element $\sqrt{\rho}$ (and total volume, if finite) on the crease will be of order $\ep$. Again, this will be important shortly. By \cite{Lehner:2016vdi,Carmi:2016wjl}, the joint term takes the form:
\begin{align}
\mA_{joint}=\frac{1}{8\pi G_N}\int_{\mJ} \eta_{\mJ} \sqrt{\rho}dx,
\end{align}
with integrand
\begin{align}
\eta_{\mJ} =-sign(k\cdot k')sign(\hat{k}_1\cdot k')\log\left(\frac{1}{2}|k\cdot k'|\right).
\label{nnintegrand}
\end{align}
Herein, $k$ and $k'$ are the outward-pointing normal one-forms associated with the two lightfronts that meet on the null-null joint from its two sides. $\hat{k}_1$ is an auxiliary vector, colinear to $k_1^\mu$, but oriented such that it points away from $\mW$ and the null-null joint. We have $sign(k\cdot k')=-1$ and $\hat{k}_1^\mu=-k_1^\mu$, hence $sign(\hat{k}_1\cdot k')=+1$.\footnote{This will apply to all three cases studied in this section.}

We do not know $\sctf$ analytically (not even perturbatively) for the regime in which the $z$-coordinate is larger than the $z$-coordinate (of order $\mO(1/\ep)$) at which caustics appear, and of course this is exactly the regime in which the creases will be located. 
However, as seen in section \ref{sec::affine}, in \Poincare-coordinates the components of $k_\mu$ are constant along each lightray, hence 
\begin{align}
k_\mu(x^P(x^B_1),z^P(x^B_1))=k_{\mu}(x^B_1,0),
\end{align}
which can be evaluated as in \eqref{kmuatbdy}, as we know that \eqref{tseries} is valid near the boundary. So it will be possible for us to evaluate \eqref{nnintegrand} on the caustic. We find
\begin{align}
k\cdot k'=-2 x^B_1{}^2-\frac{7 \ep ^2 x^B_1{}^4}{\left(x^B_1{}^2+1\right)^4}+\mO\left(\ep ^3\right).
\label{scalarproduct}
\end{align}
Let us comment on this result for a moment: The two null-vectors (or one forms) $k$ and $k'$ are oriented with respect to the future lightcones in the same way, so their scalar product is negative, as said above. In \eqref{nnintegrand}, we see there would be a logarithmic divergence if ever $k\cdot k'=0$. This could happen in two ways: 

Firstly, when setting $\ep=0$, $k_\mu$ in \eqref{kmuatbdy} becomes independent of $x$, and hence $k=k'$. So we might naively expect that $k\cdot k'\rightarrow k\cdot k=0$ as $\ep\rightarrow0$, which is clearly \textit{not} true in \eqref{scalarproduct}. Why?
Because \eqref{scalarproduct} was evaluated \textit{at the null-null joint}, where as we know by now the $z$-coordinate will be of order $\mO(1/\ep)$ (at least), and consequently in the expression $k\cdot k'=g^{\mu\nu}k_\mu k'_{\nu}$ the inverse metric will contribute such that the overall result has the expansion in $\ep$ shown in \eqref{scalarproduct}, with a non-zero term at order $\mO(\ep^0)$. So it is important to note here that in an expansion in $\ep$, the integrand \eqref{nnintegrand} will have an order $\mO(\ep^0)$ term.

Secondly, from e.g.~figure \ref{fig::rays} we see that as we move towards the caustic point along the worldline of the null-null joint, we also expect that $k'\rightarrow k$, and hence $k\cdot k'\rightarrow0$. Why does this not cause problems? Because as we had noted above, in this limit the induced volume element on the worldline of the null-null joint, $\sqrt{\rho}$, will also vanish like a power law, i.e.~faster than the divergence of the $\log$. 

Hence we expect to find an overall finite result for the null-null joint term. In fact, we can now calculate
\begin{align}
\mA_{joint}=\frac{1}{8\pi G_N}\int_{0}^{+\infty} \frac{16 \sigma  x^B_1{}^2\log\left(x^B_1\right)}{\left(x^B_1{}^2+1\right)^3} dx^B_1 +\mO(\ep^3)=0+\mO(\ep^3).
\label{res1}
\end{align}
So in this specific and simple case, the term on the null-null joint vanishes identically. However, by the arguments above, we expect generically $\sqrt{\rho}\sim\mO(\ep^1)$ and $\eta_{\mJ}\sim\mO(\ep^0)$, so it looks like the null-null joint terms will contribute at order $\mO(\ep)$ to the change of the action under conformal transformations. We will indeed see this on our next examples.

\subsubsection*{Case $\sctbdy=\frac{-\ep}{1+x^2}$}

See the upper right figure in figure \ref{fig::rays}, the specific embedding of this curve is given in \eqref{caus2}. The range of $x^B_1$ is $x_H\geq x^B_1\geq-x_H$ with $x_H\approx 1-\ep^2/8$.\footnote{$\pm x_H$ are the coordinates of the \textit{hyperbolic points}, a concept introduced in appendix \ref{sec::caustics}, see also the caption of figure \ref{fig::rays} for an explanation.} We find the volume element
\begin{align}
\sqrt{\rho}=\frac{2 \ep  \left(x^B_1{}^2-1\right)^2}{\left(x^B_1{}^2+1\right)^3}+\mO(\ep^3).
\label{rho2}
\end{align}
Interestingly, the combined volume of both arcs of the null-null joint will hence be
\begin{align}
\int_{-x_H}^{+x_H}\sqrt{\rho}dx^B_1\approx\frac{\pi\ep}{2},
\label{vol2}
\end{align}
just as in the previous case. Note that ostensibly we are only studying the creases of future lightfronts, $\sctf$, but the future lightfront $\sctf$ with boundary slice $\sctbdy$ is related to the past lightfront $\sctfm$ with boundary slice $-\sctbdy$ by simple time-reflection. So the two cases $\sctbdy=\frac{\pm\ep}{1+x^2}$ are intimately related.

The scalar product turns out to be
\begin{align}
k\cdot k'=-\frac{\left(x^B_1{}^2-1\right)^2}{2 x^B_1{}^2}+\frac{\ep ^2 \left(x^B_1{}^6-4 x^B_1{}^4+3 x^B_1{}^2+1\right)}{\left(x^B_1{}^2+1\right)^4}+\mO(\ep^3),
\label{scalarproduct2}
\end{align}
and the same overall remarks apply as in the previous case: As expected, the quantity is negative and has a term of order $\mO(\ep^0)$. Consequently\footnote{Technically, we should integrate from $-x_H$ to $-\varepsilon$ and from $\varepsilon$ to $x_H$, for some finite but infinitesimal $\varepsilon$. The integrations for negative and positive $x^B_1$ would then correspond to integrations along the two arcs of the crease. The lightray leaving the boundary at exactly $x^B_1=0$ does not reach either of the arcs of the crease (by symmetry under $x\rightarrow-x$), but goes to the \Poincare-horizon, as can be gleaned from figure \ref{fig::rays} (upper right corner). However, in our integrals the limit $\varepsilon\rightarrow0$ can be taken and yields the finite result presented below.} 
\begin{align}
\mA_{joint}=\frac{1}{8\pi G_N}\int_{-x_H}^{+x_H} \frac{2 \ep  \left(x^B_1{}^2-1\right)^2 }{\left(x^B_1{}^2+1\right)^3}\log \left(\frac{\left(x^B_1{}^2-1\right)^2}{4 x^B_1{}^2}\right) dx^B_1 +\mO(\ep^3)=\frac{1}{8\pi G_N}\pi\ep+\mO(\ep^3).
\label{res2}
\end{align}
So we obtain a term of order $\mO(\ep)$ in the change of the action under one of our infinitesimal conformal transformations. The existence of contributions at this order is one of the main results of this paper.

\subsubsection*{Case $\sctbdy=\frac{\ep x}{1+x^2}$}

See the lower left corner of figure \ref{fig::rays}, the specific embedding of this curve is given in \eqref{caus3}. The range of $x^B_1$ is $x^B_1\in]-\infty,x_{H-}]\cup[x_{H+},+\infty[$ with $x_{H+}\approx \sqrt{2}-1-\frac{1}{32} \left(3 \sqrt{2}+4\right) \ep ^2$ and $x_{H-}\approx-\sqrt{2}-1+ \frac{1}{32} \left(3 \sqrt{2}-4\right) \ep ^2$. In analogy to the previous cases, we find
\begin{align}
\sqrt{\rho}&=\frac{\ep  \left(x^B_1{}^2+2 x^B_1{}-1\right)^2}{\left(x^B_1{}^2+1\right)^3}+\mO(\ep^3),
\label{rho3}
\end{align}
\begin{align}
\left(\int_{\infty}^{x_{H-}}\cup\int_{x_{H+}}^{+\infty}\right)\sqrt{\rho}dx^B_1&\approx \frac{1}{8} (\pi -2) \ep+\frac{1}{8} (2+3 \pi ) \ep=\frac{\pi  \ep }{2} ,
\label{vol3}
\end{align}
\begin{align}
k\cdot k'&=-\frac{\left(x^B_1{}^2+2 x^B_1{}-1\right)^2}{2 (x^B_1{}+1)^2}
\label{scalarproduct3}
\\
&-\frac{\ep^2 \left(x^B_1{}^2+2 x^B_1{}-1\right) \left(x^B_1{}^6+6 x^B_1{}^5+13 x^B_1{}^4-4 x^B_1{}^3-x^B_1{}^2+6 x^B_1{}+3\right)}{8 \left(x^B_1{}^2+1\right)^4}+\mO(\ep^3),
\nonumber
\end{align}
and hence
\begin{align}
\mA_{joint}&=\frac{1}{8\pi G_N}\left(\int_{\infty}^{x_{H-}}\cup\int_{x_{H+}}^{+\infty}\right)\frac{\ep  \left(x^B_1{}^2+2 x^B_1{}-1\right)^2}{\left(x^B_1{}^2+1\right)^3}\log\left(\frac{\left(x^B_1{}^2+2 x^B_1{}-1\right)^2}{4 (x^B_1{}+1)^2}\right)
dx^B_1+\mO(\ep^3)
\\
&\approx\frac{\ep}{8\pi G_N}\left(0.5240+1.0468\right)\approx\frac{\ep}{8\pi G_N} 1.5708.
\label{res3}
\end{align}
The results in the last line come from a numerical integration. Curiously, $0.5240+1.0468\approx1.5708$ might be a numerical expression of $\pi/6+\pi/3=\pi/2$, so just as in the previous case it seems that we obtain a term at order $\mO(\ep)$ with a very nice mathematical form.

\section{Counter terms}
\label{sec::countersigma}

We are left with calculating the counter-terms which, for the \Poincare-case, had already been discussed in section \ref{sec::counter}. We would like to remind the reader that given in the form \eqref{counter}, these terms would have to be evaluated on the entire null-boundaries (i.e.~lightfronts) $\mN_i$. 
However in section \ref{sec::counter} we showed, using \cite{Reynolds:2016rvl} and in addition Raychaudhuri's equation \eqref{R}, that for our cases these terms are total derivatives, and hence boil down to expressions \eqref{newcounter}
\begin{align}
\mA_{counter}
=\frac{\pm1}{8\pi G_N}\int_{-\infty}^{+\infty}\sqrt{\rho}\log(|\theta \ell'_c|) dx
\end{align}
to be evaluated on the joints where the null-boundaries start ($-$ sign) and end ($+$ sign). For the expansion $\theta$, we will make use of the explicit equation \eqref{theta} presented in appendix
\ref{sec::codim2}.\footnote{Specifically, we will use the last expression in this equation, which is formulated in terms of the embedding of the joint-curve into the ambient \Poincare-space and the null-vector $k_\mu$, without the need to apply covariant derivatives to $k_\mu$. Of course, all expressions for $\theta$ given in section \ref{sec::codim2} are equivalent, but especially for large $z$ when we do not know the lightfronts $\sct^\pm(x,z)$ analytically it is convenient in practice to avoid having to act on $k_\mu$ with covariant derivatives.}

\subsection{Counter terms near boundary}
\label{sec::counterbdy}

Just as in section \ref{sec::TNjoints}, we will focus on the intersection between the UV-cutoff surface and the future lightfront $\sctf$. The embedding and induced volume element on this joint-curve are already given in equations \eqref{TNembedding} and \eqref{TNvol}. Note that the joint-curve is one-dimensional, so its induced metric is a $1\times1$-matrix with $\rho_{ij}=1/\rho^{ij}=(\sqrt{\rho})^2$. We hence find
\begin{align}
\theta(x)= \epsilon +\mO(\epsilon^2,\ep^3),
\end{align}
and
\begin{align}
\mA_{counter,1}&\approx \frac{+1}{8\pi G_N}\int_{-\infty}^{\infty} \left(
\frac{\log (\ell_c' \epsilon )}{\epsilon }-\delta  g_+''(x) \log (\ell_c' \epsilon )
\right)  dx
\end{align}
where again the integral over $g_+''(x) $ vanishes. Hence
\begin{align}
\delta\mA_{counter,1}=\mO(\ep^3).
\end{align}

\subsection{Counter terms at null-null joints}
\label{sec::counterjoint}

In dealing with the counter-terms induced on the null-null joints, it is important to notice that each null-null joint is the end-surface for \textit{two} types of lightrays, coming from both of its sides, with normal forms $k$ and $k'$. Hence on each of these joints, we will have to integrate two terms, one with $\theta$ (of $k$) and one with $\theta'$ (of $k'$). 
Again, we will do this on a case by case basis for the specific examples where we have identified the locations of the creases in appendix \ref{sec::creases}. The volume-forms $\sqrt{\rho}$ can be found in section \ref{sec::NNjoints}.

\subsubsection*{Case $\sctbdy=\frac{\ep}{1+x^2}$}

From equation \eqref{theta}, we can derive
\begin{align}
\theta=\theta'=\frac{\left(x^B_1{}^2+1\right)^3}{8 \ep  x^B_1{}^2}-\frac{\ep  \left(11 x^B_1{}^4+3 x^B_1{}^2-2\right)}{16 \left(x^B_1{}^3+x^B_1{}\right)^2}+\mO\left(\ep ^3\right).
\label{thetaonjoint1}
\end{align}
Let us comment on the qualitative features of this result: First of all, we see that it diverges as $x^B_1\rightarrow0$. This is to be expected, because on the worldline of the crease, taking the coordinate $x^B_1$ towards zero corresponds to moving toward the caustic point at which the crease starts. At a caustic point, the expansion of lightrays diverges by definition, as discussed in appendix
\ref{sec::caustics}\footnote{The divergence here is towards $+\infty$, because as in section \ref{sec::counter} we have effectively chosen the affine parameter to increase when going from the bulk towards the boundary.}. 
However, this divergence will not cause a divergence of the integrand of \eqref{newcounter}, as the volume element $\sqrt{\rho}$ vanishes in this limit, too. This is similar to how divergences are avoided in the integrand of the null-null joint terms, as discussed in section \ref{sec::NNjoints}. 

Another noteworthy aspect of the above equation is that its leading order is $\mO(1/\ep)$. Perhaps this should not be surprising to us. In section \ref{sec::counter}, we had seen that in the usual \Poincare-case with $\ep=0$, $\theta\sim z$. Now equation \eqref{thetaonjoint1} has to be evaluated \textit{at the location of the null-null joint}, and as we are not saying for the first time, these joints will generically start at $z$-coordinates of order $\mO(1/\ep)$, and from there on move out towards the \Poincare-horizon. Hence $\theta\sim z\sim1/\ep$ along the crease was to be expected. 
Remember also that in the $\ep=0$ case, the intersection between the lightfronts $\sct^\pm=\pm z$ and the \Poincare-horizon is also nothing but a caustic when mapped to global AdS. So it is sensible to expect a divergence in $\theta$ (evaluated at the crease) when taking the limit $\ep\rightarrow 0$, as in this limit the crease itself moves towards the \Poincare-horizon.   

We are hence left with\ckd
\begin{align}
\mA_{counter}&=2\times\frac{-1}{8\pi G_N}\int_{0}^{+\infty} \frac{8 \ep  x^B_1{}^2 }{\left(x^B_1{}^2+1\right)^3} \log \left(\frac{\ell_c' \left(x^B_1{}^2+1\right)^3}{8 \ep  x^B_1{}^2}\right)dx^B_1 +\mO(\ep^3)
\\
&=\frac{-1}{16\pi G_N} \pi  \sigma  \left(2 \log \left(\frac{\ell_c'}{\sigma }\right)-3+\log (64)\right)+\mO(\ep^3),
\end{align}
i.e.~the counter terms provide us with contributions at orders $\ep$ and even $\ep\log(\ep)$.

\subsubsection*{Case $\sctbdy=\frac{-\ep}{1+x^2}$}

In this case, we find $\theta$ and $\theta'$ as given in equations \eqref{thetaonjoint2a}, \eqref{thetaonjoint2b}, appendix \ref{sec::aux}. The integration of both counter terms (one for $\theta$, one for $\theta'$) along both arcs of the caustic then yields\ckd
\begin{align}
\mA_{counter}&\propto \frac{-1}{16\pi G_N} \pi  \sigma  \left(2 \log \left(\frac{\ell_c'}{\sigma }\right)-1+\log (64)\right)+\mO(\ep^3).
\end{align}

\subsubsection*{Case $\sctbdy=\frac{\ep x}{1+x^2}$}

The expansions $\theta$ and $\theta'$ for this case are given in equations \eqref{thetaonjoint3a}, \eqref{thetaonjoint3b}, appendix \ref{sec::aux}. We obtain
\begin{align}
\mA_{counter}&\propto \frac{-\ep}{8\pi G_N}\left(\pi \log \left(\frac{\ell_c'}{\sigma}\right) +3.39117  \right)+\mO(\ep^3),
\end{align}
where the $\mO(\ep)$ term comes from a numerical integration.

\section{Summary and conclusion}
\label{sec::conc}

Before summarising the results of this paper, let us first look at the results of \cite{Flory:2018akz} again. In this paper, together with N.~Miekley, we studied the change of complexity under infinitesimal conformal transformations according to the volume proposal \eqref{complexity}. The basic result was
\begin{align}
\mV(\Sigma) = \left.\mV\right|_{\ep=0 }+  \ep ^2\mV_{(2)}( g_\pm) + \mO (\ep^3), \text{ with }\ \mV_{(2)}( g_\pm)>0 \text{ and }\ \mV_{(2)}(- g_\pm)=\mV_{(2)}(g_\pm).
\label{mVseries}
\end{align}
This implied that, according to the volume proposal, \Poincare-AdS is, among the Ba\~nados geometries, a local minimum of complexity, with the change of complexity under an infinitesimal conformal transformation being of second order in $\ep$. 
It should also be stated that $\mV_{(2)}$ was independent of the UV cutoff $\epsilon$ and the infinite volume $V=\int dx$. The feature $\mV_{(2)}(-g_\pm)=\mV_{(2)}(g_\pm)$ was particularly interesting, as at lowest order in $\ep$, this sign change corresponds to the inverse conformal transformation. See the appendix of \cite{Flory:2018akz} for a discussion on the operators $U_\pm(\ep g_\pm)$ that implement the conformal transformation corresponding to $\ep g_\pm$ in terms of field theory expressions, such as the Virasoro generators or the field theory energy-momentum tensor.  

Let us now compare these results to the ones obtained in this paper. First of all, from the sections \ref{sec::bulk}, \ref{sec::surfaceresults}, \ref{sec::TNjoints} and \ref{sec::counterbdy}, we see that the change of the action $\mA$ integrated over the WdW-patch $\mW$ does not receive any terms depending on the UV-cutoff $\epsilon$ or $V=\int dx$, i.e.~$\delta \mA$ is finite up to $\mO(\ep^2)$.
This is a similarity between the action proposal and the volume proposal, which holds for generic functions $g_+$ subject to our assumptions concerning finiteness and falloff stated in section \ref{sec::SGD}. 
In fact, for the examples of \eqref{bulkex1}, \eqref{bulkex2}, these terms didn't lead to a change of action up to order $\mO(\ep^2)$ at all. 
A full evaluation of the finite contributions to $\delta\mA$ requires the evaluation of joint and counter terms at the null-null joints of the lightfronts $\sct^\pm(x,z)$. This is very demanding to do in general, however for some simple examples of functions $g_\pm$ (always assuming \eqref{gpm} and \eqref{t0}) we were able to calculate the necessary terms in sections \ref{sec::NNjoints} and \ref{sec::counterjoint}.  
Taking these results together now (and including the correct terms for the past lighfronts $\sctfm$, too), we find\ckd
\begin{align}
\delta\mA\left(g_+=\frac{1}{1+x^2}\right)&=\frac{-1}{4\pi G_N} \pi  \sigma  \log \left(\frac{\ell_c'}{\sigma }\right)+\frac{1}{8\pi G_N}\left(3-\log(64)\right)\pi\ep
+\mO(\ep^3),
\\
\delta\mA\left(g_+=\frac{-1}{1+x^2}\right)&=\frac{-1}{4\pi G_N} \pi  \sigma  \log \left(\frac{\ell_c'}{\sigma }\right)+\frac{1}{8\pi G_N}\left(3-\log(64)\right)\pi\ep
+\mO(\ep^3),
\\
\delta\mA\left(g_+=\frac{\pm x}{1+x^2}\right)&\approx\frac{-1}{4\pi G_N} \pi  \sigma  \log \left(\frac{\ell_c'}{\sigma }\right)-\frac{1}{8\pi G_N}3.64074\ep+\mO(\ep^3).
\end{align}
So again, the change in complexity is invariant under inversion of the conformal transformation, which is a natural consequence of time-reflection invariance of AdS-space.\footnote{Another curious fact is that $(3-\log(64))\pi\approx-3.64074$, so it seems that the change of complexity induced by the conformal transformations $g_+=\frac{1}{1+x^2}$ and $g_+=\frac{ x}{1+x^2}$ is identical subject to the assumptions \eqref{gpm} and \eqref{t0}. See also \eqref{bulkex1} and \eqref{bulkex2}. This equivalence was already a feature of the results for the volume proposal \cite{Flory:2018akz}, but we don't currently understand why this fact should hold generally for any holographic complexity proposal.  } 

\begin{figure}[htb]
	\centering
	\def\svgwidth{0.7\columnwidth}
\executeiffilenewer{Manhattan.svg}{Manhattan.pdf}%
{inkscape -z -D --file=Manhattan.svg %
--export-pdf=Manhattan.pdf --export-latex}%
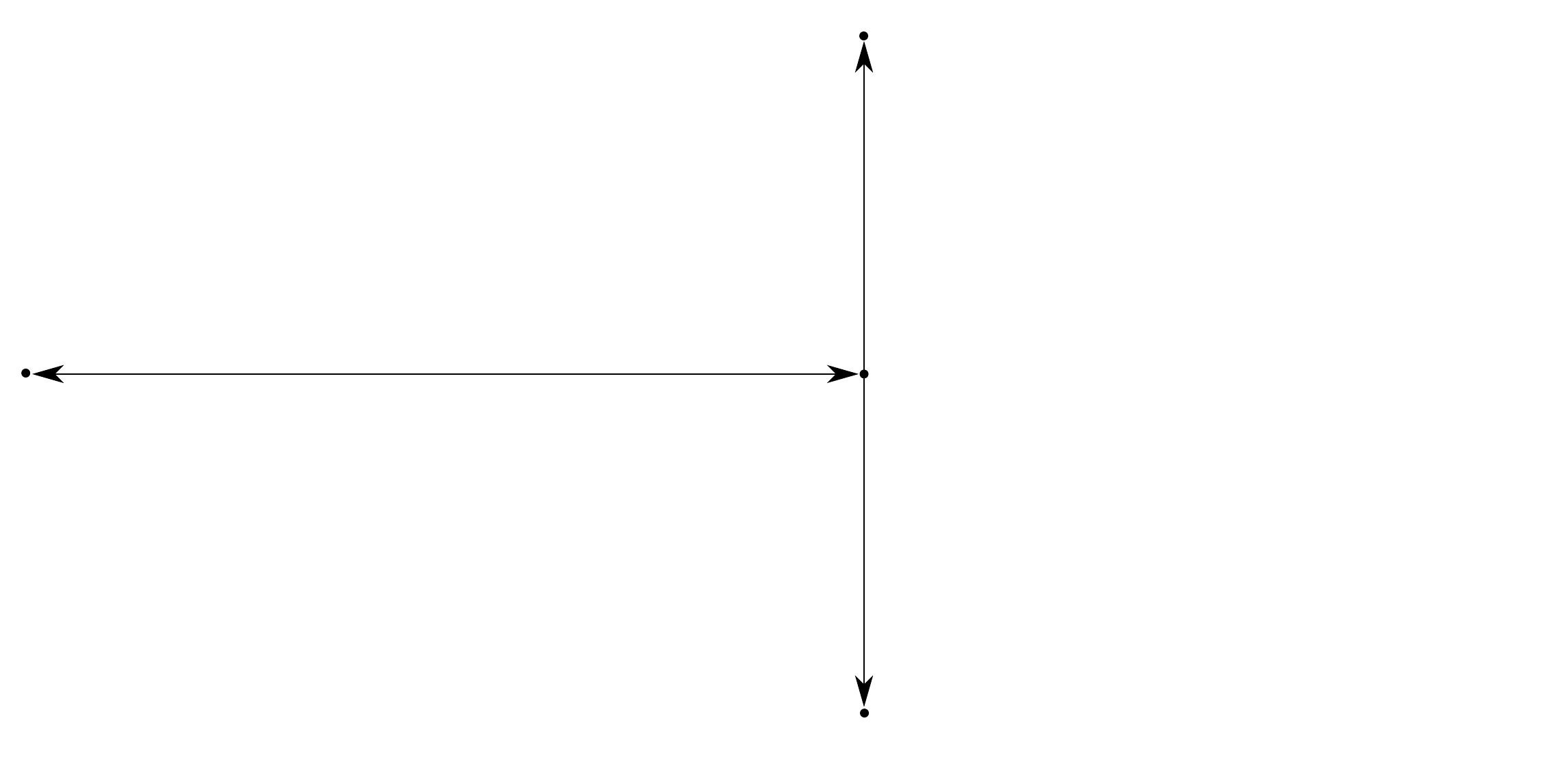%

	\caption{Possible interpretation of order $\mO(\ep^2)$ and $\mO(\ep)$ terms in $\delta\mC$}
	\label{fig::Manhattan}
\end{figure}

Of course, the elephant in the room is that $\delta\mA$ contains terms of orders $\ep$ and even $\ep\log(\ep)$. This is very hard to interpret in terms of what a physical definition of complexity might look like on the field theory side, see figure \ref{fig::Manhattan}. Complexity is meant to provide a distance measure between states, and we are essentially working with the triangle spanned by the groundstate $\left|0\right>$, the state after an infinitesimal conformal transformation $U(\ep)\left|0\right>$, and the implicit \textit{reference state} $\left|\mR\right>$. As the change of complexity caused by $U(\ep)$ and $U(-\ep)\approx U(\ep)^{-1}$ is the same, it seems in a naive geometrical picture that the line of states $U(\ep)\left|0\right>$ is perpendicular to the line between $\left|0\right>$ and $\left|\mR\right>$, so the three states under consideration form a right triangle. One of the sides of this triangle will also be of infinitesimal length, which we call 
$\mC\left(\left|0\right\rangle,U(\ep)\left|0\right\rangle\right)=b$ and assume $b\propto \ep$. If the metric defined by the complexity functional was a flat metric, then we could use the Pythagorean theorem to solve for the change of complexity and find $\delta \mC\propto\ep^2$. Even if a Riemannian metric defined by the complexity functional on the Hilbert space is curved, we might still expect a similar result. This would qualitatively correspond to the result \eqref{mVseries} of the volume proposal. 

Suppose now we had obtained only the terms of order $\mO(\ep)$ in the action proposal. Those could have a very simple interpretation if we assume that the distance measure defined on the Hilbert space by complexity is more akin to a \textit{Manhattan-metric}, where instead of $a^2+b^2=c^2$ the distance when moving along two perpendicular axes is defined as $|a|+|b|=|c|$. This could naturally lead to $\delta\mC\propto\ep$ in our naive geometrical picture. A change of order $\delta\mC\propto\ep\log(\ep)$ however would seem very hard to interpret in terms of a plausible distance measure on the Hilbert space, especially as it would mean $\delta\mC<0$ to lowest order, with an initial decrease with infinite negative slope.\footnote{A somewhat similar behaviour of complexity decrease with infinite slope was observed in \cite{Carmi:2017jqz} in the time evolution of complexity in black hole backgrounds.}
Above we have made the assumption that the relative complexity between $\left|0\right\rangle$ and $U(\ep)\left|0\right\rangle$, $\mC\left(\left|0\right\rangle,U(\ep)\left|0\right\rangle\right)=b$, is of order $\sigma$. By the relation between operator-complexity and (relative) complexity of states outlined in section \ref{sec::Intro}, we also have $\mC(U(\ep))\geq b$. Furthermore, with the notation of figure \ref{fig::Manhattan}, the \textit{triangle inequality} would imply $b\geq |a-c|$. With $|a-c|\approx|\delta\mC|$ and our results from above, for $\ep\rightarrow0$ this would mean
\begin{align}
\mC(U(\ep))\geq |\ep\log(\ep)|\cdot \mK
\label{triangle}
\end{align}
with some positive finite constant $\mK$. Note that for small $\ep$, $\ep \mK'<|\ep\log(\ep)|\mK$ for any positive constants $\mK,\mK'$, as $\lim_{\ep\rightarrow0}\partial_{\ep}\left(-\ep\log(\ep)\right)=+\infty$. Hence \eqref{triangle} and our results imply the following statement:
\begin{center}
	\textit{
		Any definition of field-theory complexity (for both operators and states as discussed in section \ref{sec::Intro}) that utilises a unique reference state $\left|\mR\right>$, satisfies the triangle inequality and assigns to any operator of the form $U(\ep)=\mathbbm{1}+\ep V +\mO(\ep^2)$\footnote{See e.g.~the appendices of \cite{Flory:2018akz} and \cite{Mandal:2014wfa} for how to write the generators of conformal transformations in this form.}
		a complexity of the form $\mC(U(\ep))=\ep \mK'+\mO(\ep^2)$ (for sufficiently small $\ep$ and a finite constant $\mK'$ depending on $V$) can not possibly be dual to the CA proposal \eqref{action} in AdS$_3$/CFT$_2$ with the counter-terms chosen as in \eqref{counter}. 
	}
\end{center}

The existence of the $\mO(\ep\log(\ep))$ terms is the central result of this paper: Despite the fact that we were only able to explicitly compute them for three concrete examples, we have provided arguments throughout the paper that these terms should generally be expected to contribute with the orders that they do. Let us repeat: For non-constant $\sctbdy$, we generically expect caustics and creases to emerge in the lightfronts bounding the WdW-patch \cite{Akers:2017nrr}. 
The focusing theorem implies that the caustics will have $z$-coordinates of order $\mO(1/\ep)$ (section \ref{sec::caustics}), and consequently the creases starting there will too. So the (codimension-2) creases, on which joint- and counter terms will have to be evaluated, will have induced volume elements $\sqrt{\rho}\sim\sigma$ due to the factors of $z$ induced by the ambient metric \eqref{Poincare}. The integrands to be evaluated on these creases will have the form $\sqrt{\rho}\log(...)$, see sections \ref{sec::NNjoints} and \ref{sec::countersigma}. As argued in section \ref{sec::NNjoints}, the term $k\cdot k'$ will be of order $\mO(\ep^0)$ and hence lead to a term $\delta\mA\sim\ep$. However, the expansion $\theta$ of the lightfronts evaluated at the crease will diverge as $1/\ep$. This gives rise to the $\ep\log(\ep)$-terms, however as explained in section \ref{sec::countersigma}, this divergence has to be expected: In global AdS-coordinates the intersection between the lightfront and the \Poincare-horizon is also just a caustic point, thus $\theta$ diverges when approaching it. Hence, with our present hindsight and understanding of the topic, the terms of order $\ep$ and $\ep\log(\ep)$ seem almost inevitable. 

We leave a further discussion of what possible implications this has for the CA-conjecture (or the terms required in \eqref{CAbulk}-\eqref{counter}) and proposed field-theory definitions of complexity to the future. In any case, our results show a significant qualitative difference between volume proposal \eqref{complexity}, action proposal \eqref{action}, and also the volume 2.0 proposal \eqref{Vol2.0}, for which our results implied $\delta\mC=\mO(\ep^3)$ for the $g_+$ of \eqref{bulkex1} and \eqref{bulkex2}. Other papers in which qualitative differences between these proposals where found are \cite{Kim:2017qrq,Reynolds:2017jfs,Fu:2018kcp,Agon:2018zso,Fan:2018xwf,Chapman:2018bqj}\footnote{The paper \cite{Chapman:2018bqj} dealt with complexity of AdS/BCFT models, a topic also studied in \cite{Flory:2017ftd}.} Which of the proposals is the ``better" one according to these comparisons still seems to be an open question, to which we hope to have made a contribution with this paper. 

Despite there being already considerable theoretical knowledge concerning the geometry of lightfronts (see the discussion in appendix \ref{sec::details}), some of our ideas outlined there may be helpful in practice for dealing with WdW-patches in generic cases, i.e.~when the background-spacetime is not translation invariant or when the boundary-conditions on the lightfront are nontrivial. This may be useful for further investigations along the lines of \cite{Takayanagi:2018pml} or \cite{Caceres:2018luq,Bakhshaei:2019ope}, although in \cite{Caceres:2018luq} it was shown that the caustics would not play a role. 

\section*{Acknowledgements}

I am particularly grateful to Nina Miekley for many discussions and initial collaboration on this project. I would also like to thank Shira Chapman, Zach Fisher, Federico Galli, Hugo Marrochio, Rob Myers, Shan-Ming Ruan and Alvaro Veliz Osorio for helpful discussions. This research was supported by the Polish National Science Centre (NCN) grant 2017/24/C/ST2/00469. This research was supported in part by Perimeter Institute for Theoretical Physics. Research at
Perimeter Institute is supported by the Government of Canada through the Department of
Innovation, Science and Economic Development and by the Province of Ontario through the
Ministry of Research, Innovation and Science.

\appendix

\section{Explicit expressions for extrinsic curvature and geodesic expansion}
\label{sec::equations}

In this appendix we will collect a number of explicit expressions useful in calculating geometrical quantities such as extrinsic curvatures or null expansions.

\subsection{Codimension-1 extrinsic curvature}
\label{sec::codim1}

We begin with a codimension-1 surface $\Sigma$, which is either timelike or spacelike, i.e.~which has a nondegenerate induced metric of definite sign. Then, there exists a normal vector which can be normalised so that
\begin{align}
n_\mu n^\mu = \pm 1,
\end{align}
where $n^\mu$ is spacelike for timelike $\Sigma$ and vice versa. One can then define a degenerate tensor
\begin{align}
\gamma_{\mu\nu}=g_{\mu\nu}\mp n_\mu n_\nu.
\end{align}
which can be used to project quantities into the tangent-space of $\Sigma$ after raising one of its indices. Alternatively, for coordinates $X^\mu$ in the spacetime manifold and coordinates $y^i$ in the worldsheet of $\Sigma$, we can define the \textit{induced metric} on $\Sigma$,
\begin{align}
\gamma_{ij}=g_{\mu\nu}\frac{\partial X^\mu}{\partial y^i}\frac{\partial X^\nu}{\partial y^j}
\end{align}
The \textit{extrinsic curvature tensor} or \textit{second fundamental form}, in $y^i$-coordinates, is then given by \cite{Johnson:2003gi}\footnote{There is an overall ambiguity of sign choice in the definition of the extrinsic curvature, which is related to the ambiguity of choosing the orientation of $n^\mu$. For the timelike Gibbons-Hawking type boundary terms, we chose the normal vector to be pointing outward of $\mW$ \cite{PhysRevD.15.2752}.}
\begin{align}
K_{ij}=\frac{\partial X^\mu}{\partial y^i}\frac{\partial X^\nu}{\partial y^j}\nabla_\mu n_\nu = -n_\mu \left(\frac{\partial^2 X^\mu}{\partial y^i\partial y^j}+\Gamma^\mu_{\alpha\beta}\frac{\partial X^\alpha}{\partial y^i}\frac{\partial X^\beta}{\partial y^j}\right),
\end{align}
and its trace is
\begin{align}
K=\gamma^{ij}K_{ij}.
\end{align}

\subsection{Codimension-2 extrinsic curvatures and null expansion}
\label{sec::codim2}

We will now turn to a codimension-2 surface $\Sigma$, which we assume to be spacelike. One can then choose two normal vectors, one timelike and one spacelike, subject to the normalisation and orthogonality conditions
\begin{align}
n_\mu^{(1)} n^{(1)\mu} = -1,\ n_\mu^{(2)} n^{(2)\mu} = 1,\ n_\mu^{(1)} n^{(2)\mu} = 0.
\end{align}
Similar to the previous subsection, we can then introduce the projector
\begin{align}
\rho_{\mu\nu}=g_{\mu\nu} + n_\mu^{(1)} n_\nu^{(1)}-n_\mu^{(2)} n_\nu^{(2)}
\label{projector}
\end{align}
and the induced metric
\begin{align}
\rho_{ij}=g_{\mu\nu}\frac{\partial X^\mu}{\partial y^i}\frac{\partial X^\nu}{\partial y^j}.
\end{align}
For each normal direction, it is now possible to define an extrinsic curvature tensor (or second fundamental form) by
\begin{align}
K_{ij}^{(i)}=\frac{\partial X^\mu}{\partial y^i}\frac{\partial X^\nu}{\partial y^j}\nabla_\mu n_\nu^{(i)} = -n_\mu^{(i)} \left(\frac{\partial^2 X^\mu}{\partial y^i\partial y^j}+\Gamma^\mu_{\alpha\beta}\frac{\partial X^\alpha}{\partial y^i}\frac{\partial X^\beta}{\partial y^j}\right),
\end{align}
and
\begin{align}
K^{(i)}=\rho^{ij}K_{ij}^{(i)}.
\end{align}
Another interesting aspect of the geometry of spacelike codimension-2 surfaces are the properties of the lightfronts emanating from them. To understand this better, we will collect a few more equations, following mostly \cite{Parattu:2015gga} (see also \cite{Gourgoulhon:2005ng}). In general, there will be four lightfronts emanating from a codimension-2 spacelike surface, two towards the future and two towards the past. 
Assume that we pick one of them, and its null-normal one-form is given by $k_\mu$, just as in section \ref{sec::surface}. We introduce an auxiliary null-vector $l^\mu$ such that
\begin{align}
l^\mu l_\mu=0, l^\mu k_\mu=-1.
\end{align} 
So although the null vectors $l^\mu$, $k^\mu$ cannot be normalised individually, they are normalised with respect to each other. The tensor of \eqref{projector} then takes the form 
\begin{align}
\rho_{\mu\nu}=g_{\mu\nu} + l_\mu k_\nu+ k_\mu l_\nu,
\end{align}
which easily follows by rewriting the null-normals as linear combinations of the time- and spacelike normals. An important geometrical quantity of the lightfront in question is its \textit{expansion} $\theta$. It is intuitively appealing, because it measures the normalised change of the volume element $\sqrt{\rho}$ of $\Sigma$ as we make a step $d\lambda$ of affine parameter away from the surface along the light rays:  
\begin{flalign}
&&\theta=\frac{1}{\sqrt{\rho}}\partial_\lambda \sqrt{\rho}.\text{\hspace{5.3cm} \eqref{theta1}}
\nonumber
\end{flalign}
It can be shown \cite{Parattu:2015gga,Hubeny:2007xt} that this is simply the trace of the extrinsic curvature with respect to the null vector $k^\mu$:
\begin{align}
\theta =K^{(k)}
=
\rho^{ij}\frac{\partial X^\mu}{\partial y^i}\frac{\partial X^\nu}{\partial y^j}\nabla_\mu k_\nu
=-k_\mu\rho^{ij}\left(\frac{\partial^2 X^\mu}{\partial y^i\partial y^j}+\Gamma^\mu_{\alpha\beta}\frac{\partial X^\alpha}{\partial y^i}\frac{\partial X^\beta}{\partial y^j}\right).
\label{theta}
\end{align}
The overall freedom of rescaling $k^\mu$ hereby corresponds to the freedom of rescaling the affine parameter $\lambda$ in \eqref{theta1}, so $\theta$ transforms under these rescalings in the expected way.

\subsection{The Raychaudhuri equation}

In the previous subsection, we saw how the expansion $\theta$ of a lightfront originating from a spacelike codimension-2 surface is determined, \textit{at this surface}, by its geometry and embedding into the ambient space. Now, we would like to understand how this expansion will evolve along the lightfront, as a function of the affine parameter of the lightrays. To this end, we introduce the important \textit{Raychaudhuri equation}. A general overview is given for example in \cite{Witten:2019qhl}, but here we will only need the case relevant for null-geodesics in $2+1$-dimensions, where shear and twist automatically vanish. Assuming Einsteins equations, we are then left with
\begin{align}
\dot{\theta}=-\theta^2-T_{\mu\nu}\dot{\gamma}^\mu\dot{\gamma}^\nu,
\label{R}
\end{align}
where $\theta$ is the expansion of a family of lightrays with tangent vectors $\dot{\gamma}^\mu$ and $\dot{\theta}$ is the derivative of the expansion with respect to the affine parameter.

\section{Details on WdW-patches in AdS$_3$}
\label{sec::details}


\subsection{Numerical method}
\label{sec::numerics}

\begin{figure}[htb]
	\centering
	\def\svgwidth{0.6\columnwidth}
\executeiffilenewer{lightcones.svg}{lightcones.pdf}%
{inkscape -z -D --file=lightcones.svg %
--export-pdf=lightcones.pdf --export-latex}%
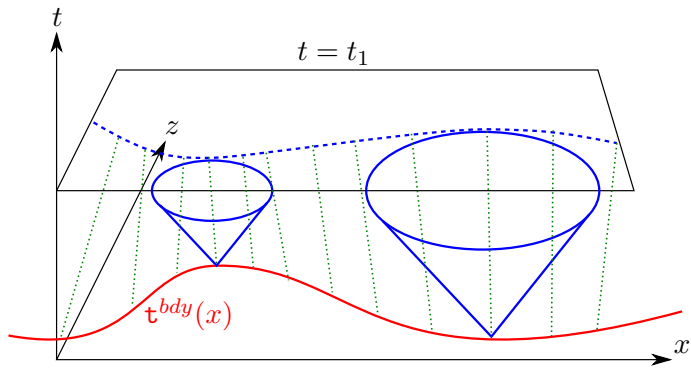%

	\caption{The solid (red) line is the equal time slice $\sctbdy$ on the boundary, and the dotted (green) lines are the ligthrays emanating from this slice, forming the lightfront that is the boundary of $\mW$ to the future. The dashed (blue) line is the intersection of the lightfront with the bulk equal-time slice at $t=t_1$. Two lightcones are sketched with solid (blue) lines.}
	\label{fig::lightcones}
\end{figure}

In this section, we present our numerical method for finding (physical) solutions to \eqref{pde}. A basic illustration for this is given in figure \ref{fig::lightcones}. We assume that we have given the boundary slice $\sct^{bdy}(x)$, and we want to calculate the intersection of the lightfront $\sctf$ with a bulk equal-time slice at $t=t_1$, as a function $z=\scz_{t_1}(x)$.

\begin{figure}[htb]
	\centering
	\includegraphics[width=0.46\textwidth]{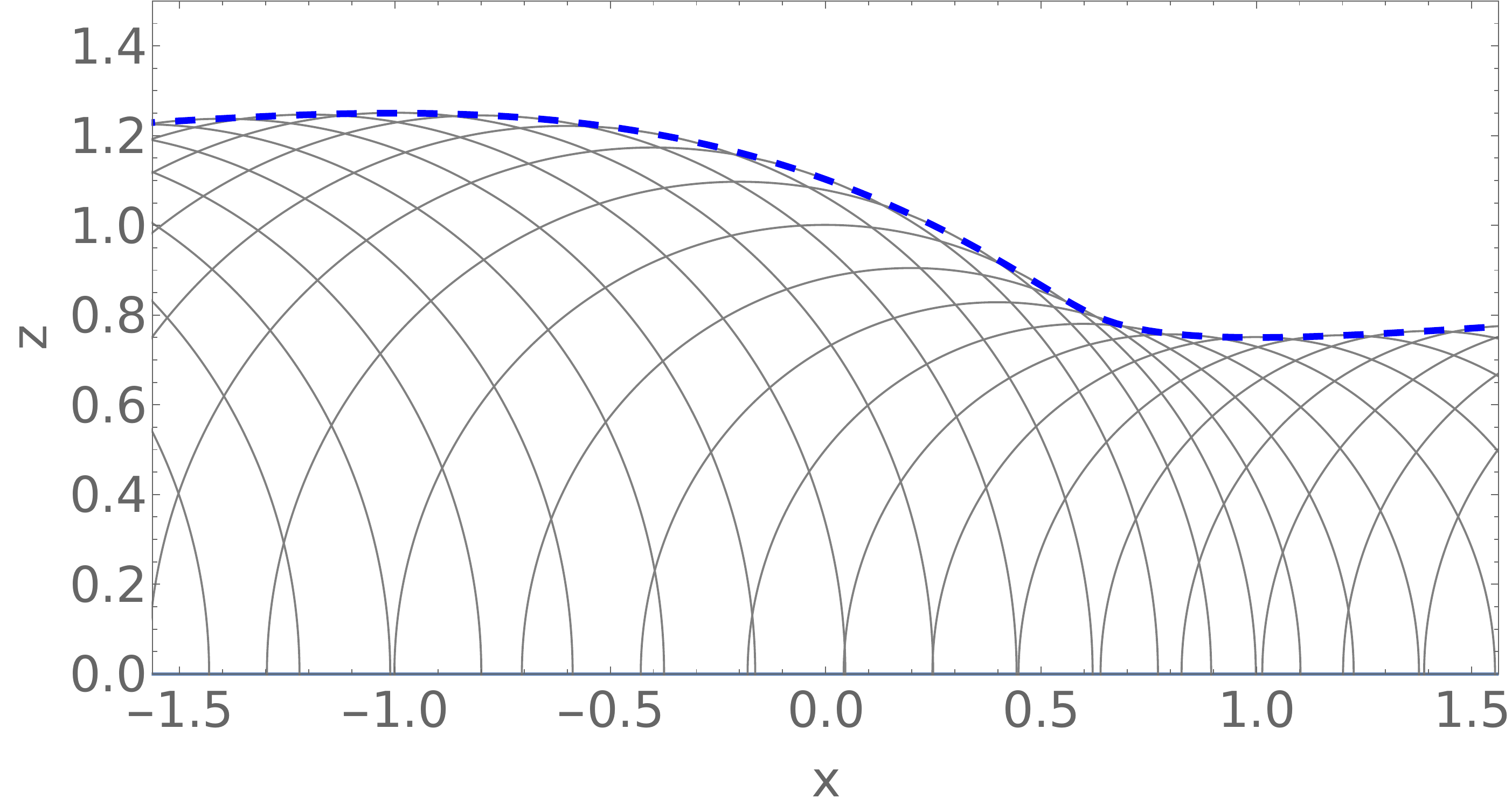}
	\includegraphics[width=0.46\textwidth]{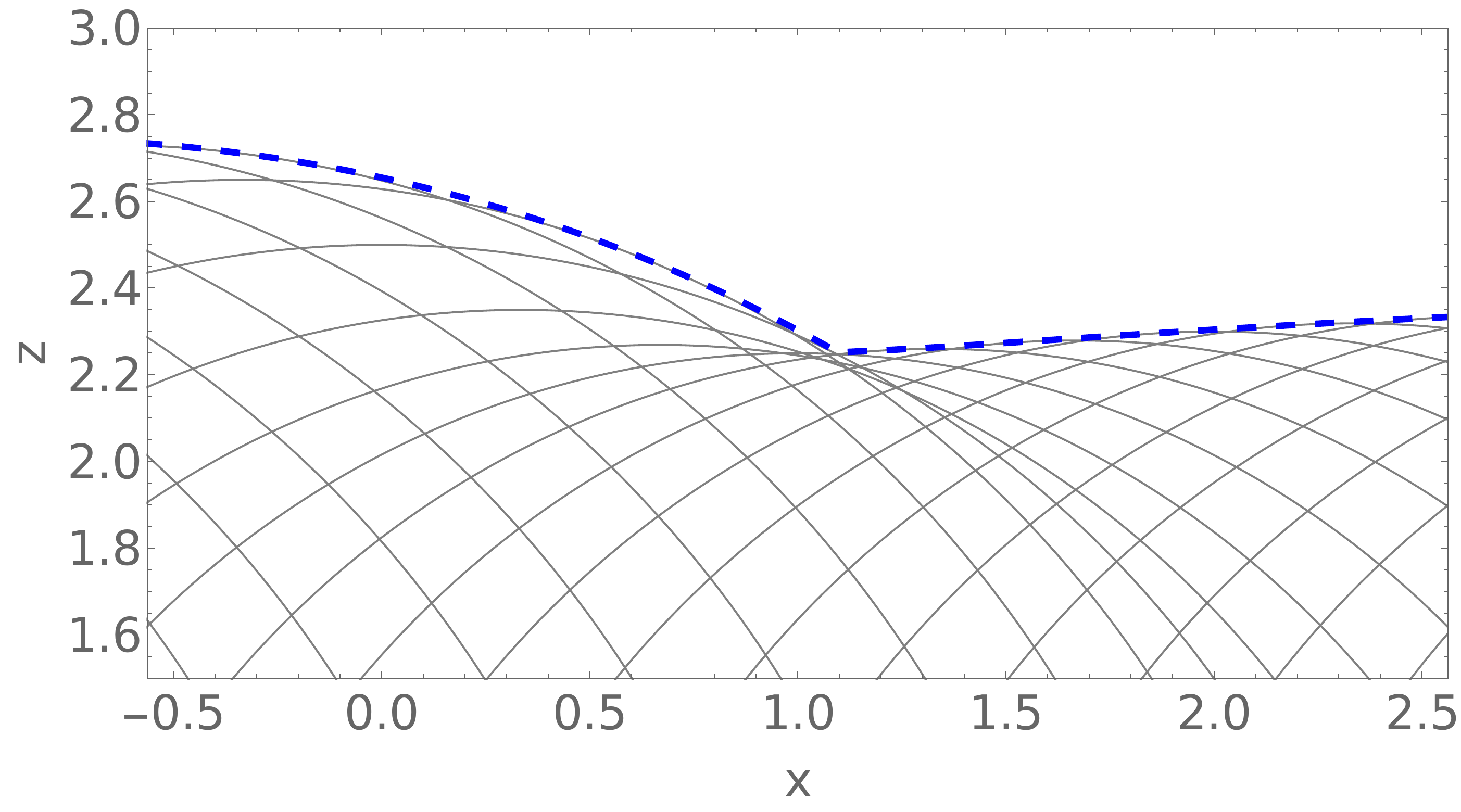}
	\caption{Construction of the enveloping function $\scz_{t_1}(x)$ as in figure \ref{fig::lightcones} for $\sctbdy=\frac{x}{2+2x^2}$ and $t_1=1$ (left) and $t_1=2.5$ (right). We see that $\scz_{1}(x)$ (left) is a smooth function while $\scz_{2.5}(x)$ (right) has developed a kink. This indicates the presence of a caustic point on the lightfront somewhere between $t_1=1$ and $t_1=2.5$. }
	\label{fig::envelope}
\end{figure}

\begin{figure}[htb]
	\centering
	\includegraphics[width=0.46\textwidth]{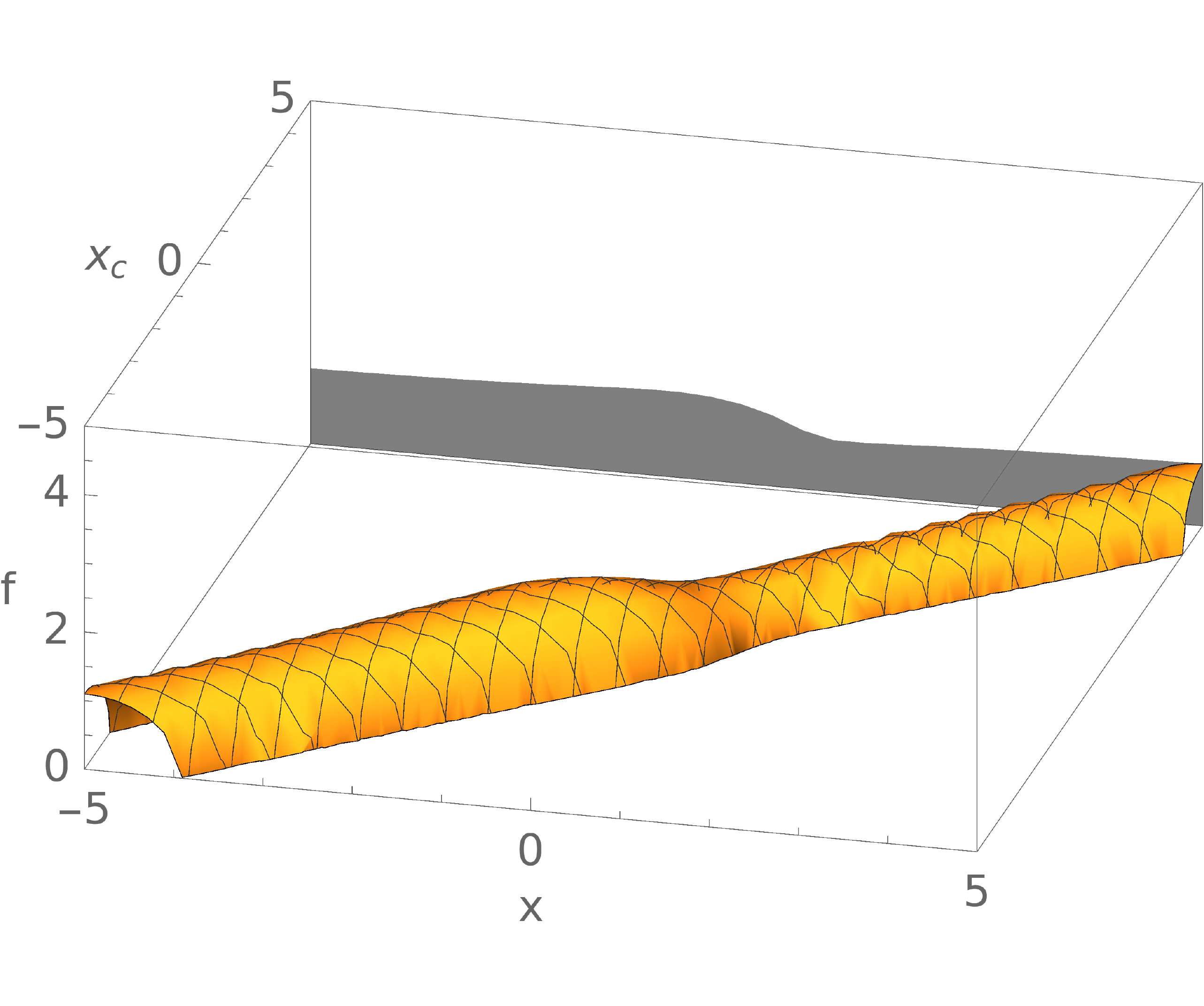}
	\includegraphics[width=0.46\textwidth]{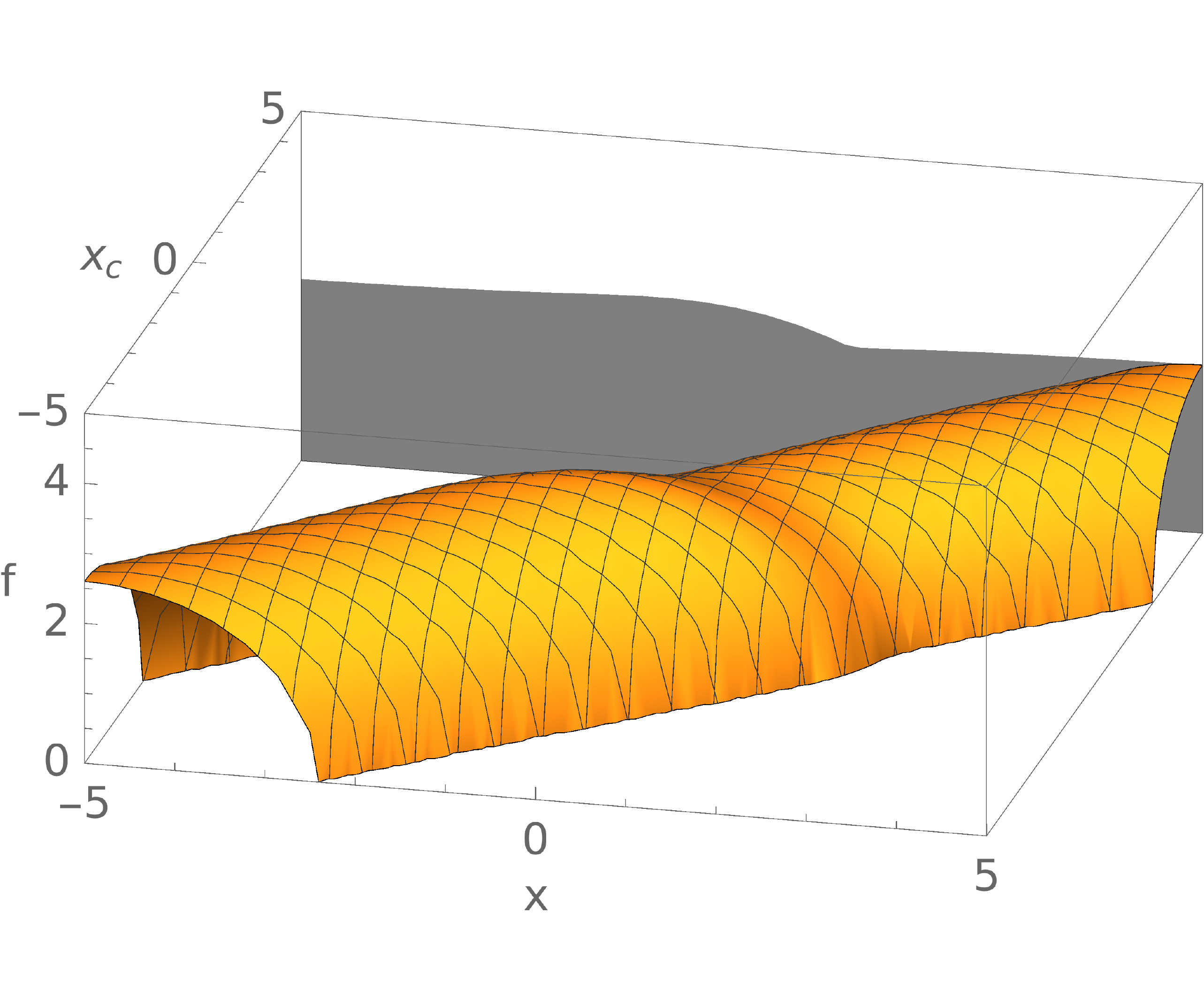}
	\caption{Fictitious three-dimensional bodies. When viewed along the $x_c$-axis, the silhouette of these bodies (shown in gray) corresponds to the functions $\scz_{t_1}(x)$ shown in figure \ref{fig::envelope}. }
	\label{fig::wurst}
\end{figure}

A point inside of $\mW$ by definition is not in causal contact with any point on the boundary slice $t=\sctbdy$, and hence is outside of any lightcone emanating from such a point. Consequently, the function $\scz_{t_1}(x)$, i.e. the intersection of the lightfront with the bulk slice $t=t_1$, will be the enveloping function of the circular intersections of the bulk slice $t=t_1$ with all the lightcones emanating from a point on the boundary slice, see figure \ref{fig::envelope}. 
How could we derive this enveloping function? Again, the explicit conformal flatness of \eqref{Poincare} is of help here, because it means that in $t,z,x$-coordinates, the lightcones will just be straight undeformed cones with $90^{\circ} $ opening angle. The intersection between any of the lightcones with the $t=t_1$ bulk slice ($t_1>t_0+\mO(\ep)$) will hence be a (semi)-circle with center at coordinate $x=x_c,z=0$ and radius
\begin{align}
r(t_1,x_c)=t_1-\sct^{bdy}(x_c).
\end{align}
This defines the family of circles shown in figure \ref{fig::envelope}. For a fixed center $x_c$, the functional form of these semi-circles will then be 
\begin{align}
f(t_1,x_c,x)=\sqrt{-(x-x_c)^2+r(t_1,x_c)^2}.
\end{align}
This defines a fictitious three dimensional surface, shown in figure \ref{fig::wurst}, which is generated by smearing out the circles of figure \ref{fig::envelope} along the $x_c$ axis. The silhouette of this surface, when viewed along the $x_c$ axis, is precisely given by the enveloping function $\scz_{t_1}(x)$ that we are trying to calculate. This means that for any given $x$ and $t_1$, we need to maximize $f(t_1,x_c,x)$ as a function of $x_c$ in order to obtain the value $\scz_{t_1}(x)$. 
This will in general have to be done numerically, and doing so on a grid of points in the $x,t_1$-plane will give us, by numerical interpolation, the function $\sctf$. It can then be checked that these numerical solutions will indeed, within numerical errors, satisfy equation \eqref{pde}. Drawing the contours along which the quantities $\partial_z\sct^+(z,x)$ and $\partial_x\sct^+(z,x)$ are constant does, as expected due to the discussion in section \ref{sec::WdWs}, yield (identical) straight lines which are the projections to the $z,x$-plane of the light-rays which foliate the lightfront, see figure \ref{fig::rays}.

In this context it has to be pointed out that for AdS$_3$, causal wedges and entanglement wedges for intervals on the boundary are identical \cite{Hubeny:2012wa,Headrick:2014cta}. In other words, the half-circles that we dealt with above, which were of interest to us because they are intersections of lightcones with the equal time slice $t=t_1$, were also geodesics describing the entanglement entropy of a given boundary interval via the Ryu-Takayanagi formula. 
There is hence an overlap between our calculations above and results concerning \textit{hole-ography} and \textit{differential entropy} \cite{Balasubramanian:2013lsa,Czech:2014ppa}, see especially \cite{Myers:2014jia,Headrick:2014eia}. In the nomenclature of \cite{Myers:2014jia}, the function $\scz_{t_1}(x)$ was the \textit{outer envelope} of a given set of intervals that can be derived from $\sctbdy$ and $t_1$. Also, the swallow-tail like feature shown in figure 6b of \cite{Headrick:2014eia} is related to the emergence of a caustic and null-null joint in the case $\sctbdy=\frac{\ep}{1+x^2}$ which we study throughout this paper, see e.g.~figure \ref{fig::rays}, upper left corner. We leave it to the future to study in more generality the possible relations between differential entropy and WdW-patches, respectively complexity. \todo{Talk with Michal about this? See also \cite{Alishahiha:2018lfv}.}

\subsection{Identifying caustics}
\label{sec::caustics}

As visible in figure \ref{fig::rays}, for generic functions $\sctbdy$ the lightfronts will, at finite $z$ (for finite $\ep$), develop caustics from which null-null joints emerge.
This is a well known consequence of the \textit{focusing theorem}, which can be derived by integrating the Raychaudhuri equation \eqref{R}, either in vacuum or assuming the null energy condition (see  e.g.~\cite{Witten:2019qhl,Bousso:2015mna,Akers:2017nrr}). 
As we are working with vacuum-solutions in which $T_{\mu\nu}=0$, it is easy to solve \eqref{R} and prove that generically, whenever the expansion $\theta$ is negative near the boundary, it will diverge to minus infinity after a finite (positive) affine parameter, signaling that the lightrays have met a \textit{caustic}, i.e.~that they have been focused to a point.

In the remainder of this section, instead of integrating equation \eqref{R}, we will show how the emergence of such caustics can be predicted directly from the shape of the boundary slice $\sctbdy$. As can be seen from figure \ref{fig::rays}, the shape of $\sctbdy$ determines in which direction the lightrays emanating from the boundary timeslice initially go, before at some point lightrays start to collide forming caustics and null-null joints. Depending on the curvature of $\sctbdy$, these lightrays can be initially focused or defocused A caustic is a point where \textit{neighbouring} lightrays first collide, and hence locally looks like the tip of a past lightcone. The past lightcone of the caustic point at bulk coordinates $t_c,x_c,z_c$ will intersect the boundary in a hyperbolic curve of the form
\begin{align}
h(x)=t_c-\sqrt{z_c^2+(x-x_c)^2}.
\label{hyperbola}
\end{align}
Consequently, in order to find the (infinitesimal) section of $\sctbdy$ which focuses lightrays such that they meet in a caustic, we need to find the point $x$ at which $\sctbdy$ locally looks like a hyperbola \eqref{hyperbola}. Given the number of free parameters in \eqref{hyperbola}, fitting a hyperbola to $\sctbdy$ at any point $x$ is always possible to second order in a Taylor expansion around $x$, but nontrivial to third or higher order. The hyperbola \eqref{hyperbola} satisfies the characteristic third order differential equation
\begin{align}
\frac{h'(x)h''(x)^2}{-1+h'(x)^2}-\frac{1}{3}h'''(x)=0,
\end{align}
so any boundary point $x_H$ at which $\sctbdy$ satisfies
\begin{align}
\frac{\sct^{bdy}{}'(x_H)\sct^{bdy}{}''(x_H)^2}{-1+\sct^{bdy}{}'(x_H)^2}-\frac{1}{3}\sct^{bdy}{}'''(x_H)=0,\ \sct^{bdy}{}''(x_H)<0
\label{hyperbolicequation}
\end{align}
will generate a caustic in $\sctf$ at some point in the bulk.\footnote{Points with $\sct^{bdy}{}''(x_H)>0$ can be fitted by a hyperbola that is opened upwards, and hence generate a caustic in the past lightfront $\sct^{-}(z,x)$. Also, we can point out that to first order in $\ep$, equation \eqref{hyperbolicequation} boils down to $\sct^{bdy}{}'''(x_H)=0$.} We will call such a point $x_H$ a \textit{hyperbolic point}. From the fitting of the parameters of the hyperbola \eqref{hyperbola} to $\sctbdy$ at $x_H$, we can then also read off the location of the caustic in the bulk. 
\begin{align}
x_c&=x_H+\frac{\sct^{bdy}{}'(x_H)\left(-1+\sct^{bdy}{}'(x_H)^2\right)}{\sct^{bdy}{}''(x_H)},
\\
z_c&=\frac{\left(1-\sct^{bdy}{}'(x_H)^2\right)^{3/2}}{-\sct^{bdy}{}''(x_H)},
\\
t_c&=\sct^{bdy}{}(x_H)+\frac{-1+\sct^{bdy}{}'(x_H)^2}{\sct^{bdy}{}''(x_H)}
\end{align}
The most important lesson from this is that for $\sctbdy\sim\mO(\ep)$, the $z$-coordinate of the caustic will generically be of order $\mO(1/\ep)$.

\subsection{Identifying null-null joints or ``creases"}
\label{sec::creases}

\begin{figure}[htb]
	\centering
	\def\svgwidth{0.6\columnwidth}
\executeiffilenewer{crease3.svg}{crease3.pdf}%
{inkscape -z -D --file=crease3.svg %
--export-pdf=crease3.pdf --export-latex}%
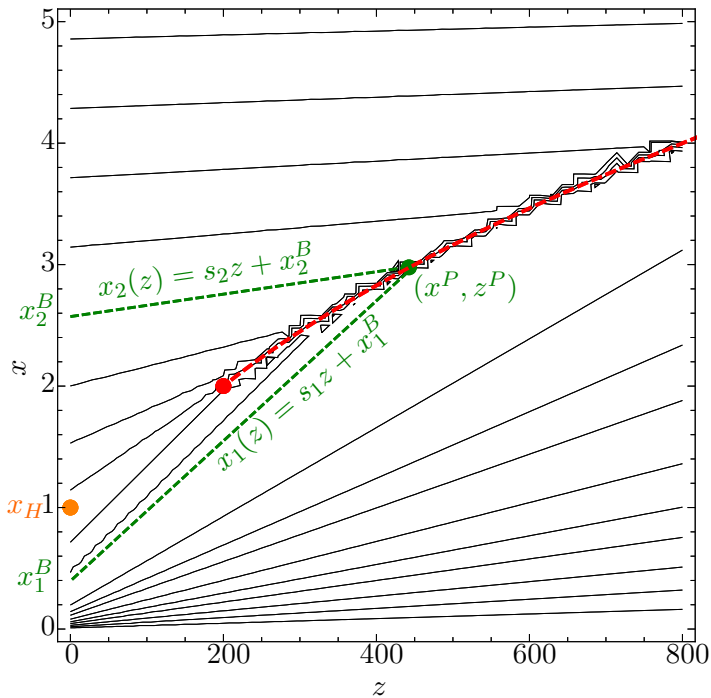%

	\caption{This figure is essentially a reproduction of the figure in the top right corner of figure \ref{fig::rays}, which depicts the situation for $\sctbdy=\frac{-0.01}{1+x^2}$. The lines are projections of the null rays forming the lightfront down to the $x,z$-plane, and should hence be perfectly straight. Any deviation from straight line behaviour is due to numerical inaccuracies. The red point is the caustic and the orange point is the hyperbolic point, both as defined in section \ref{sec::caustics}. The change compared to figure \ref{fig::rays} is that we have plotted fewer lightrays overall, and highlighted two specific lightrays emanating from the boundary points $x^B_1$ and $x^B_2$ as green dashed lines. These two lightrays meet at the same point with coordinates $(x^P,z^P)$ of the null-null joint.}
	\label{fig::crease}
\end{figure}

In this section we will explain how to analytically calculate the position of the null-null joints which where depicted as dashed red lines in figure \ref{fig::rays}. By definition, these null-null joints are spacelike curves in the lightfront on which two lightrays foliating the lightfront will meet coming from different directions (see \cite{Akers:2017nrr} for a related discussion). 
We will refer to the $x$-coordinates from which these two light-rays emanate on the boundary as $x^B_1$ and $x^B_2$, respectively. See figure \ref{fig::crease}. As is clear by the conformal flatness of the \Poincare-metric \eqref{Poincare}, lightrays in this spacetime will be straight lines in the space spanned by the coordinates $t,x,z$, and their projections to the $x,z$-plane will also be straight lines 
\begin{align}
x_{1/2}(z)=s_{1/2} z +x^B_{1/2}
\end{align}
with slopes $s_{1/2}$. These slopes are entirely determined by the function $\sctbdy$, and read
\begin{align}
s_i = \frac{\sct^{bdy}{}'(x^B_i)}{\sqrt{1-\sct^{bdy}{}'(x^B_i)^2}}.
\label{slopes}
\end{align}
This is easy to derive: It is clear that the slopes $s_i$ should be locally determined by the choice of the boundary slice, i.e.~that they will be a function of $\sct^{bdy}(x^B_i)$ and its derivatives only. For $\sctbdy=const.$, we find $s_1=0$, and for the boosted case $\sctbdy=const_1 x +const_2$ ($|const_1|<1$) it is easy to derive \eqref{slopes} explicitly from the analytical solution of \eqref{pde} which can be found in this case. Now, for general smooth $\sctbdy$, if we zoom in close enough around any $x^B_i$, the setup should be well approximated by $\sctbdy=const_1 x +const_2$, and hence \eqref{slopes} is the general result.

We will now assume that these two straight lines (projections of the two lightrays to the $x,z$-plane) cross in a point with coordinates $(x^P,z^P)$ on the $x,z$-plane. This implies the set of equations
\begin{align}
x^P-x^B_1=s_1 z^P,\ \ x^P-x^B_2=s_2 z^P,
\end{align}
which has the solution
\begin{align}
x^P=\frac{s_1x^B_2-s_2x^B_1}{s_1-s_2},\ \ z^P=\frac{x^B_2-x^B_1}{s_1-s_2}.
\end{align}
For the point at $(x^P,z^P)$ to truly lie on the crease, it is not enough that the projections of the lightrays to the $x,z$-plane meet each other at this point, the lightrays themselves also need to have the same $t$-coordinate $t^P$ there. In the three-dimensional coordinate space spanned by $t,x,z$ the slope of the lightrays is 1, i.e.~$\Delta t=\sqrt{\Delta x^2+\Delta z^2}$, and this yields the additional equation
\begin{align}
t^P=\sct^{bdy}(x^B_1)+z^P\sqrt{1+s_1^2}\equiv\sct^{bdy}(x^B_2)+z^P\sqrt{1+s_2^2}.
\label{locator}
\end{align}
This equation is important because if we could solve it, then for any given $x^B_1$ it would tell us the coordinate $x^B_2$ from which a second lightray would have to emerge from the boundary in order to intercept the ray emanating from $x^B_1$ at the crease.\footnote{An additional physical assumption $z^P>0$ has to be imposed.}
Unfortunately, for generic $\sctbdy$ this equation cannot be solved analytically. It is possible to treat \eqref{locator} perturbatively in $\ep$, but this is best done on a case by basis for $\sctbdy$. So in the following we will study a few specific examples which are of relevance in this paper.

\subsubsection*{Case $\sctbdy=\frac{\ep}{1+x^2}$}

This was plotted in the upper left corner of figure \ref{fig::rays}. By symmetry, it is obvious that the solution to \eqref{locator} is $x_2^B=-x_1^B$. Consequently, the crease can be parametrized as
\begin{align}
t^P(x^B_1)=\frac{\left(1+x^B_1{}^2\right)^2}{2\ep }+\frac{\ep }{1+x^B_1{}^2},\ 
x^P(x^B_1)=0,\ 
z^P(x^B_1)=\frac{\left(x^B_1{}^2+1\right)^2 \sqrt{1-\frac{4 \ep^2 x^B_1{}^2}{\left(x^B_1{}^2+1\right)^4}}}{2 \ep }.
\label{caus1}
\end{align} 
Taking the limit $x^B_1\rightarrow0=x_H$ reproduces the coordinates of the caustic point which we could also identify with the methods of section \ref{sec::caustics}.
So as expected we see that the creases will always emerge at a caustic point, which will have a $z$-coordinate of order $1/\ep$. It would also be possible to invert the expression $z^P(x^B_1)$ in \eqref{caus1} perturbatively in $\ep$ and then calculate $t^P(z^P)$ along the caustic perturbatively in $\ep$, however for most applications the expressions in \eqref{caus1} are sufficient, i.e.~we can view the crease as a spacelike curve parametrised by a coordinate $x^B_1\in[0,+\infty[$. 

\subsubsection*{Case $\sctbdy=\frac{-\ep}{1+x^2}$}

This was plotted in the upper right corner of figure \ref{fig::rays}. 
This case is related to the previous one in that the creases of the \textit{past} lightfront of the case $\sctbdy=\frac{\ep}{1+x^2}$ are related to the creases of the \textit{future} lightfront of this case by simple time inversion. There will now be two arc-shaped creases, one in the region $x>0$ and, by symmetry, one in the region $x<0$. We will only focus on the case $x>0$ now.
Of course $x_1^B=-x_2^B$ would still be a solution to \eqref{locator}, but one that would imply $z^P<0$. There are however also nontrivial solutions for the physical regime $z^P>0$ which can be found perturbatively in $\ep$. Assuming $x^B_2\geq x_H\geq x^B_1>0$ with $x_H\approx1-\frac{\ep ^2}{8}$\ckd being the hyperbolic point as defined in section \ref{sec::caustics}, we find
\begin{align}
x^B_2=\frac{1}{x^B_1}-\frac{\ep^2 x^B_1}{\left(x^B_1{}^2+1\right)^2}+\mO(\ep^3).
\end{align}
We then find
\begin{align}
t^P(x^B_1)&\approx\frac{\left(x^B_1{}^2+1\right)^2}{2 \ep  x^B_1{}^2}-\frac{\ep  \left(x^B_1{}^2+3\right)}{2 \left(x^B_1{}^2+1\right)},\ \ 
x^P(x^B_1)\approx x^B_1{}+\frac{1}{x^B_1{}}-\frac{\ep ^2 x^B_1{}}{\left(x^B_1{}^2+1\right)^2},\ 
\label{caus2}
\\
z^P(x^B_1)&\approx\frac{\left(x^B_1{}^2+1\right)^2}{2 \ep  x^B_1{}^2}-\frac{\ep  \left(x^B_1{}^4+2 x^B_1{}^2+3\right)}{2 \left(x^B_1{}^2+1\right)^2},
\nonumber
\end{align}
where the crease is parametrised by $x_H\geq x^B_1>0$. In fact, \eqref{caus2} parametrises both arcs of the caustic if we allow for $x_H\geq x^B_1\geq-x_H$. Curiously, we see that at the very lowest order in $\ep$, the embedding functions in \eqref{caus2} satisfy the relation $\ep z^P\approx x^P{}^2/2$, so the crease plotted in figure \ref{fig::rays} (upper right corner) is approximately a parabolic arc.

\subsubsection*{Case $\sctbdy=\frac{\ep x}{1+x^2}$}

This was plotted in the lower left corner of figure \ref{fig::rays}. We now see that there are two asymmetric creases, one in the $x>0$ region, and one in the $x<0$ region. Correspondingly, there are also two hyperbolic points $\sqrt{2}-1-\frac{1}{32} \left(3 \sqrt{2}+4\right) \ep ^2\approx x_{H+}>0>x_{H-}\approx-\sqrt{2}-1+ \frac{1}{32} \left(3 \sqrt{2}-4\right) \ep ^2$\ckd. Assuming $x_{H+}\geq x^B_2\geq x_{H-}$ and $x^B_1\in]-\infty,x_{H-}]\cup[x_{H+},+\infty[$, the perturbative solution of \eqref{locator} is 
\begin{align}
x^B_2= \frac{1-x^B_1{}}{x^B_1{}+1}-\ep\frac{x^B_1{}+1}{2 \left(x^B_1{}^2+1\right)^2}+\mO(\ep^3).
\end{align}
Consequently
\begin{align}
t^P(x^B_1)&\approx \frac{\left(x^B_1{}^2+1\right)^2}{\ep  (x^B_1{}+1)^2}-\frac{\ep  \left(x^B_1{}^2-4 x^B_1{}+1\right)}{4 \left(x^B_1{}^2+1\right)},
\\
x^P(x^B_1)&\approx \frac{x^B_1{}^2+1}{x^B_1{}+1}+\frac{\ep ^2 (x^B_1{}^2-1)}{4 \left(x^B_1{}^2+1\right)^2},\ 
\\
z^P(x^B_1)&\approx \frac{\left(x^B_1{}^2+1\right)^2}{\ep  (x^B_1{}+1)^2}-\frac{\ep  \left(x^B_1{}^4+4 x^B_1{}^2-4 x^B_1{}+3\right)}{4 \left(x^B_1{}^2+1\right)^2},
\label{caus3}
\end{align}
where the two arcs of the crease are parametrised by $x^B_1\in]-\infty,x_{H-}]$ and $x^B_1\in[x_{H+},+\infty[$. Again, as in the previous case, we can note that at the very lowest order in $\ep$, the embedding functions in \eqref{caus3} satisfy the relation $\ep z^P\approx x^P{}^2$, so the two parts of the crease plotted in figure \ref{fig::rays} (lower left corner) are approximately arcs of the same parabola.

\subsection{Auxiliary results}
\label{sec::aux}

\subsubsection*{Results for section \ref{sec::bulk}}

Here, we write down the analogue of the perturbative results \eqref{tseries}, \eqref{tseriesm} in tilded coordinates:\ckd
\begin{align}
\tilde{\sct}^\pm(\tx,\tz)=&\pm\tz+\frac{\ep}{2} \left(-\tz g_+'(\tx-\tz)+\tz g_+'(\tx+\tz)-g_+(\tx-\tz)-g_+(\tx+\tz)+2 g_+(\tx)\right)
\label{tildetseries}
\\
&\pm \ep^2\Big(
\frac{1}{2} \tz g_+(\tx) g_+''(\tx-\tz)-\frac{1}{4} \tz g_+(\tx-\tz) g_+''(\tx-\tz)-\frac{1}{4} \tz g_+(\tx+\tz) g_+''(\tx-\tz)
\nonumber
\\
&+\frac{1}{2} \tz g_+(\tx) g_+''(\tx+\tz)-\frac{1}{4} \tz g_+(\tx-\tz) g_+''(\tx+\tz)-\frac{1}{4} \tz g_+(\tx+\tz) g_+''(\tx+\tz)
\nonumber
\\
&-\frac{1}{2} \tz g_+'(\tx)^2-\frac{3}{8} \tz g_+'(\tx-\tz)^2-\frac{3}{8} \tz g_+'(\tx+\tz)^2+\frac{1}{4} \tz g_+'(\tx-\tz) g_+'(\tx+\tz)
\nonumber
\\
&-\frac{1}{2} g_+(\tx-\tz) g_+'(\tx)+\frac{1}{2} g_+(\tx+\tz) g_+'(\tx)+\frac{1}{2} g_+(\tx) g_+'(\tx-\tz)-\frac{1}{4} g_+(\tx-\tz) g_+'(\tx-\tz)
\nonumber
\\
&-\frac{1}{4} g_+(\tx+\tz) g_+'(\tx-\tz)-\frac{1}{2} g_+(\tx) g_+'(\tx+\tz)+\frac{1}{4} g_+(\tx-\tz) g_+'(\tx+\tz)
\nonumber
\\
&+\frac{1}{4} g_+(\tx+\tz) g_+'(\tx+\tz)-\frac{1}{4} \tz^2 g_+'(\tx-\tz) g_+''(\tx-\tz)+\frac{1}{4} \tz^2 g_+'(\tx+\tz) g_+''(\tx-\tz)
\nonumber
\\
&-\frac{1}{4} \tz^2 g_+'(\tx-\tz) g_+''(\tx+\tz)+\frac{1}{4} \tz^2 g_+'(\tx+\tz) g_+''(\tx+\tz)
\Big)
+\mO(\ep^3).
\nonumber
\end{align}

\subsubsection*{Results for section \ref{sec::counterjoint}}

Some useful expressions of interest in section \ref{sec::counterjoint} are
\begin{align}
\theta&=\frac{\left(x^B_1{}^2+1\right)^3}{2 \ep  \left(x^B_1{}^2-1\right)^2}
\label{thetaonjoint2a}
\\
&+\frac{\ep \left(-x^B_1{}^{14}-3 x^B_1{}^{12}-35 x^B_1{}^{10}-21 x^B_1{}^8+85 x^B_1{}^6-65 x^B_1{}^4+7 x^B_1{}^2+1\right)}{2 \left(x^B_1{}^2-1\right)^5 \left(x^B_1{}^2+1\right)^2}+\mO(\ep^3)
\nonumber
\\
\theta'&=\frac{\left(x^B_1{}^2+1\right)^3}{2 \ep  x^B_1{}^2 \left(x^B_1{}^2-1\right)^2}
\label{thetaonjoint2b}
\\
&+\frac{\ep \left(-3 x^B_1{}^{14}+7 x^B_1{}^{12}-29 x^B_1{}^{10}-11 x^B_1{}^8+67 x^B_1{}^6-87 x^B_1{}^4+21 x^B_1{}^2+3\right)}{2 \left(x^B_1{}^2-1\right)^5 \left(x^B_1{}^2+1\right)^2}+\mO(\ep^3)
\nonumber
\end{align}
for the case $\sctbdy=\frac{-\ep}{1+x^2}$ and 
\begin{align}
\theta&=\frac{\left(x^B_1{}^2+1\right)^3}{\ep  \left(x^B_1{}^2+2 x^B_1{}-1\right)^2}
+\mO(\ep^1)
\label{thetaonjoint3a}
\\
\theta'&=\frac{2 \left(x^B_1{}^2+1\right)^3}{\ep  (x^B_1{}+1)^2 \left(x^B_1{}^2+2 x^B_1{}-1\right)^2}
+\mO(\ep^1)
\label{thetaonjoint3b}
\end{align}
for the case $\sctbdy=\frac{\ep x}{1+x^2}$. Here, although the $\mO(\ep^1)$-terms might in principle be relevant, we have not explicitly given them for the sake of brevity.


\begin{thebibliography}{10}

\bibitem{2005quant.ph..2070N}
M.~A. {Nielsen}, ``{A geometric approach to quantum circuit lower bounds},''
  {\em eprint arXiv:quant-ph/0502070} (Feb., 2005) ,
  \href{http://arxiv.org/abs/quant-ph/0502070}{{\ttfamily quant-ph/0502070}}.

\bibitem{2006Sci...311.1133N}
M.~A. {Nielsen}, M.~R. {Dowling}, M.~{Gu}, and A.~C. {Doherty}, ``{Quantum
  Computation as Geometry},''
  \href{http://dx.doi.org/10.1126/science.1121541}{{\em Science} {\bfseries
  311} (Feb., 2006) 1133--1135},
  \href{http://arxiv.org/abs/quant-ph/0603161}{{\ttfamily quant-ph/0603161}}.

\bibitem{Harlow:2013tf}
D.~Harlow and P.~Hayden, ``{Quantum Computation vs. Firewalls},''
  \href{http://dx.doi.org/10.1007/JHEP06(2013)085}{{\em JHEP} {\bfseries 06}
  (2013) 085},
\href{http://arxiv.org/abs/1301.4504}{{\ttfamily arXiv:1301.4504 [hep-th]}}.

\bibitem{Susskind:2013aaa}
L.~Susskind, ``{Butterflies on the Stretched Horizon},''
\href{http://arxiv.org/abs/1311.7379}{{\ttfamily arXiv:1311.7379 [hep-th]}}.

\bibitem{Susskind:2014rva}
L.~Susskind, ``{Computational Complexity and Black Hole Horizons},''
  \href{http://dx.doi.org/10.1002/prop.201500093, 10.1002/prop.201500092}{{\em
  Fortsch. Phys.} {\bfseries 64} (2016) 44--48},
  \href{http://arxiv.org/abs/1403.5695}{{\ttfamily arXiv:1403.5695 [hep-th]}}.
[Fortsch. Phys.64,24(2016)].

\bibitem{Susskind:2018pmk}
L.~Susskind, ``{Three Lectures on Complexity and Black Holes},''
\newblock 2018.
\newblock
\href{http://arxiv.org/abs/1810.11563}{{\ttfamily arXiv:1810.11563 [hep-th]}}.
\newblock

\bibitem{Susskind:2014moa}
L.~Susskind, ``{Entanglement is not enough},''
  \href{http://dx.doi.org/10.1002/prop.201500095}{{\em Fortsch. Phys.}
  {\bfseries 64} (2016) 49--71},
\href{http://arxiv.org/abs/1411.0690}{{\ttfamily arXiv:1411.0690 [hep-th]}}.

\bibitem{Stanford:2014jda}
D.~Stanford and L.~Susskind, ``{Complexity and Shock Wave Geometries},''
  \href{http://dx.doi.org/10.1103/PhysRevD.90.126007}{{\em Phys. Rev.}
  {\bfseries D90} no.~12, (2014) 126007},
\href{http://arxiv.org/abs/1406.2678}{{\ttfamily arXiv:1406.2678 [hep-th]}}.

\bibitem{Susskind:2014jwa}
L.~Susskind and Y.~Zhao, ``{Switchbacks and the Bridge to Nowhere},''
\href{http://arxiv.org/abs/1408.2823}{{\ttfamily arXiv:1408.2823 [hep-th]}}.

\bibitem{Brown:2015bva}
A.~R. Brown, D.~A. Roberts, L.~Susskind, B.~Swingle, and Y.~Zhao,
  ``{Holographic Complexity Equals Bulk Action?},''
  \href{http://dx.doi.org/10.1103/PhysRevLett.116.191301}{{\em Phys. Rev.
  Lett.} {\bfseries 116} no.~19, (2016) 191301},
\href{http://arxiv.org/abs/1509.07876}{{\ttfamily arXiv:1509.07876 [hep-th]}}.

\bibitem{Brown:2015lvg}
A.~R. Brown, D.~A. Roberts, L.~Susskind, B.~Swingle, and Y.~Zhao,
  ``{Complexity, action, and black holes},''
  \href{http://dx.doi.org/10.1103/PhysRevD.93.086006}{{\em Phys. Rev.}
  {\bfseries D93} no.~8, (2016) 086006},
\href{http://arxiv.org/abs/1512.04993}{{\ttfamily arXiv:1512.04993 [hep-th]}}.

\bibitem{Couch:2016exn}
J.~Couch, W.~Fischler, and P.~H. Nguyen, ``{Noether charge, black hole volume,
  and complexity},'' \href{http://dx.doi.org/10.1007/JHEP03(2017)119}{{\em
  JHEP} {\bfseries 03} (2017) 119},
\href{http://arxiv.org/abs/1610.02038}{{\ttfamily arXiv:1610.02038 [hep-th]}}.

\bibitem{Jefferson:2017sdb}
R.~Jefferson and R.~C. Myers, ``{Circuit complexity in quantum field theory},''
  \href{http://dx.doi.org/10.1007/JHEP10(2017)107}{{\em JHEP} {\bfseries 10}
  (2017) 107},
\href{http://arxiv.org/abs/1707.08570}{{\ttfamily arXiv:1707.08570 [hep-th]}}.

\bibitem{Yang:2017nfn}
R.-Q. Yang, ``{Complexity for quantum field theory states and applications to
  thermofield double states},''
  \href{http://dx.doi.org/10.1103/PhysRevD.97.066004}{{\em Phys. Rev.}
  {\bfseries D97} no.~6, (2018) 066004},
\href{http://arxiv.org/abs/1709.00921}{{\ttfamily arXiv:1709.00921 [hep-th]}}.

\bibitem{Hackl:2018ptj}
L.~Hackl and R.~C. Myers, ``{Circuit complexity for free fermions},''
  \href{http://dx.doi.org/10.1007/JHEP07(2018)139}{{\em JHEP} {\bfseries 07}
  (2018) 139},
\href{http://arxiv.org/abs/1803.10638}{{\ttfamily arXiv:1803.10638 [hep-th]}}.

\bibitem{Ali:2018fcz}
T.~Ali, A.~Bhattacharyya, S.~Shajidul~Haque, E.~H. Kim, and N.~Moynihan,
  ``{Time Evolution of Complexity: A Critique of Three Methods},''
\href{http://arxiv.org/abs/1810.02734}{{\ttfamily arXiv:1810.02734 [hep-th]}}.

\bibitem{Chapman:2018hou}
S.~Chapman, J.~Eisert, L.~Hackl, M.~P. Heller, R.~Jefferson, H.~Marrochio, and
  R.~C. Myers, ``{Complexity and entanglement for thermofield double states},''
\href{http://arxiv.org/abs/1810.05151}{{\ttfamily arXiv:1810.05151 [hep-th]}}.

\bibitem{Magan:2018nmu}
J.~M. Magán, ``{Black holes, complexity and quantum chaos},''
  \href{http://dx.doi.org/10.1007/JHEP09(2018)043}{{\em JHEP} {\bfseries 09}
  (2018) 043},
\href{http://arxiv.org/abs/1805.05839}{{\ttfamily arXiv:1805.05839 [hep-th]}}.

\bibitem{Khan:2018rzm}
R.~Khan, C.~Krishnan, and S.~Sharma, ``{Circuit Complexity in Fermionic Field
  Theory},''
\href{http://arxiv.org/abs/1801.07620}{{\ttfamily arXiv:1801.07620 [hep-th]}}.

\bibitem{Caputa:2018kdj}
P.~Caputa and J.~M. Magan, ``{Quantum Computation as Gravity},''
\href{http://arxiv.org/abs/1807.04422}{{\ttfamily arXiv:1807.04422 [hep-th]}}.

\bibitem{Bhattacharyya:2018bbv}
A.~Bhattacharyya, A.~Shekar, and A.~Sinha, ``{Circuit complexity in interacting
  QFTs and RG flows},'' \href{http://dx.doi.org/10.1007/JHEP10(2018)140}{{\em
  JHEP} {\bfseries 10} (2018) 140},
\href{http://arxiv.org/abs/1808.03105}{{\ttfamily arXiv:1808.03105 [hep-th]}}.

\bibitem{Guo:2018kzl}
M.~Guo, J.~Hernandez, R.~C. Myers, and S.-M. Ruan, ``{Circuit Complexity for
  Coherent States},'' \href{http://dx.doi.org/10.1007/JHEP10(2018)011}{{\em
  JHEP} {\bfseries 10} (2018) 011},
\href{http://arxiv.org/abs/1807.07677}{{\ttfamily arXiv:1807.07677 [hep-th]}}.

\bibitem{Caputa:2017urj}
P.~Caputa, N.~Kundu, M.~Miyaji, T.~Takayanagi, and K.~Watanabe, ``{Anti-de
  Sitter Space from Optimization of Path Integrals in Conformal Field
  Theories},'' \href{http://dx.doi.org/10.1103/PhysRevLett.119.071602}{{\em
  Phys. Rev. Lett.} {\bfseries 119} no.~7, (2017) 071602},
\href{http://arxiv.org/abs/1703.00456}{{\ttfamily arXiv:1703.00456 [hep-th]}}.

\bibitem{Caputa:2017yrh}
P.~Caputa, N.~Kundu, M.~Miyaji, T.~Takayanagi, and K.~Watanabe, ``{Liouville
  Action as Path-Integral Complexity: From Continuous Tensor Networks to
  AdS/CFT},'' \href{http://dx.doi.org/10.1007/JHEP11(2017)097}{{\em JHEP}
  {\bfseries 11} (2017) 097},
\href{http://arxiv.org/abs/1706.07056}{{\ttfamily arXiv:1706.07056 [hep-th]}}.

\bibitem{Czech:2017ryf}
B.~Czech, ``{Einstein Equations from Varying Complexity},''
  \href{http://dx.doi.org/10.1103/PhysRevLett.120.031601}{{\em Phys. Rev.
  Lett.} {\bfseries 120} no.~3, (2018) 031601},
\href{http://arxiv.org/abs/1706.00965}{{\ttfamily arXiv:1706.00965 [hep-th]}}.

\bibitem{Bhattacharyya:2018wym}
A.~Bhattacharyya, P.~Caputa, S.~R. Das, N.~Kundu, M.~Miyaji, and T.~Takayanagi,
  ``{Path-Integral Complexity for Perturbed CFTs},''
  \href{http://dx.doi.org/10.1007/JHEP07(2018)086}{{\em JHEP} {\bfseries 07}
  (2018) 086},
\href{http://arxiv.org/abs/1804.01999}{{\ttfamily arXiv:1804.01999 [hep-th]}}.

\bibitem{Yang:2018nda}
R.-Q. Yang, Y.-S. An, C.~Niu, C.-Y. Zhang, and K.-Y. Kim, ``{Principles and
  symmetries of complexity in quantum field theory},''
\href{http://arxiv.org/abs/1803.01797}{{\ttfamily arXiv:1803.01797 [hep-th]}}.

\bibitem{Yang:2018tpo}
R.-Q. Yang, Y.-S. An, C.~Niu, C.-Y. Zhang, and K.-Y. Kim, ``{More on complexity
  of operators in quantum field theory},''
\href{http://arxiv.org/abs/1809.06678}{{\ttfamily arXiv:1809.06678 [hep-th]}}.

\bibitem{Lin:2018cbk}
H.~W. Lin, ``{Cayley graphs and complexity geometry},''
\href{http://arxiv.org/abs/1808.06620}{{\ttfamily arXiv:1808.06620 [hep-th]}}.

\bibitem{Chapman:2017rqy}
S.~Chapman, M.~P. Heller, H.~Marrochio, and F.~Pastawski, ``{Toward a
  Definition of Complexity for Quantum Field Theory States},''
  \href{http://dx.doi.org/10.1103/PhysRevLett.120.121602}{{\em Phys. Rev.
  Lett.} {\bfseries 120} no.~12, (2018) 121602},
\href{http://arxiv.org/abs/1707.08582}{{\ttfamily arXiv:1707.08582 [hep-th]}}.

\bibitem{Hashimoto:2017fga}
K.~Hashimoto, N.~Iizuka, and S.~Sugishita, ``{Time evolution of complexity in
  Abelian gauge theories},''
  \href{http://dx.doi.org/10.1103/PhysRevD.96.126001}{{\em Phys. Rev.}
  {\bfseries D96} no.~12, (2017) 126001},
\href{http://arxiv.org/abs/1707.03840}{{\ttfamily arXiv:1707.03840 [hep-th]}}.

\bibitem{Hashimoto:2018bmb}
K.~Hashimoto, N.~Iizuka, and S.~Sugishita, ``{Thoughts on Holographic
  Complexity and its Basis-dependence},''
  \href{http://dx.doi.org/10.1103/PhysRevD.98.046002}{{\em Phys. Rev.}
  {\bfseries D98} no.~4, (2018) 046002},
\href{http://arxiv.org/abs/1805.04226}{{\ttfamily arXiv:1805.04226 [hep-th]}}.

\bibitem{Abt:2017pmf}
R.~Abt, J.~Erdmenger, H.~Hinrichsen, C.~M. Melby-Thompson, R.~Meyer, C.~Northe,
  and I.~A. Reyes, ``{Topological Complexity in AdS$_3$/CFT$_2$},''
  \href{http://dx.doi.org/10.1002/prop.201800034}{{\em Fortsch. Phys.}
  {\bfseries 66} no.~6, (2018) 1800034},
\href{http://arxiv.org/abs/1710.01327}{{\ttfamily arXiv:1710.01327 [hep-th]}}.

\bibitem{Abt:2018ywl}
R.~Abt, J.~Erdmenger, M.~Gerbershagen, C.~M. Melby-Thompson, and C.~Northe,
  ``{Holographic Subregion Complexity from Kinematic Space},''
\href{http://arxiv.org/abs/1805.10298}{{\ttfamily arXiv:1805.10298 [hep-th]}}.

\bibitem{Yang:2018cgx}
R.-Q. Yang and K.-Y. Kim, ``{Complexity of operators generated by quantum
  mechanical Hamiltonians},''
\href{http://arxiv.org/abs/1810.09405}{{\ttfamily arXiv:1810.09405 [hep-th]}}.

\bibitem{Reynolds:2017jfs}
A.~P. Reynolds and S.~F. Ross, ``{Complexity of the AdS Soliton},''
  \href{http://dx.doi.org/10.1088/1361-6382/aab32d}{{\em Class. Quant. Grav.}
  {\bfseries 35} no.~9, (2018) 095006},
\href{http://arxiv.org/abs/1712.03732}{{\ttfamily arXiv:1712.03732 [hep-th]}}.

\bibitem{Kim:2017qrq}
R.-Q. Yang, C.~Niu, C.-Y. Zhang, and K.-Y. Kim, ``{Comparison of holographic
  and field theoretic complexities for time dependent thermofield double
  states},'' \href{http://dx.doi.org/10.1007/JHEP02(2018)082}{{\em JHEP}
  {\bfseries 02} (2018) 082},
\href{http://arxiv.org/abs/1710.00600}{{\ttfamily arXiv:1710.00600 [hep-th]}}.

\bibitem{Fu:2018kcp}
Z.~Fu, A.~Maloney, D.~Marolf, H.~Maxfield, and Z.~Wang, ``{Holographic
  complexity is nonlocal},''
  \href{http://dx.doi.org/10.1007/JHEP02(2018)072}{{\em JHEP} {\bfseries 02}
  (2018) 072},
\href{http://arxiv.org/abs/1801.01137}{{\ttfamily arXiv:1801.01137 [hep-th]}}.

\bibitem{Agon:2018zso}
C.~A. Agón, M.~Headrick, and B.~Swingle, ``{Subsystem Complexity and
  Holography},''
\href{http://arxiv.org/abs/1804.01561}{{\ttfamily arXiv:1804.01561 [hep-th]}}.

\bibitem{Flory:2018akz}
M.~Flory and N.~Miekley, ``{Complexity change under conformal transformations
  in AdS$_{3}$/CFT$_{2}$},''
\href{http://arxiv.org/abs/1806.08376}{{\ttfamily arXiv:1806.08376 [hep-th]}}.

\bibitem{Numasawa:2018grg}
T.~Numasawa, ``{Holographic Complexity for disentangled states},''
\href{http://arxiv.org/abs/1811.03597}{{\ttfamily arXiv:1811.03597 [hep-th]}}.

\bibitem{Belin:2018bpg}
A.~Belin, A.~Lewkowycz, and G.~S\'{a}rosi, ``{Complexity and the bulk volume, a
  new York time story},''
\href{http://arxiv.org/abs/1811.03097}{{\ttfamily arXiv:1811.03097 [hep-th]}}.

\bibitem{MIyaji:2015mia}
M.~Miyaji, T.~Numasawa, N.~Shiba, T.~Takayanagi, and K.~Watanabe, ``{Distance
  between Quantum States and Gauge-Gravity Duality},''
  \href{http://dx.doi.org/10.1103/PhysRevLett.115.261602}{{\em Phys. Rev.
  Lett.} {\bfseries 115} no.~26, (2015) 261602},
\href{http://arxiv.org/abs/1507.07555}{{\ttfamily arXiv:1507.07555 [hep-th]}}.

\bibitem{Moosa:2018mik}
M.~Moosa and I.~Shehzad, ``{Is volume the holographic dual of fidelity
  susceptibility?},''
\href{http://arxiv.org/abs/1809.10169}{{\ttfamily arXiv:1809.10169 [hep-th]}}.

\bibitem{Mandal:2014wfa}
G.~Mandal, R.~Sinha, and N.~Sorokhaibam, ``{The inside outs of
  AdS$_{3}$/CFT$_{2}$: exact AdS wormholes with entangled CFT duals},''
  \href{http://dx.doi.org/10.1007/JHEP01(2015)036}{{\em JHEP} {\bfseries 01}
  (2015) 036},
\href{http://arxiv.org/abs/1405.6695}{{\ttfamily arXiv:1405.6695 [hep-th]}}.

\bibitem{Chapman:2016hwi}
S.~Chapman, H.~Marrochio, and R.~C. Myers, ``{Complexity of Formation in
  Holography},'' \href{http://dx.doi.org/10.1007/JHEP01(2017)062}{{\em JHEP}
  {\bfseries 01} (2017) 062},
\href{http://arxiv.org/abs/1610.08063}{{\ttfamily arXiv:1610.08063 [hep-th]}}.

\bibitem{Bousso:1999xy}
R.~Bousso, ``{A Covariant entropy conjecture},''
  \href{http://dx.doi.org/10.1088/1126-6708/1999/07/004}{{\em JHEP} {\bfseries
  07} (1999) 004},
\href{http://arxiv.org/abs/hep-th/9905177}{{\ttfamily arXiv:hep-th/9905177
  [hep-th]}}.

\bibitem{Lehner:2016vdi}
L.~Lehner, R.~C. Myers, E.~Poisson, and R.~D. Sorkin, ``{Gravitational action
  with null boundaries},''
  \href{http://dx.doi.org/10.1103/PhysRevD.94.084046}{{\em Phys. Rev.}
  {\bfseries D94} no.~8, (2016) 084046},
\href{http://arxiv.org/abs/1609.00207}{{\ttfamily arXiv:1609.00207 [hep-th]}}.

\bibitem{Carmi:2016wjl}
D.~Carmi, R.~C. Myers, and P.~Rath, ``{Comments on Holographic Complexity},''
  \href{http://dx.doi.org/10.1007/JHEP03(2017)118}{{\em JHEP} {\bfseries 03}
  (2017) 118},
\href{http://arxiv.org/abs/1612.00433}{{\ttfamily arXiv:1612.00433 [hep-th]}}.

\bibitem{Hilbert:1915tx}
D.~Hilbert, ``{Die Grundlagen der Physik. 1.},'' {\em Gott. Nachr.} {\bfseries
  27} (1915) 395--407.
[,120(1915)].

\bibitem{1917SPAW.......142E}
A.~{Einstein}, ``{Kosmologische Betrachtungen zur allgemeinen
  Relativit{\"a}tstheorie},'' {\em Sitzungsberichte der K{\"o}niglich
  Preu{\ss}ischen Akademie der Wissenschaften (Berlin), Seite 142-152.} (1917)
  .

\bibitem{PhysRevLett.28.1082}
J.~W. York, ``Role of conformal three-geometry in the dynamics of
  gravitation,'' \href{http://dx.doi.org/10.1103/PhysRevLett.28.1082}{{\em
  Phys. Rev. Lett.} {\bfseries 28} (Apr, 1972) 1082--1085}.

\bibitem{PhysRevD.15.2752}
G.~W. Gibbons and S.~W. Hawking, ``Action integrals and partition functions in
  quantum gravity,'' \href{http://dx.doi.org/10.1103/PhysRevD.15.2752}{{\em
  Phys. Rev. D} {\bfseries 15} (May, 1977) 2752--2756}.

\bibitem{PhysRevD.47.3275}
G.~Hayward, ``Gravitational action for spacetimes with nonsmooth boundaries,''
  \href{http://dx.doi.org/10.1103/PhysRevD.47.3275}{{\em Phys. Rev. D}
  {\bfseries 47} (Apr, 1993) 3275--3280}.

\bibitem{Brill:1994mb}
D.~Brill and G.~Hayward, ``{Is the gravitational action additive?},''
  \href{http://dx.doi.org/10.1103/PhysRevD.50.4914}{{\em Phys. Rev.} {\bfseries
  D50} (1994) 4914--4919},
\href{http://arxiv.org/abs/gr-qc/9403018}{{\ttfamily arXiv:gr-qc/9403018
  [gr-qc]}}.

\bibitem{Parattu:2015gga}
K.~Parattu, S.~Chakraborty, B.~R. Majhi, and T.~Padmanabhan, ``{A Boundary Term
  for the Gravitational Action with Null Boundaries},''
  \href{http://dx.doi.org/10.1007/s10714-016-2093-7}{{\em Gen. Rel. Grav.}
  {\bfseries 48} no.~7, (2016) 94},
\href{http://arxiv.org/abs/1501.01053}{{\ttfamily arXiv:1501.01053 [gr-qc]}}.

\bibitem{Reynolds:2016rvl}
A.~Reynolds and S.~F. Ross, ``{Divergences in Holographic Complexity},''
  \href{http://dx.doi.org/10.1088/1361-6382/aa6925}{{\em Class. Quant. Grav.}
  {\bfseries 34} no.~10, (2017) 105004},
\href{http://arxiv.org/abs/1612.05439}{{\ttfamily arXiv:1612.05439 [hep-th]}}.

\bibitem{Chapman:2018dem}
S.~Chapman, H.~Marrochio, and R.~C. Myers, ``{Holographic complexity in Vaidya
  spacetimes. Part I},'' \href{http://dx.doi.org/10.1007/JHEP06(2018)046}{{\em
  JHEP} {\bfseries 06} (2018) 046},
\href{http://arxiv.org/abs/1804.07410}{{\ttfamily arXiv:1804.07410 [hep-th]}}.

\bibitem{Yang:2016awy}
R.-Q. Yang, ``{Strong energy condition and complexity growth bound in
  holography},'' \href{http://dx.doi.org/10.1103/PhysRevD.95.086017}{{\em Phys.
  Rev.} {\bfseries D95} no.~8, (2017) 086017},
\href{http://arxiv.org/abs/1610.05090}{{\ttfamily arXiv:1610.05090 [gr-qc]}}.

\bibitem{Chapman:2018lsv}
S.~Chapman, H.~Marrochio, and R.~C. Myers, ``{Holographic complexity in Vaidya
  spacetimes. Part II},'' \href{http://dx.doi.org/10.1007/JHEP06(2018)114}{{\em
  JHEP} {\bfseries 06} (2018) 114},
\href{http://arxiv.org/abs/1805.07262}{{\ttfamily arXiv:1805.07262 [hep-th]}}.

\bibitem{Jubb:2016qzt}
I.~Jubb, J.~Samuel, R.~Sorkin, and S.~Surya, ``{Boundary and Corner Terms in
  the Action for General Relativity},''
  \href{http://dx.doi.org/10.1088/1361-6382/aa6014}{{\em Class. Quant. Grav.}
  {\bfseries 34} no.~6, (2017) 065006},
\href{http://arxiv.org/abs/1612.00149}{{\ttfamily arXiv:1612.00149 [gr-qc]}}.

\bibitem{Alishahiha:2018lfv}
M.~Alishahiha, K.~Babaei~Velni, and M.~R. Mohammadi~Mozaffar, ``{Subregion
  Action and Complexity},''
\href{http://arxiv.org/abs/1809.06031}{{\ttfamily arXiv:1809.06031 [hep-th]}}.

\bibitem{Banados:1998gg}
M.~Banados, ``{Three-dimensional quantum geometry and black holes},''
  \href{http://dx.doi.org/10.1063/1.59661}{{\em AIP Conf. Proc.} {\bfseries
  484} no.~1, (1999) 147--169},
\href{http://arxiv.org/abs/hep-th/9901148}{{\ttfamily arXiv:hep-th/9901148
  [hep-th]}}.

\bibitem{Balasubramanian:1999re}
V.~Balasubramanian and P.~Kraus, ``{A Stress tensor for Anti-de Sitter
  gravity},'' \href{http://dx.doi.org/10.1007/s002200050764}{{\em Commun. Math.
  Phys.} {\bfseries 208} (1999) 413--428},
\href{http://arxiv.org/abs/hep-th/9902121}{{\ttfamily arXiv:hep-th/9902121
  [hep-th]}}.

\bibitem{Sheikh-Jabbari:2014nya}
M.~M. Sheikh-Jabbari and H.~Yavartanoo, ``{On quantization of AdS$_{3}$ gravity
  I: semi-classical analysis},''
  \href{http://dx.doi.org/10.1007/JHEP07(2014)104}{{\em JHEP} {\bfseries 07}
  (2014) 104},
\href{http://arxiv.org/abs/1404.4472}{{\ttfamily arXiv:1404.4472 [hep-th]}}.

\bibitem{Sheikh-Jabbari:2016unm}
M.~M. Sheikh-Jabbari and H.~Yavartanoo, ``{On 3d bulk geometry of Virasoro
  coadjoint orbits: orbit invariant charges and Virasoro hair on locally
  AdS$_3$ geometries},''
  \href{http://dx.doi.org/10.1140/epjc/s10052-016-4326-z}{{\em Eur. Phys. J.}
  {\bfseries C76} no.~9, (2016) 493},
\href{http://arxiv.org/abs/1603.05272}{{\ttfamily arXiv:1603.05272 [hep-th]}}.

\bibitem{Sheikh-Jabbari:2016znt}
M.~M. Sheikh-Jabbari and H.~Yavartanoo, ``{Excitation entanglement entropy in
  two dimensional conformal field theories},''
  \href{http://dx.doi.org/10.1103/PhysRevD.94.126006}{{\em Phys. Rev.}
  {\bfseries D94} no.~12, (2016) 126006},
\href{http://arxiv.org/abs/1605.00341}{{\ttfamily arXiv:1605.00341 [hep-th]}}.

\bibitem{Bhaseen:2013ypa}
M.~J. Bhaseen, B.~Doyon, A.~Lucas, and K.~Schalm, ``{Far from equilibrium
  energy flow in quantum critical systems},''
  \href{http://arxiv.org/abs/1311.3655}{{\ttfamily arXiv:1311.3655 [hep-th]}}.
[Nature Phys.11,5(2015)].

\bibitem{Erdmenger:2017gdk}
J.~Erdmenger, D.~Fernandez, M.~Flory, E.~Megias, A.-K. Straub, and
  P.~Witkowski, ``{Time evolution of entanglement for holographic steady state
  formation},'' \href{http://dx.doi.org/10.1007/JHEP10(2017)034}{{\em JHEP}
  {\bfseries 10} (2017) 034},
\href{http://arxiv.org/abs/1705.04696}{{\ttfamily arXiv:1705.04696 [hep-th]}}.

\bibitem{Maldacena:2016upp}
J.~Maldacena, D.~Stanford, and Z.~Yang, ``{Conformal symmetry and its breaking
  in two dimensional Nearly Anti-de-Sitter space},''
  \href{http://dx.doi.org/10.1093/ptep/ptw124}{{\em PTEP} {\bfseries 2016}
  no.~12, (2016) 12C104},
\href{http://arxiv.org/abs/1606.01857}{{\ttfamily arXiv:1606.01857 [hep-th]}}.

\bibitem{Moosa:2017yiz}
M.~Moosa, ``{Divergences in the rate of complexification},''
  \href{http://dx.doi.org/10.1103/PhysRevD.97.106016}{{\em Phys. Rev.}
  {\bfseries D97} no.~10, (2018) 106016},
\href{http://arxiv.org/abs/1712.07137}{{\ttfamily arXiv:1712.07137 [hep-th]}}.

\bibitem{Carmi:2017jqz}
D.~Carmi, S.~Chapman, H.~Marrochio, R.~C. Myers, and S.~Sugishita, ``{On the
  Time Dependence of Holographic Complexity},''
  \href{http://dx.doi.org/10.1007/JHEP11(2017)188}{{\em JHEP} {\bfseries 11}
  (2017) 188},
\href{http://arxiv.org/abs/1709.10184}{{\ttfamily arXiv:1709.10184 [hep-th]}}.

\bibitem{Akers:2017nrr}
C.~Akers, R.~Bousso, I.~F. Halpern, and G.~N. Remmen, ``{Boundary of the future
  of a surface},'' \href{http://dx.doi.org/10.1103/PhysRevD.97.024018}{{\em
  Phys. Rev.} {\bfseries D97} no.~2, (2018) 024018},
\href{http://arxiv.org/abs/1711.06689}{{\ttfamily arXiv:1711.06689 [hep-th]}}.

\bibitem{Fan:2018xwf}
Z.-Y. Fan and M.~Guo, ``{Holographic complexity under a global quantum
  quench},''
\href{http://arxiv.org/abs/1811.01473}{{\ttfamily arXiv:1811.01473 [hep-th]}}.

\bibitem{Chapman:2018bqj}
S.~Chapman, D.~Ge, and G.~Policastro, ``{Holographic Complexity for Defects
  Distinguishes Action from Volume},''
\href{http://arxiv.org/abs/1811.12549}{{\ttfamily arXiv:1811.12549 [hep-th]}}.

\bibitem{Flory:2017ftd}
M.~Flory, ``{A complexity/fidelity susceptibility $g$-theorem for
  AdS$_{3}$/BCFT$_{2}$},''
  \href{http://dx.doi.org/10.1007/JHEP06(2017)131}{{\em JHEP} {\bfseries 06}
  (2017) 131},
\href{http://arxiv.org/abs/1702.06386}{{\ttfamily arXiv:1702.06386 [hep-th]}}.

\bibitem{Takayanagi:2018pml}
T.~Takayanagi, ``{Holographic Spacetimes as Quantum Circuits of
  Path-Integrations},'' \href{http://dx.doi.org/10.1007/JHEP12(2018)048}{{\em
  JHEP} {\bfseries 12} (2018) 048},
\href{http://arxiv.org/abs/1808.09072}{{\ttfamily arXiv:1808.09072 [hep-th]}}.

\bibitem{Caceres:2018luq}
E.~Caceres and M.-L. Xiao, ``{Complexity-action of singular subregions},''
\href{http://arxiv.org/abs/1809.09356}{{\ttfamily arXiv:1809.09356 [hep-th]}}.

\bibitem{Bakhshaei:2019ope}
E.~Bakhshaei and A.~Shirzad, ``{Entanglement entropy and complexity of singular
  subregions in deformed CFT},''
\href{http://arxiv.org/abs/1902.02504}{{\ttfamily arXiv:1902.02504 [hep-th]}}.

\bibitem{Johnson:2003gi}
C.~V. Johnson, \href{http://dx.doi.org/10.1017/CBO9780511606540}{{\em
  {D-branes}}}.
\newblock Cambridge Monographs on Mathematical Physics. Cambridge University
  Press, 2005.
\newblock
\url{http://books.cambridge.org/0521809126.htm}.
\newblock

\bibitem{Gourgoulhon:2005ng}
E.~Gourgoulhon and J.~L. Jaramillo, ``{A 3+1 perspective on null hypersurfaces
  and isolated horizons},''
  \href{http://dx.doi.org/10.1016/j.physrep.2005.10.005}{{\em Phys. Rept.}
  {\bfseries 423} (2006) 159--294},
\href{http://arxiv.org/abs/gr-qc/0503113}{{\ttfamily arXiv:gr-qc/0503113
  [gr-qc]}}.

\bibitem{Hubeny:2007xt}
V.~E. Hubeny, M.~Rangamani, and T.~Takayanagi, ``{A Covariant holographic
  entanglement entropy proposal},''
  \href{http://dx.doi.org/10.1088/1126-6708/2007/07/062}{{\em JHEP} {\bfseries
  07} (2007) 062},
\href{http://arxiv.org/abs/0705.0016}{{\ttfamily arXiv:0705.0016 [hep-th]}}.

\bibitem{Witten:2019qhl}
E.~Witten, ``{Light Rays, Singularities, and All That},''
\href{http://arxiv.org/abs/1901.03928}{{\ttfamily arXiv:1901.03928 [hep-th]}}.

\bibitem{Hubeny:2012wa}
V.~E. Hubeny and M.~Rangamani, ``{Causal Holographic Information},''
  \href{http://dx.doi.org/10.1007/JHEP06(2012)114}{{\em JHEP} {\bfseries 06}
  (2012) 114},
\href{http://arxiv.org/abs/1204.1698}{{\ttfamily arXiv:1204.1698 [hep-th]}}.

\bibitem{Headrick:2014cta}
M.~Headrick, V.~E. Hubeny, A.~Lawrence, and M.~Rangamani, ``{Causality \&
  holographic entanglement entropy},''
  \href{http://dx.doi.org/10.1007/JHEP12(2014)162}{{\em JHEP} {\bfseries 12}
  (2014) 162},
\href{http://arxiv.org/abs/1408.6300}{{\ttfamily arXiv:1408.6300 [hep-th]}}.

\bibitem{Balasubramanian:2013lsa}
V.~Balasubramanian, B.~D. Chowdhury, B.~Czech, J.~de~Boer, and M.~P. Heller,
  ``{Bulk curves from boundary data in holography},''
  \href{http://dx.doi.org/10.1103/PhysRevD.89.086004}{{\em Phys. Rev.}
  {\bfseries D89} no.~8, (2014) 086004},
\href{http://arxiv.org/abs/1310.4204}{{\ttfamily arXiv:1310.4204 [hep-th]}}.

\bibitem{Czech:2014ppa}
B.~Czech and L.~Lamprou, ``{Holographic definition of points and distances},''
  \href{http://dx.doi.org/10.1103/PhysRevD.90.106005}{{\em Phys. Rev.}
  {\bfseries D90} (2014) 106005},
\href{http://arxiv.org/abs/1409.4473}{{\ttfamily arXiv:1409.4473 [hep-th]}}.

\bibitem{Myers:2014jia}
R.~C. Myers, J.~Rao, and S.~Sugishita, ``{Holographic Holes in Higher
  Dimensions},'' \href{http://dx.doi.org/10.1007/JHEP06(2014)044}{{\em JHEP}
  {\bfseries 06} (2014) 044},
\href{http://arxiv.org/abs/1403.3416}{{\ttfamily arXiv:1403.3416 [hep-th]}}.

\bibitem{Headrick:2014eia}
M.~Headrick, R.~C. Myers, and J.~Wien, ``{Holographic Holes and Differential
  Entropy},'' \href{http://dx.doi.org/10.1007/JHEP10(2014)149}{{\em JHEP}
  {\bfseries 10} (2014) 149},
\href{http://arxiv.org/abs/1408.4770}{{\ttfamily arXiv:1408.4770 [hep-th]}}.

\bibitem{Bousso:2015mna}
R.~Bousso, Z.~Fisher, S.~Leichenauer, and A.~C. Wall, ``{Quantum focusing
  conjecture},'' \href{http://dx.doi.org/10.1103/PhysRevD.93.064044}{{\em Phys.
  Rev.} {\bfseries D93} no.~6, (2016) 064044},
\href{http://arxiv.org/abs/1506.02669}{{\ttfamily arXiv:1506.02669 [hep-th]}}.

\end{thebibliography}

\providecommand{\href}[2]{#2}\begingroup\raggedright\endgroup

\end{document}